\documentclass[a4paper,11pt]{article}
\pdfoutput=1 

\usepackage{jheppub} 

\usepackage[T1]{fontenc} 
\usepackage{braket}
\newcommand{\Tr}{\text{Tr}}
\usepackage{tikz}
\usepackage{bbm}
\usetikzlibrary{decorations.markings}

\newcommand{\pra}{\text{Physical Review A}}

\newcommand{\prd}{\text{Physical Review D}}
\newcommand{\pre}{\text{Physical Review E}}

\newcommand{\prl}{\text{Physical Review Letters}}
\title{Distinguishing Random and Black Hole Microstates
}

\author[a]{Jonah Kudler-Flam,}
\author[b]{Vladimir Narovlansky,}
\author[b,c]{Shinsei Ryu}

\affiliation[a]{Kadanoff Center for Theoretical Physics, University of Chicago, Chicago, IL~60637, USA}
\affiliation[b]{Princeton Center for Theoretical Science,
Princeton University, Princeton, NJ 08544, USA}
\affiliation[c]{Department of Physics, Princeton University, Princeton, NJ 08544, 08540, USA}
\emailAdd{jkudlerflam@uchicago.edu}
\emailAdd{narovlansky@princeton.edu}
\emailAdd{shinseir@princeton.edu}

\abstract{This is an expanded version of the short report [Phys.~Rev.~Lett.~126, 171603 (2021)], where the relative entropy was used to distinguish random states drawn from the Wishart ensemble as well as black hole microstates. In this work, we expand these ideas by computing many generalizations including the Petz R\'enyi relative entropy, sandwiched R\'enyi relative entropy, fidelities, and trace distances. These generalized quantities are able to teach us about new structures in the space of random states and black hole microstates where the von Neumann and relative entropies were insufficient. We further generalize to generic random tensor networks where new phenomena arise due to the locality in the networks. These phenomena sharpen the relationship between holographic states and random tensor networks. We discuss the implications of our results on the black hole information problem using replica wormholes, specifically the state dependence (hair) in Hawking radiation. Understanding the differences between Hawking radiation of distinct evaporating black holes is an important piece of the information problem that was not addressed by entropy calculations using the island formula. We interpret our results in the language of quantum hypothesis testing and the subsystem eigenstate thermalization hypothesis (ETH), deriving that chaotic (including holographic) systems obey subsystem ETH for all subsystems less than half the total system size.}

\begin{document} 
\maketitle
\flushbottom

\section{Introduction and summary of results}

A unifying idea spanning quantum information theory, quantum chaos and thermalization, and black hole physics is that of (in)distinguishability of quantum states. In quantum information theory, we would like to understand what the space of quantum states are. In particular, how can we characterize which states are close or far away and endow the Hilbert space with a geometry?
This notion of distinguishability is critical for storing and processing quantum information.

Quantum chaos and thermalization is all about distinguishibility. A natural \textit{definition} of quantum thermalization is that the state is indistinguishable (up to a certain error) from a completely thermal e.g.~Gibbs state. It is then important to characterize which systems thermalize and the mechanism for thermalization to occur. We can begin with two states that are easily distinguishable e.g.~the ``all spin up'' and ``all spin down'' states of a quantum spin chain. If we evolve these states with a thermalizing Hamiltonian, the states will become indistinguishable using ``simple'' measurements.

Similarly, the black hole information problem is most naturally framed in terms of thermalization and indistinguishability. Black holes can be formed in many different ways. Moreover, they have an extraordinary number of microstates \cite{Bekenstein:1973ur,cmp/1103899181}. Even so, using semiclassical calculations, Hawking showed that all black holes with identical thermodynamic quantities (mass, charge, and angular momentum) will radiate \textit{thermal} radiation \cite{cmp/1103899181}. This means that at late times, after the black hole has evaporated, all of these microstates are completely indistinguishable, which is in sharp tension with the unitarity of quantum mechanics. To resolve this apparent paradox, different black hole microstates must be made distinguishable directly from the radiation.

Remarkably, all three of these broad problems may be addressed using random matrix theory calculations of distinguishability measures. The purpose of this paper is to make this statement precise and elucidate the intricate and surprising connections, extending the analysis of Ref.~\cite{2021PhRvL.126q1603K}. 

The techniques of random matrix theory have become ubiquitous across far-ranging fields of physics. Originally used to characterize the spectra of heavy nuclei \cite{10.2307/1970079}, random matrix theory has flourished in its applications in quantum information theory \cite{2016JMP....57a5215C}, quantum chaos and thermalization \cite{2016AdPhy..65..239D}, and black hole physics \cite{1993PhRvL..71.3743P,2007JHEP...09..120H}. What's more is that these fields are now understood to be deeply related to one another and, to some extent, inseparable. 

In Section \ref{disting_sec}, we lay the foundation by reviewing the precise definitions of distinguishability in quantum information theory. In particular, we review various distinguishability measures that are used to characterize how well different states can be discriminated between. This is formalized by the operational tasks of \textit{quantum hypothesis testing} and \textit{state discrimination}. These are the most fundamental information processing tasks and make precise the operational meaning of our subsequent results.

In Section \ref{randomstate_sec}, we undertake our main technical computations. We introduce the ensemble of random mixed states (Wishart ensemble) and a diagrammatic approach in evaluating moments of the reduced density matrices. We exactly compute the relative entropy, Petz R\'enyi relative entropies, sandwiched R\'enyi relative entropies, 
fidelities, and trace distance of random states in the limit of large Hilbert space dimensions. This characterizes the space of generic quantum states. In particular, we find that when the logarithm of the dimension of the sub-Hilbert space that we consider is less than half of the total Hilbert space, then generic states are indistinguishable up to exponentially small terms in the system size. When the sub-Hilbert space is larger, we find the states are completely distinguishable up to exponentially small terms. We find interesting $O(1)$ crossover behavior. We compare these ``large-$N$'' results to finite size numerics and find precise agreement.

In Section \ref{black_hole_sec}, we begin to interpret the random matrix theory results in the language of gravity. First, we show that in AdS/CFT, if one considers two different black hole microstates, the evaluation of distinguishability measures between the black holes microstates
when an observer only has access to a subregion of the boundary is formally identical to the formulas present in Section \ref{randomstate_sec}. This occurs in special states called ``fixed-area states'' \cite{2019JHEP...10..240D,2019JHEP...05..052A}. 
With this realization, we characterize the distinguishability of black hole microstates in Anti deSitter space or equivalently, high-energy states in conformal field theories. We subsequently apply this formalism to a toy model of an evaporating black hole \cite{2019arXiv191111977P}.
We conclude that before the Page time, an observer of an evaporating black hole is only able to distinguish different black holes microstates
if one has an $O(e^{1/G_N})$ number of copies of the radiation, meaning that
the microstates are nearly completely indistinguishable. After the Page time, an observer of an evaporating black hole can easily distinguish microstates with a single copy of the radiation, though the needed measurement will be quite complex. In these calculations, replica wormholes play a central role.

In Section \ref{sec_tensor}, we generalize our computations to random tensor networks. These represent new ensembles of random matrix theory and introduce a notion of locality into the quantum state. We find qualitatively new features in distinguishability with larger tensor networks being more distinguishable than smaller tensor networks and the Haar random states of Section \ref{randomstate_sec}.

In Section \ref{sETH_sec}, we discuss thermalization in chaotic quantum many-body systems. Using an ansatz for the structure of high energy eigenstates in chaotic systems \cite{2010NJPh...12g5021D,2019PhRvE.100b2131M,2017arXiv170908784L}, 
we evaluate the various distinguishability measures. We interpret these results using the subsystem eigenstate thermalization hypothesis \cite{2018PhRvE..97a2140D}, a very strong version of thermalization. We determine that chaotic systems obeying the ansatz also obey subsystem ETH for subsystem sizes that are less than
half the total system, 
after which subsystem ETH is violated. Furthermore, we find similar structures in gravity, generalizing the calculations of Section \ref{black_hole_sec} to black hole microstates without fixed-areas. Analogous conclusions apply and we conclude that holographic CFTs in generic dimensions obey subsystem ETH.

We relegate certain details and extensions to the appendices including alternative derivations using free probability theory in Appendix \ref{free_prob_app}.

\section{Distinguishability measures and their use}
\label{disting_sec}

\subsection{Review of distinguishibility measures}


In this section, we review various distinguishability measures commonly used in quantum information theory. Each measure has an operational meaning and there are various relations between the measures. Readers familiar with distinguishability measures and hypothesis testing may move to Section \ref{randomstate_sec} as there are no new results in this section. 

\paragraph{Relative entropy}
We begin with the quantum relative entropy which is arguably the most important quantity in quantum information theory as many of the deepest results in the field are directly derivable from its fundamental properties. The classical relative entropy or Kullback–Leibler divergence is defined as
\begin{align}
    D_{KL}(P||Q):= \log\left[ \sum_{x\in X} P(x)\log \left( \frac{P(x)}{Q(x)}\right)\right],
\end{align}
where $P$ and $Q$ are classical probability distributions over a set $X$.
The quantum relative entropy is the noncommutative analog defined for two density matrices, $\rho$ and $\sigma$ as
\begin{align}
    D(\rho || \sigma) := \Tr \left[\rho \log\left[ \rho\right] - \rho \log \left[\sigma\right]\right].
\end{align}
This is only well-defined when the support of $\rho$ is contained within the support of $\sigma$. Otherwise, the relative entropy is infinite. 

The relative entropy acts as a distinguishability measure as can be seen from its basic properties. The first is positivity, $D(\rho || \sigma)  \geq  0$, with the inequality saturated if and only if $\rho = \sigma$. The second is referred to as the data processing inequality or monotonicity of relative entropy\footnote{Positivity can actually be derived from monotonicity, though we choose to separate these conditions for added clarity.} which states that the relative entropy is non-increasing under completely-positive trace-preserving (CPTP) quantum channels, $\mathcal{N}$ \cite{lindblad1975}
\begin{align}
    D(\mathcal{N}(\rho)|| \mathcal{N}(\sigma)) \leq D(\rho|| \sigma) .
\end{align}
This property is crucial for a distinguishability measure because it asserts that if you are given two quantum states, after performing operations on them, they can never become easier to distinguish.

A particularly important quantum channel is the partial trace operation on a bipartite Hilbert space
\begin{align}
    \mathcal{N}: \mathcal{H}_A \otimes \mathcal{H}_B &\rightarrow \mathcal{H}_A
    \\
    \rho &\mapsto \Tr_B \left[\rho\right]:=\rho_A.
\end{align}
Under the partial trace, we lose all information about region $B$, making $\rho$ harder to distinguish from other states that look similar on $A$. The partial trace will play a central role throughout the rest of the paper because we are generally interested in how to distinguish states when only having access to a subregion.

While the relative entropy characterizes the structure of the space of quantum states, importantly, it is not a metric. This is most obviously seen from the definition which is not symmetric under exchange of $\rho$ and $\sigma$. This is a feature and not a bug as can be seen by its operational meaning that we will soon explore.

The relative entropy is a parent quantity to many other central information-theoretic quantities, such as the von Neumann entropy
\begin{align}
    S_{vN}(\rho_A) = \log \left[d_A\right] - D\left(\rho_A || \frac{\mathbbm{1}}{d_A}\right),
\end{align}
where $d_A$ is the Hilbert space dimension, the mutual information
\begin{align}
    I(A,B) = D(\rho_{AB}|| \rho_A \otimes \rho_B),
\end{align}
and conditional entropy
\begin{align}
    S(B|A) = \log \left[d_B\right] - D\left(\rho_{AB} || \rho_A \otimes \frac{\mathbbm{1}}{d_B}\right).
\end{align}
In these terms, the strong subadditivity of von Neumann entropy
\begin{align}
    S_{vN}(\rho_B) + S_{vN}(\rho_{ABC})\leq S_{vN}(\rho_{AB}) + S_{vN}(\rho_{BC})
\end{align}
is a straightforward consequence of the data processing inequality
\begin{align}
    D(\Tr_C\left[ \rho_{ABC}\right] || \Tr_C\left[\rho_A \otimes \rho_{BC}\right]) \leq D( \rho_{ABC} || \rho_A \otimes \rho_{BC}) .
\end{align}

\paragraph{R\'enyi relative entropies} Like the Kullback-Leibler divergence, the relative entropy can be generalized
into R\'enyi relative entropies. However, because of the noncommutativity of density matrices, there are many inequivalent ways to generalize the relative entropy such that it reduces to the classical $\alpha$-R\'enyi divergences, the unique set of quantities satisfying the five axioms of a generalized divergence \cite{10020820209}
\begin{align}
    D_{KL,\alpha}(P||Q):= \frac{1}{\alpha -1 }\log\left[ \sum_{x\in X} P(x)^{\alpha} Q(x)^{1-\alpha}\right],
    \label{renyiKL}
\end{align}
where $\alpha$ is a positive semi-definite real variable.
We will study two complementary families which have served the most uses in quantum information theory. 

The first is the most obvious quantum analog of \eqref{renyiKL} and is referred to as the Petz R\'enyi relative entropy (PRRE) \cite{1986RpMP...23...57P}
\begin{align}
    D_{\alpha}(\rho || \sigma) := \frac{1}{\alpha - 1}\log\left[ \Tr \left[\rho^{\alpha} \sigma^{1-\alpha} \right]\right].
\end{align}
The PRRE satisfies various nice properties, such as reduction to the von Neumann relative entropy when $\alpha \rightarrow 1$. For $\alpha \in [0,1)$, the PRRE is finite even when the support of $\rho$ is larger than the support of $\sigma$. Most importantly, the PRRE satisfies the data processing inequality when $\alpha \in [0, 2]$ \cite{LIEB1973267,cmp/1103900757,1986RpMP...23...57P}. One particularly useful case is at $\alpha = 1/2$, which defines what has been called Holevo's ``just-as-good fidelity'' \cite{2018arXiv180102800W} or affinity \cite{1972TMP....13.1071K}
\begin{align}
    F_H(\rho || \sigma) := \left(\Tr\left[\sqrt{\rho} \sqrt{\sigma} \right]\right)^2 =  e^{-D_{1/2}(\rho || \sigma)},
\end{align}
which, for most purposes is just as (if not more) useful as the more widely used Uhlmann fidelity
\begin{align}
    F(\rho || \sigma) := \left(\Tr \left[\sqrt{\sqrt{\sigma} \rho \sqrt{\sigma}}\right]\right)^2.
\end{align}
Both satisfy all of Jozsa's axioms for distinguishability measures \cite{doi:10.1080/09500349414552171} and define metrics on the space of quantum states
\begin{align}
    D_B(\rho || \sigma) &:= \sqrt{2\left(1-\sqrt{F(\rho ||\sigma )}\right)}, \nonumber
    \\
    D_A(\rho || \sigma) &:= \arccos\left[\sqrt{F(\rho ||\sigma )}\right], \nonumber
    \\
    D_H(\rho || \sigma) &:= 2\left(1-\sqrt{F_H(\rho ||\sigma )}\right),
\end{align}
called the Bures distance, Bures angle, and Hellinger distance repectively.

The other quantum generalization of \eqref{renyiKL} we will study is the sandwiched R\'enyi relative entropy\footnote{We note that both PRRE and SRRE can be described as specific cases of the $\alpha$-$z$-relative entropies defined as \cite{2015JMP....56b2202A}
\begin{align}
    D_{\alpha,z}(\rho || \sigma) :=\frac{1}{\alpha - 1}\log \left[\Tr\left[\left(\sigma^{\frac{1-\alpha}{2z}} \rho^{\frac{\alpha}{z}} \sigma^{\frac{1-\alpha}{2z}} \right)^z\right] \right],
\end{align}
though we will not discuss this more general quantity.
} (SRRE) \cite{2013JMP....54l2203M,2014CMaPh.331..593W}
\begin{align}
    \tilde{D}_{\alpha}(\rho || \sigma) := \frac{1}{\alpha - 1}\log\left[ \Tr\left[ \left(\sigma^{\frac{1-\alpha}{2\alpha}}\rho\sigma^{\frac{1-\alpha}{2\alpha}} \right)^{\alpha}\right]\right].
\end{align}
It is clear that this is equivalent to the PRRE when $\rho$ and $\sigma$ commute and reduces to the Uhlmann fidelity at $\alpha = 1/2$. Like the PRRE, the SRRE reduces to the von Neumann relative entropy in the $\alpha \rightarrow 1$ limit and is only finite if either $\alpha \in [0,1)$ or the support of $\rho$ is contained within the support of $\sigma$. The most important property of SRRE is that it satisfies the data-processing inequality for $\alpha \in [1/2, \infty)$. In this way, it is complementary to the PRRE. Similar formulas for R\'enyi analogs of entropy, mutual information, and conditional entropies can be written in terms of the R\'enyi relative entropies.


\paragraph{Trace distance} The final distinguishability measure that we will study is the trace distance, defined as 
\begin{align}
    T(\rho || \sigma) := \frac{1}{2}\left|\rho-\sigma\right|_1,
    \label{trace_norm_def}
\end{align}
where $|\cdot|_1$ is the trace norm. The trace distance defines a metric on the space of quantum states and takes values between zero and one. However, unlike Holevo's just-as-good and Uhlmann fidelities, it does not descend from a relative entropy. The trace distance is monotonically decreasing under quantum operations. It will play a central role in our discussion of eigenstate thermalization in Section \ref{sETH_sec}.

There are various useful relations between the above distinguishability measures that we now list. First, we note that both PRRE and SRRE are monotonic in $\alpha$
\begin{align}
    {D}_{\alpha_1}(\rho || \sigma) \leq {D}_{\alpha_2}(\rho || \sigma), \quad \tilde{D}_{\alpha_1}(\rho || \sigma) \leq \tilde{D}_{\alpha_2}(\rho || \sigma), \quad \alpha_1 < \alpha_2,
    \label{RRE_ordering}
\end{align}
while the SRRE lower bounds the PRRE
\begin{align}
    \tilde{D}_{\alpha}(\rho || \sigma)\leq {D}_{\alpha}(\rho || \sigma), \quad \alpha \geq 0.
\end{align}
By Pinsker's inequality, the von Neumann relative entropy upper bounds the trace distance \cite{ohya2004quantum}
\begin{align}
      \frac{1}{2}T(\rho|| \sigma)^2 \leq D(\rho \lvert \rvert\sigma) ,
      \label{pinsker_eq}
\end{align}
while the Fuchs-van de Graaf inequalities assert that both fidelities place upper and lower bound the trace distance \cite{1997quant.ph.12042F}
\begin{align}
    1-\sqrt{F(\rho ||\sigma)}\leq 1-\sqrt{F_H(\rho ||\sigma)}\leq T(\rho ||\sigma)
 \leq \sqrt{1-{F(\rho ||\sigma)}}\leq \sqrt{1-{F_H(\rho ||\sigma)}}.
\end{align}
These are strong results that we will use throughout the paper due to the difficulty in directly computing the trace distance. They are also important, nontrivial consistency checks of our results.

\subsection{Operational interpretations in hypothesis testing}

The most fundamental information processing processes are quantum state discrimination (QSD) and hypothesis testing (QHT). It should then be no surprise that this is where the most fundamental quantities, relative entropy and trace distance, find their operational meanings. In this section, we make precise what it means for states to be distinguishable by first introducing QSD and QHT, then stating what the distinguishability measures say about our ability to perform these tasks. For more details, we refer the reader to the literature e.g.~Refs.~\cite{Hayashi:1338967,2020arXiv201104672K}.

The general set-up is that we are given a state on $\mathcal{H}$ that is either $\rho$ or $\sigma$ and we wish to determine which state we were given. We are allowed to use any positive operator-valued measure (POVM) 
which is a collection of positive semi-definite operators, $\{ M_i\}$, that sum to the identity operator on $\mathcal{H}$. Each subscript, $i$, corresponds to a measurement outcome. Because we are looking for a binary outcome (is our state $\rho$ or $\sigma$?), we can consolidate the $M_i$'s into just two elements. 
For outcomes $i\in \mathcal{A}$, we conclude the state is $\rho$, while for outcomes $i \notin \mathcal{A}$, we conclude the state is $\sigma$. Our POVM is then $\{ A, \mathbbm{1}-A\}$ where $A := \sum_{i\in\mathcal{A}} M_i $. There are many choices for $A$ and we want to optimize this choice as to have the least error in our conclusions. There are two types of errors. The probability of mistakenly concluding that we have $\sigma$ when we were really given $\rho$ is given by
\begin{align}
    \alpha(A):= \Tr \left[ (\mathbbm{1} - A)\rho\right], 
\end{align}
while the probability of mistakenly concluding that we have $\rho$ when we were really given $\sigma$ is given by
\begin{align}
    \beta(A):= \Tr \left[ A \sigma \right].
\end{align}
These are referred to as the error probabilities of the first and second kind respectively (or type I and II).

There are various ways of optimizing these errors\footnote{Recently, an interpolation QSD and QHT has bee introduced \cite{2021arXiv210409553S}. We comment on this interpolation in Appendix \ref{interp_app}.}. The symmetric way is called state discrimination. The smallest combined error is given by the trace distance between the states \cite{1969JSP.....1..231H, Helstrom:110988}
\begin{align}
    \min_{A}\left[ \alpha(A) + \beta(A) \right] = 1 - T(\rho || \sigma),
    \label{trace_distance_operational}
\end{align}
where the optimization is taken over all POVM. If the trace distance is very large (close to one), we are able to choose a POVM that has very small error probabilities. If the trace distance is small (close to zero), then the combined error is close to one, the maximal optimized error which can be saturated by taking $A = \mathbbm{1}$. Likewise, the probability that we correctly discriminate, $P_+$, is also given by the trace distance
\begin{align}
     P_+(A) := \frac{1}{2}\max_A\left[\Tr \left[ A \rho \right] + \Tr \left[ (\mathbbm{1} - A)\sigma\right]\right] = \frac{1}{2} \left(1+ T(\rho || \sigma)\right).
\end{align}

State discrimination can be made easier if instead of given one copy of the state, we are given multiple, $n$, copies. This is the topic of asymptotic state discrimination. With these $n$ copies, we can ask what is the optimal POVM on $\mathcal{H}^{\otimes n}$. The error probabilities are generalized in the obvious way
\begin{align}
    \alpha_n(A):= \Tr \left[ (\mathbbm{1} - A)\rho^{\otimes n}\right], \quad \beta_n(A):= \Tr \left[ A \sigma^{\otimes n} \right].
\end{align}
The sum of the errors can be shown to be bounded above by Holevo's just-as-good fidelity
\begin{align}
    \min_{A}\left[ \alpha_n(A) + \beta_n(A) \right] \leq  F_H(\rho || \sigma)^{\frac{n}{2}}.
\end{align}
Unless the states are identical ($F_H = 1$), the error rate exponentially decays to zero as we are given a large number of copies. If the fidelity is small, we may only need one (or very few) copies to confidently discriminate the states. Asymptotically ($n \rightarrow \infty$), this is strengthened to an equality by the \textit{quantum Chernoff bound} \cite{2007PhRvL..98p0501A,2006quant.ph..7216N}
\begin{align}
    \lim_{n\rightarrow \infty}- \frac{\log\left[\min_{A}\left[ \alpha_n(A) + \beta_n(A) \right]\right] }{n}=\max_{\alpha \in (0, 1)}(1-\alpha)D_{\alpha}(\rho || \sigma):= \xi(\rho||\sigma).
    \label{chernoff_bound}
\end{align}
The quantity on the right-hand side of this equation is called the \textit{quantum Chernoff distance}.

We progress to the asymmetric treatment of this problem, quantum hypothesis testing. The asymmetric optimization is the task of minimizing one of the errors while keeping the other error below some fixed, finite threshold $\epsilon$. We define
\begin{align}
    \alpha_n^*(\epsilon) := \min_{\beta_n(A) \leq \epsilon}\left[\alpha_n(A) \right], \quad \beta_n^*(\epsilon) := \min_{\alpha_n(A) \leq \epsilon}\left[\beta_n(A) \right].
\end{align}
Quantum Stein's Lemma \cite{cmp/1104248844,2005atqs.book...28O} asserts that for \textit{any} $\epsilon \in (0,1)$, the type II error decreases exponentially with the rate given by the relative entropy
\begin{align}
    \lim_{n\rightarrow \infty} -\frac{\log\left[ \beta_n^*(\epsilon)\right]}{n} = D(\rho || \sigma).
\end{align}
Quantum Stein's Lemma can be further refined to optimize the error of the first kind assuming the error of the second kind decays exponentially. Defining
\begin{align}
    \tilde{\alpha}_{n,r} := \min_{\beta_n(A) \leq e^{-nr}}\left[\alpha_n(A) \right],
\end{align}
the PRRE determines 
this error rate if $r < D(\rho || \sigma)$ \cite{2007PhRvA..76f2301H,2006quant.ph.11289N,2009arXiv0912.1286M}
\begin{align}
    \lim_{n \rightarrow \infty} -\frac{\log\left[\tilde{\alpha}_{n,r}\right]}{n} = \max_{\alpha \in (0,1)}\left[ \frac{\alpha - 1}{\alpha} \left( r- D_{\alpha}(\rho || \sigma)\right)\right],
\end{align}
while SRRE determines 
this error rate if $r > D(\rho || \sigma)$ \cite{2015CMaPh.334.1617M}
\begin{align}
    \lim_{n \rightarrow \infty} -\frac{\log\left[1-\tilde{\alpha}_{n,r}\right]}{n} = \max_{\alpha \in (1,\infty)}\left[ \frac{\alpha - 1}{\alpha} \left( r- \tilde{D}_{\alpha}(\rho || \sigma)\right)\right].
\end{align}


With the above review, we have a thorough understanding of how to quantify the ability to discriminate between \textit{two} quantum states. It would be desirable to generalize this to an arbitrary, finite number of states $\{ \rho_i\}$. As we will see in section \ref{black_hole_sec}, this is particularly important for the black hole information problem. In the case that we are discriminating many states, we no longer consolidate the $M_i$'s into $A$ and $\mathbbm{1} - A$. Rather, each measurement outcome $i$ can lead us to conclude that we have state $\rho_i$. If we are given the state $\rho_i$ with probability $p_i$, the error probability is given by
\begin{align}
    P_{err}(\rho_i,M_i) := \sum_{i = 1}\Tr\left[p_i \rho_i(\mathbbm{1} -M_i) \right],
\end{align}
whose optimized value we define as
\begin{align}
    P_{err}^*(\rho_i) :=\min_{\{M_i\}}\left[P_{err}(\rho_i,M_i)\right].
\end{align}
Rather remarkably, building on the work of Refs.~\cite{2010JMP....51g2203N, 10.1007/978-3-642-18073-6_1, 2011arXiv1112.1529N,2014JMP....55j2201A}, the quantum Chernoff bound was generalized in the multiple state case, referred to as the \textit{multiple quantum Chernoff bound}\footnote{The quantum Sanov’s lemma provides the analogous asymmetric multiple state hypothesis testing result \cite{2002JPhA...3510759H,2005CMaPh.260..659B}. We also note intriguing new multiple state divergences obeying the data processing inequality whose operational meaning is not yet fully understood \cite{2021arXiv210309893F}.} \cite{2015arXiv150806624L}
\begin{align}
    \lim_{n\rightarrow \infty}-\frac{\log\left[ P_{err}^*(\rho_i) \right]}{n} = \min_{i\neq j}\left[\max_{\alpha \in (0,1)}(1-\alpha )D_{\alpha}(\rho_i||\rho_j)\right].
    \label{mult_chernoff}
\end{align}
The value on the right hand side is referred to as the \textit{multiple quantum Chernoff distance}. When comparing to \eqref{chernoff_bound}, it is surprising that when discriminating between arbitrarily many more states, all one needs to do is apply a global minimum.

In the one-shot case, bounds can be placed on $P_{err}^*(\rho_i)$, though, to our knowledge, an equality is not known. If we take the spectral decompositions of our POVM as $M_i := \sum_i^{T_i}\lambda_{ik}Q_{ik}$, an upper and  bound is given by \cite{2015arXiv150806624L}
\begin{align}
    \frac{\sum_{i<j}\sum_{k,l}\min\left[\lambda_{ik},\lambda_{jl} \right]\Tr \left[Q_{ik} Q_{jl} \right]}{2(r-1)} &\leq P_{err}^*(\rho_i)\nonumber
    \\
    &\leq 10(r-1)^2 T^2\sum_{i<j}\sum_{k,l}\min\left[\lambda_{ik},\lambda_{jl} \right]\Tr \left[Q_{ik} Q_{jl} \right],
\end{align}
where $r$ is the total number of states and $T:= \max\left[T_i\right]$. The upper bound can be made more intuitive, though generally weaker, by noting
\begin{align}
    \sum_{k,l}\min\left[\lambda_{ik},\lambda_{jl} \right]\Tr \left[Q_{ik} Q_{jl} \right] \leq \min_{\alpha \in (0,1)}(1-\alpha )D_{\alpha}(\rho_i ||\rho_j),
\end{align}
leading to 
\begin{align}
    P_{err}^*(\rho_i) &\leq 10(r-1)^2 T^2\sum_{i<j}\min_{\alpha \in (0,1)}(1-\alpha )D_{\alpha}(\rho_i ||\rho_j)
    \nonumber
    \\
    &\leq 5(r-1)^3r T^2\max_{i\neq j}\left[ \min_{\alpha \in (0,1)}(1-\alpha )D_{\alpha}(\rho_i ||\rho_j)\right],
\end{align}
where in the second line, we have removed the remaining sum to mimic the form of \eqref{mult_chernoff} even though this formula is strictly weaker when setting $r = 2$.
In passing, we note that determining $P^*_{err}$ is a computation can be formulated as a semi-definite program \cite{1055351, 2020arXiv201104672K}, which means that it may be efficiently evaluated.

\section{Distinguishing random states}

\label{randomstate_sec}

In this section, we consider Haar random states. This ensemble can be described in several ways. Perhaps the simplest is to consider an arbitrary reference state $\ket{0} \in \mathcal{H}$ and act with a random unitary matrix drawn from the Haar measure, the unique left-right invariant measure over $U(\mathrm{dim}\, \mathcal{H})$
: $\ket{\Psi }= U\ket{0}$. This ensemble is particularly nice because the averages over the $\alpha$ copies of Haar random states are sums of permutations, $\tau$, of the $\alpha$ copies
\begin{align}
    \overline{\ket{\Psi}\bra{\Psi}^{\otimes \alpha}} = \frac{\sum_{\tau \in S_\alpha} g_{\tau}}{\sum_{\tau \in S_\alpha} \Tr\left[g_\tau\right]},
    \label{haar_avg}
\end{align}
where $g_{\tau}$ is the matrix representation of $\tau$ and the denominator ensures that the state has unit norm. We are generally interested in ensembles for mixed states that are induced from taking a partial trace over a sub-Hilbert space. If $\mathcal{H} = \mathcal{H}_A \otimes \mathcal{H}_B$, the reduced density matrix on $A$ is given by
\begin{align}
    \rho_A^{\otimes \alpha} := \Tr_B\left[\overline{\ket{\Psi}\bra{\Psi}^{\otimes \alpha}}\right] = \frac{\sum_{\tau \in S_\alpha} g_{\tau_A} \Tr\left[ g_{\tau_B}\right]}{\sum_{\tau \in S_\alpha} \Tr\left[g_{\tau}\right]},
    \label{haar_avg_dens}
\end{align}
where the subscript on the permutation elements mean that they only permute within
a sub-Hilbert space. The trace of a permutation element is straightforward to work out, equaling the dimension of the Hilbert space, $d_A d_B$, to the number of cycles in the permutation, $C(\tau)$. The denominator can then be written as
\begin{align}
    \sum_{\tau \in S_\alpha} \Tr\left[g_{\tau}\right] = \sum_{\tau \in S_\alpha} \left(d_A d_B\right)^{C(\tau)},
    \label{normalization_haar}
\end{align}
which can be summed exactly because we know that the number of permutations of $\alpha$ elements with $k$ cycles is given by the Stirling number of the first kind. However, we can easily avoid this technical point because we will be interested in the regime where the Hilbert space dimension is large. Therefore, only the permutations that maximize $C(\tau)$ will contribute at leading order. The unique permutation that maximizes $C(\tau)$ is the identity permutation which has $\alpha$ cycles, so throughout this paper, we will approximate the denominator as $(d_A d_B)^\alpha$.

There is an alternative description of the same induced ensemble of density matrices that will be useful for us when generalizing to tensor networks. Rather than starting with fiducial state $\ket{0}$, we begin with complete bases, 
$\ket{i}_A$ and $\ket{J}_B$, 
on $\mathcal{H}_A$ and $\mathcal{H}_B$ respectively. The Haar random state is then represented as
\begin{align}
    \ket{\Psi } = \mathcal{N}\sum_{iJ}X_{iJ}\ket{i}_A\otimes \ket{J}_B,
\end{align}
where $X_{iJ}$ are complex Gaussian i.i.d.~matrix elements of $d_A \times d_B$ matrix $X$ with (unnormalized) joint probability distribution
\begin{align}
    P(\{ X_{iJ}\}) \propto \exp\left[-d_Ad_B \Tr \left(X X^{\dagger}\right)\right],
\end{align}
and $\mathcal{N}$ is a normalization constant. The reduced density matrix on $\mathcal{H}_A$ is then \cite{2001JPhA...34.7111Z,2004JPhA...37.8457S,2011JMP....52f2201Z}
\begin{align}
    \rho_A = \frac{X X^{\dagger}}{\Tr\left[X X^{\dagger} \right]}.
\end{align}
Ensemble averages over $n$ copies are given by the same formula in terms of permutations, so at large dimensions, $\rho_A \simeq X X^{\dagger}$. This is the famous Wishart-Laguerre ensemble and is equivalent to the previously introduced Haar random states \cite{2007AnHP....8.1521N,2009arXiv0910.1768C}. The advantage of working with random Gaussian states instead of Haar random states is due to ``Wick calculus'' being simpler than ``Weingarten calculus.'' The difference will appear for random tensor networks although the ensembles will still be equivalent at large Hilbert space dimension \cite{2010JPhA...43A5303C}. Moreover, the class of random tensor networks used for holography involving projected Haar random states \cite{2016JHEP...11..009H} precisely correspond to the states we study even at finite Hilbert space dimension, as explained in Appendix \ref{equiv_haar_wick_app}.

We now introduce a diagrammatic approach for computations of certain moments of the Wishart ensemble involving multiple states, building on Refs.~\cite{1995NuPhB.453..531B,2008AcPPB..39..799J,2020arXiv201101277S,2021PhRvL.126q1603K}. This will prove invaluable in the following calculations.

We represent the elements of the random global pure state as two vertical lines
\begin{align}
    \ket{\Psi}_{iJ}= X_{iJ} :=
    \,
    \tikz[baseline=-0.5ex]{
    \draw[dashed] (0,0.2) node[align=center, above] {\footnotesize $J$} -- (0,-0.2);
    \draw (-0.2,0.2) node[align=center, above] {\footnotesize $i$} -- (-0.2,-0.15);
    }\ ,
\end{align}
where the solid line represents $\mathcal{H}_A$ and the dashed line $\mathcal{H}_B$. To form the density matrix, we take the outer product
\begin{align}
    \left[\ket{\Psi}\bra{\Psi}\right]_{iJ,jK}= X_{iJ} X^\ast_{jK}
    =
    \,
    \tikz[baseline=-0.5ex]{
    \draw[dashed] (0,0.2) node[align=center, above] {\footnotesize $J$} -- (0,-0.2);
    \draw[dashed] (1,0.2) node[align=center, above] {\footnotesize $K$} -- (1,-0.2);
    \draw (-0.2,0.2) node[align=center, above] {\footnotesize $i$} -- (-0.2,-0.15);
    \draw (1.2,0.2) node[align=center, above] {\footnotesize $j$} -- (1.2,-0.15);
    }\ .
\end{align}
We will usually drop the index labeling of the lines to avoid cumbersome notation. All matrix manipulations are done on the lower ends of the lines. For example, we can take a partial trace over $\mathcal{H}_B$ by connecting the dashed line
\begin{align}
    \label{eq:rho_diag}
    [\rho_A]_{i,j}= 
    {\sum_{J=1}^{d_B} 
    X_{iJ} X^\ast_{jJ}}
:= 
\,
    \tikz[scale=1.0,baseline=-0.5ex]{
    \draw[dashed] (0,0.2) node[align=center, above] {\footnotesize $J$} -- (0,0);
    \draw[dashed] (0,0)  -- (1,0);
    \draw[dashed]  (1,0.2) node[align=center, above] {\footnotesize $J$} -- (1,0);
    \draw (-0.2,0.2) node[align=center, above] {\footnotesize $i$} -- (-0.2,-0.15);
    \draw (1.2,0.2) node[align=center, above] {\footnotesize $j$} -- (1.2,-0.15);
    }\ ,
\end{align} 
square the matrix by taking two copies and connecting the \textit{bra} of the first matrix with the \textit{ket} of the second
\begin{align}
\label{eq:tr_r2}
    [\rho_A^2]_{i,j}=  
    \,
    \tikz[scale=1.0,baseline=0ex]{
    \draw[dashed] (0,0.2)-- (0,0);
    \draw[dashed] (0,0) -- (1,0);
    \draw[dashed]  (1,0.2) -- (1,0);
    \draw (-0.2,0.2)-- (-0.2,-0.25);
    \draw (1.2,-0.1)-- (1.2,0.2);
    \draw[dashed] (2,0.2)-- (2,0);
    \draw[dashed] (2,0) -- (3,0);
    \draw[dashed]  (3,0.2) -- (3,0);
    \draw (1.8,0.2)-- (1.8,-0.1);
    \draw (3.2,-0.25)-- (3.2,0.2);
    \draw (1.2,-0.1)-- (1.8,-0.1);
    }\ ,
\end{align}
then take a trace by connecting the remaining solid lines to determine the purity
\begin{align}
    \Tr\left[\rho_A^2\right]= 
    \,
    \tikz[scale=1.0,baseline=0ex]{
    \draw[dashed] (0,0.2)-- (0,0);
    \draw[dashed] (0,0) -- (1,0);
    \draw[dashed]  (1,0.2) -- (1,0);
    \draw (-0.2,0.2)-- (-0.2,-0.25);
    \draw (1.2,-0.1)-- (1.2,0.2);
    \draw[dashed] (2,0.2)-- (2,0);
    \draw[dashed] (2,0) -- (3,0);
    \draw[dashed]  (3,0.2) -- (3,0);
    \draw (1.8,0.2)-- (1.8,-0.1);
    \draw (3.2,-0.25)-- (3.2,0.2);
    \draw (1.2,-0.1)-- (1.8,-0.1);
    \draw (-0.2,-0.25) -- (3.2,-0.25);
    }\ .
    \label{purity_diag}
\end{align}
For every insertion of the density matrix, we include a factor of $(d_A d_B)^{-1}$. This will give the normalization factor that we computed from \eqref{normalization_haar}.

The ensemble averaging of the states are done on the upper ends of the lines. The rule here is that we must add up all diagrams contracting the any \textit{bra} with any \textit{ket}. For $\alpha$ insertions of the density matrix, there will be $\alpha!$ diagrams, corresponding to the $\alpha!$ allowed permutations. Within each diagram, we count the number of loops with each loop giving a factor of the Hilbert space dimension. One can see that this diagrammatic sum is precisely the numerator of \eqref{haar_avg}.

We can now practice by taking the ensemble averaged purity. There are two ($2!$) diagrams descending from \eqref{purity_diag}
\label{purity_diagram}
\begin{align}
    \overline{\Tr\left[\rho_A^2\right]} &=
    \,
    \tikz[scale=0.75,baseline=0.5ex]{
    \draw[dashed] (0,0) -- (1,0);
    \draw (-0.2,0.)-- (-0.2,-0.25);
    \draw (1.2,-0.1)-- (1.2,0.);
    \draw[dashed] (2,0) -- (3,0);
    \draw (1.8,0.)-- (1.8,-0.1);
    \draw (3.2,-0.25)-- (3.2,0.);
    \draw (1.2,-0.1)-- (1.8,-0.1);
    \draw (-0.2,-0.25) -- (3.2,-0.25);
    \draw[dashed] (1.0,0) arc (0:180:0.5);
    \draw[dashed] (3.0,0) arc (0:180:0.5);
    \draw (1.2,0) arc (0:180:0.7);
    \draw (3.2,0) arc (0:180:0.7);
    }
\ \,
+
\ \,
    \tikz[scale=1.0,baseline=0.5ex]{
    \draw[dashed] (0,0) -- (1,0);
    \draw (-0.2,0.)-- (-0.2,-0.25);
    \draw (1.2,-0.1)-- (1.2,0.);
    \draw[dashed] (2,0) -- (3,0);
    \draw (1.8,0.)-- (1.8,-0.1);
    \draw (3.2,-0.25)-- (3.2,0.);
    \draw (1.2,-0.1)-- (1.8,-0.1);
    \draw (-0.2,-0.25) -- (3.2,-0.25);
    \draw[dashed] (2.0,0) arc (0:180:0.5);
    \draw[dashed] (3.0,0) arc (0:180:1.5);
    \draw (1.8,0) arc (0:180:0.3);
    \draw (3.2,0) arc (0:180:1.7);
    }
\ 
,
\end{align}
immediately leading to $d_A^{-1} + d_B^{-1}$.

Because we are interested in distinguishing density matrices that are \textit{independently} sampled from the ensemble, we must extend the diagrammatic technique. We do this by introducing different colors for different density matrices. When ensemble averaging, \textit{bra}'s of one color can only contract with \textit{ket}'s of the same color. For example, the overlap between independent induced states $\rho_A$ (black) and $\sigma_A$ (red) looks similar to the purity
\begin{align}
    \Tr \left[\rho_A \sigma_A\right] = 
    \,
    \tikz[scale=1.0,baseline=0ex]{
    \draw[dashed] (0,0.2)-- (0,0);
    \draw[dashed] (0,0) -- (1,0);
    \draw[dashed]  (1,0.2) -- (1,0);
    \draw (-0.2,0.2)-- (-0.2,-0.25);
    \draw (1.2,-0.1)-- (1.2,0.2);
    \draw[dashed,red] (2,0.2)-- (2,0);
    \draw[dashed,red] (2,0) -- (3,0);
    \draw[dashed,red]  (3,0.2) -- (3,0);
    \draw[red] (1.8,0.2)-- (1.8,-0.1);
    \draw[red] (3.2,-0.25)-- (3.2,0.2);
    \draw (1.2,-0.1)-- (1.8,-0.1);
    \draw (-0.2,-0.25) -- (1.8,-0.25);
    \draw[red] (1.8,-0.25) -- (3.2,-0.25);
    }\ ,
\end{align}
but the ensemble averaging will only include a single diagram
\begin{align}
    \overline{\Tr \left[\rho_A \sigma_A\right]} &=
    \,
    \tikz[scale=1.0,baseline=0.5ex]{
    \draw[dashed] (0,0) -- (1,0);
    \draw (-0.2,0.)-- (-0.2,-0.25);
    \draw (1.2,-0.1)-- (1.2,0.);
    \draw[dashed,red] (2,0) -- (3,0);
    \draw[red] (1.8,0.)-- (1.8,-0.1);
    \draw[red] (3.2,-0.25)-- (3.2,0.);
    \draw (1.2,-0.1)-- (1.8,-0.1);
    \draw[black] (-0.2,-0.25) -- (1.8,-0.25);
    \draw[red] (1.8,-0.25) -- (3.2,-0.25);
    \draw[dashed] (1.0,0) arc (0:180:0.5);
    \draw[dashed,red] (3.0,0) arc (0:180:0.5);
    \draw (1.2,0) arc (0:180:0.7);
    \draw[red] (3.2,0) arc (0:180:0.7);
    }
\ \, = \frac{1}{d_A}
\label{overlap_diagram}
\end{align}
because the second diagram would have connected the black and red indices which is disallowed. With this formalism, we are now ready to compute each distinguishability measures using a replica trick. 

\subsection{Relative entropy}

We begin with the von Neumann relative entropy, the topic of Ref.~\cite{2021PhRvL.126q1603K}, both because it is the most fundamental quantity and the simplest to compute using our techniques. This will illustrate our strategy that will be used throughout. The relative entropy may be computed using a replica trick. That is, we first compute a certain series of moments of the ensemble and then analytically continue to arrive at the desired quantity.
The replica trick for the relative entropy is given by
\cite{2016PhRvL.117d1601L}
\begin{align}
    D(\rho_A \lvert \rvert \sigma_A) = \lim_{\alpha \rightarrow1}\frac{1}{\alpha -1} \left(\log\left[ \Tr \left[\rho_A^\alpha \right]\right]- \log \left[\Tr\left[ \rho^{\ }_A  \sigma_A^{\alpha -1}\right]\right]\right).
    \label{S_replica}
\end{align}
We compute the ensemble average of the two terms separately. The first term is the R\'enyi entropy and, as a diagram, looks like
\begin{align}
    \label{renyi_diag}
    \Tr \left[\rho_{A}^{\alpha} \right]=
    \tikz[scale=0.75,baseline=-0.5ex]{
    \draw[dashed] (0,0) -- (1,0);
    \draw (0,0)-- (0,.15);
    \draw (1,0)-- (1,.15);
    \draw (-0.2,0.15)-- (-0.2,-0.35);
    \draw (1.2,-0.15)-- (1.2,0.15);
    \draw[dashed,black] (2,0) -- (3,0);
    \draw[black] (2,0)-- (2,0.15);
    \draw[black] (3,0)-- (3,0.15);
    \draw[black] (1.8,-.15)-- (1.8,0.15);
    \draw[black] (3.2,-0.15)-- (3.2,0.15);
    \draw[dashed,black] (5,0) -- (6,0);
    \draw[black] (5,0)-- (5,0.15);
    \draw[black] (6,0)-- (6,0.15);
    \draw[black] (4.8,0.15)-- (4.8,-0.15);
    \draw[black] (6.2,-0.35)-- (6.2,0.15);
    \draw (1.2,-0.15)--(1.8,-0.15);
    \draw[black] (3.2,-0.15)--(3.45,-0.15);
    \draw (-0.2,-0.35)-- (6.2,-0.35);
    \draw[black] (4.5,-0.15)--(4.8,-0.15);
    \node[] at (4.,0.2) {$\cdots$};
    }
    \, .
\end{align}
While we make the dimensions of the sub-Hilbert spaces large, $d_A, d_B \propto N\rightarrow \infty$, we keep their relative sizes, $d_A/d_B$, finite.
The leading diagrams maximize the total number of loops.  These are the planar diagrams as this double line notation corresponds the standard large-$N$ topological expansion. Planar diagrams correspond to the \textit{non-crossing permutations}, $NC_{\alpha}$, a well-studied object in enumerative combinatorics and probability theory. The ensemble averaged R\'enyi purity is then given by
\begin{align}
    \overline{\Tr \left[\rho_{A}^{{\alpha}} \right]}= \frac{1}{(d_A d_B)^{\alpha}}\sum_{\tau \in NC_{\alpha}} d_A^{C(\eta^{-1} \circ \tau)} d_B^{C(\tau)},
    \label{renyi_sum}
\end{align}
where $\eta$ is the cyclic permutation, spawning from the matrix multiplication and trace in \eqref{renyi_diag}. 
The non-crossing permutations maximize the total exponent as $C(\eta^{-1} \circ \tau) +C(\tau)= {\alpha} + 1$. A more refined statement is that the number of non-crossing permutations with $C(\eta^{-1} \circ \tau) = k$ (and therefore $C( \tau) = \alpha + 1- k$) is given by the Narayana number
\begin{align}
    N_{{\alpha},k} := \frac{1}{\alpha}\binom{{\alpha}}{k}\binom{{\alpha}}{k-1}.
\end{align}
With this information, we can reorganize \eqref{renyi_sum} as a sum over $k$ instead of a sum over permutations
\begin{align}
    \overline{\Tr \left[\rho_{A}^{{\alpha}} \right]}= \frac{1}{(d_A d_B)^{\alpha}}\sum_{k= 1}^{\alpha} N_{{\alpha},k} d_A^{k} d_B^{{\alpha}+1-k},
    \label{renyi_sum_reorg}
\end{align}
which can be rewritten again as a hypergeometric function\footnote{The two elements of the piecewise function are equivalent on the integers. The reason why we write it as a piecewise function is for ease of analytic continuation to non-integer values because the hypergeometric functions are entire when the argument is less than one.}
\begin{align}
    \overline{\Tr \left[\rho_{A}^{{\alpha}} \right]}=\begin{cases}
        d_A^{1-{\alpha}} \, _2F_1\left(1-{\alpha},-{\alpha};2;\frac{d_A}{d_B}\right), &d_A < d_B
        \\
        d_B^{1-{\alpha}} \, _2F_1\left(1-{\alpha},-{\alpha};2;\frac{d_B}{d_A}\right), &d_A > d_B
    \end{cases} .
    \label{renyi_hyper}
\end{align}
The $A\leftrightarrow B$ symmetry of R\'enyi entropies of bipartite pure states is manifest.

Taking the logarithm and analytically continuing to $\alpha = 1$, we obtain Page's formula \cite{1993PhRvL..71.1291P}
\begin{align}
    \lim_{\alpha \rightarrow 1} \frac{1}{1 - \alpha} \overline{\log\left[\Tr \left[\rho_{A}^{{\alpha}} \right]\right]} = \begin{cases}
        \log \left[d_A\right] -\frac{d_A}{2d_B}, & d_A < d_B
        \\
        \log \left[d_B\right] -\frac{d_B }{2d_A}, & d_B < d_A
    \end{cases}.
\end{align}
In writing this formula, we have assumed that logarithm and ensemble average commute. In Appendix \ref{commute_app}, we explain why this is true when the Hilbert space dimensions are large. 

The second term in \eqref{S_replica} involves both $\rho_A$ and $\sigma_A$\footnote{This can be generalized such that the auxiliary systems for $\sigma$ and $\rho$ are of different sizes $d_{B_1}$ and $d_{B_2}$. In the diagrammatics, this corresponds to assigning different weights to the black and red dashed lines. The resulting generalized sums are still tractable, though, we do not currently have use for these calculations in our applications to black holes because $d_{B}$ corresponds to the size of the black hole, which is simple to measure by an outside observer, rendering $\sigma_A$ and $\rho_A$ easily distinguishable when $d_{B_1}\neq d_{B_2}$. Some exact results for this set up in the Wishart ensemble can be found in Refs.~\cite{2020PhRvA.102a2405K,2020JPhA...53X5202K,2021arXiv210502743L}.}
\begin{align}
    \Tr\left[ \rho_A \sigma_A^{1-\alpha}\right]
    =
    \tikz[scale=0.75,baseline=-0.5ex]{
    \draw[dashed] (0,0) -- (1,0);
    \draw (0,0)-- (0,.15);
    \draw (1,0)-- (1,.15);
    \draw (-0.2,0.15)-- (-0.2,-0.35);
    \draw (1.2,-0.15)-- (1.2,0.15);
    \draw[dashed,red] (2,0) -- (3,0);
    \draw[red] (2,0)-- (2,0.15);
    \draw[red] (3,0)-- (3,0.15);
    \draw[red] (1.8,-.15)-- (1.8,0.15);
    \draw[red] (3.2,-0.15)-- (3.2,0.15);
    \draw[dashed,red] (5,0) -- (6,0);
    \draw[red] (5,0)-- (5,0.15);
    \draw[red] (6,0)-- (6,0.15);
    \draw[red] (4.8,0.15)-- (4.8,-0.15);
    \draw[red] (6.2,-0.35)-- (6.2,0.15);
    \draw (1.2,-0.15)--(1.8,-0.15);
    \draw[red] (3.2,-0.15)--(3.45,-0.15);
    \draw[black] (-0.2,-0.35)-- (1.8,-0.35);
    \draw[red] (1.8,-0.35)-- (6.2,-0.35);
    \draw[red] (4.5,-0.15)--(4.8,-0.15);
    \node[] at (4.,0.2) {$\cdots$};
    } \, .
\end{align}
Because there is only a single copy of $\rho_A$, when ensemble averaging, we must contract the first density matrix with itself. There are no constraints on how to contract the red lines. This means that the $S_{\alpha}$ permutations are broken down to $\mathbbm{1}\times S_{\alpha-1}$
\begin{align}
    \overline{\Tr\left[ \rho_A \sigma_A^{1-\alpha}\right]} =\frac{1}{(d_A d_B)^{\alpha}} \sum_{\tau \in \mathbbm{1}\times S_{\alpha -1}} d_A^{C(\eta^{-1}\circ \tau)}d_B^{C(\tau)}.
    \label{vn_RE_sum}
\end{align}
We still need to maximize the exponent by choosing non-crossing permutations, though many such permutations are disallowed by the identity factor on the first matrix.
The diagrams are topological, so we have
\begin{align}
    \tikz[scale=0.75,baseline=-0.5ex]{
    \draw[dashed] (0,0) -- (1,0);
    \draw (-0.2,0.)-- (-0.2,-0.35);
    \draw (1.2,-0.15)-- (1.2,0.);
    \draw[dashed,red] (2,0) -- (3,0);
    \draw[red] (2,0)-- (2,0.15);
    \draw[red] (3,0)-- (3,0.15);
    \draw[red] (1.8,-.15)-- (1.8,0.15);
    \draw[red] (3.2,-0.15)-- (3.2,0.15);
    \draw[dashed,red] (5,0) -- (6,0);
    \draw[red] (5,0)-- (5,0.15);
    \draw[red] (6,0)-- (6,0.15);
    \draw[red] (4.8,0.15)-- (4.8,-0.15);
    \draw[red] (6.2,-0.35)-- (6.2,0.15);
    \draw (1.2,-0.15)--(1.8,-0.15);
    \draw[red] (3.2,-0.15)--(3.45,-0.15);
    \draw[black] (-0.2,-0.35)-- (1.8,-0.35);
    \draw[red] (1.8,-0.35)-- (6.2,-0.35);
    \draw[red] (4.5,-0.15)--(4.8,-0.15);
    \node[] at (4.,0.2) {$\cdots$};
    \draw[dashed] (1.0,0) arc (0:180:0.5);
    \draw (1.2,0) arc (0:180:0.7);
    } 
    \,
    =
    \,
    \tikz[scale=0.75,baseline=-0.5ex]{
    \draw[dashed] (0.5,0) -- (1.5,0);
    \draw[dashed,red] (2,0) -- (3,0);
    \draw[red] (2,0)-- (2,0.15);
    \draw[red] (3,0)-- (3,0.15);
    \draw[red] (1.8,-.35)-- (1.8,0.15);
    \draw[red] (3.2,-0.15)-- (3.2,0.15);
    \draw[dashed,red] (5,0) -- (6,0);
    \draw[red] (5,0)-- (5,0.15);
    \draw[red] (6,0)-- (6,0.15);
    \draw[red] (4.8,0.15)-- (4.8,-0.15);
    \draw[red] (6.2,-0.35)-- (6.2,0.15);
    \draw[red] (3.2,-0.15)--(3.45,-0.15);
    \draw[red] (1.8,-0.35)-- (6.2,-0.35);
    \draw[red] (4.5,-0.15)--(4.8,-0.15);
    \node[] at (4.,0.2) {$\cdots$};
    \draw[dashed] (1.5,0) arc (0:180:0.5);
    } \, .
\end{align}
From this diagram, it is clear that the cardinality of the intersection of $NC_{\alpha}$ and $\mathbbm{1}\times S_{\alpha - 1}$ is given by the cardinality of $NC_{\alpha-1}$ and the number of such non-crossing permutations with $C(\eta^{-1}\circ \tau) = k$ is given by Narayana number $N_{\alpha-1,k}$
\begin{align}
    \overline{\Tr\left[ \rho_A \sigma_A^{1-\alpha}\right]} =\frac{1}{(d_A d_B)^{\alpha}} \sum_{k= 1}^{\alpha} N_{{\alpha}-1,k} d_A^{k} d_B^{{\alpha}+1-k},
    \label{narayana_sum_rel}
\end{align}
which can also be represented by a hypergeometric function
\begin{align}
    \overline{\Tr \left[\rho_{A}\sigma_A^{{\alpha}-1} \right]}=\begin{cases}
        d_A^{1-{\alpha}} \, _2F_1\left(1-{\alpha},2-{\alpha};2;\frac{d_A}{d_B}\right), &d_A < d_B
        \\
        d_B^{2-{\alpha}}d_A^{-1} \, _2F_1\left(1-{\alpha},2-{\alpha};2;\frac{d_B}{d_A}\right), &d_A > d_B
    \end{cases} .
\end{align}
Taking the $\alpha \rightarrow 1$ limit, we have
\begin{align}
    \lim_{\alpha \rightarrow 1}\frac{1}{1-\alpha} \overline{\log\left[\Tr \left[\rho_{A}\sigma_A^{{\alpha}-1} \right]\right]}= \begin{cases}
        \log \left[d_A\right] +1 +\left(\frac{d_B}{d_A}-1 \right)\log\left[1-\frac{d_A}{d_B} \right], & d_A < d_B
        \\
        \infty,& d_B < d_A
    \end{cases}.
\end{align}
Therefore, we find the ensemble average of
the relative entropy to be
\begin{align}
    \overline{D(\rho_A || \sigma_A)} = \begin{cases}
        1+\frac{d_A}{2 d_B}+\left(\frac{d_B}{d_A}-1\right) \log
   \left[1-\frac{d_A}{d_B}\right], & d_A < d_B
   \\
   \infty, & d_A > d_B
    \end{cases}.
    \label{relative_random}
\end{align}
This is a satisfying, simple answer. For small $d_A/d_B$, the relative entropy is given by $d_A/d_B$. If we think in terms of ``number of qubits,'' $N_A$ and $N_B$, this is exponentially small in the difference ($N_B-N_A$), meaning that the states will be very difficult to distinguish whenever we have access to a few qubits less than half the system; the asymptotic error rate, $\beta^*_n(\epsilon)$, is very small, meaning we will need exponentially (in $N_A$) many copies of the state to identify it with confidence. \eqref{relative_random} is also monotonically increasing in $d_A/d_B$, a consequence of the data processing inequality when we take the partial trace as the quantum channel. When $d_A \rightarrow d_B$, the relative entropy approaches the curious value of $3/2$. This value of $3/2$ was also determined in Ref.~\cite{2016PhRvA..93f2112P} using very different techniques which serves as an additional consistency check of our results.

When $d_A > d_B$, every reduced state on $\mathcal{H}_A$ in the ensemble will be rank deficient with $d_A - d_B$ zero eigenvalues. This is because the Wishart ensemble has rank at most $\min(d_A,d_B)$. It is therefore overwhelmingly unlikely that two independent states, $\rho_A$ and $\sigma_A$, will have the same support. In particular, the support of $\rho_A$ will not be contained within the support of $\sigma_A$. This is the reason why the relative entropy becomes infinite in this regime; there will be a measurement we can choose that easily distinguishes $\rho_A$ and $\sigma_A$. 

\subsection{Petz R\'enyi relative entropy and Holevo's just-as-good fidelity}
\label{PRRE_sec}

To understand more sophisticated structures in Haar random states, we progress to the computation of the PRRE. The PRRE has a tricky $1-\alpha$ exponent for $\sigma_A$, so we use a replica trick with two replica parameters, $\alpha$ and $m$
\begin{align}
    D_{\alpha}(\rho_A|| \sigma_A) = \lim_{m\rightarrow 1-\alpha}\frac{1}{\alpha - 1}\log\left[ \Tr \left[ \rho_A^{\alpha} \sigma_A^{m}\right]\right].
\end{align}
We will compute this for $\alpha, m \in \mathbb{Z}^+$, only taking the limit to $\alpha, m \in \mathbb{R}$ at the end of the calculation.
The positive integer moments in diagrammatic form are
\begin{align}
    \Tr \left[ \rho_A^{\alpha} \sigma_A^{m}\right] = 
    \tikz[scale=0.75,baseline=-0.5ex]{
    \draw[dashed] (0,0) -- (1,0);
    \draw (0,0)-- (0,.15);
    \draw (1,0)-- (1,.15);
    \draw (-0.2,0.15)-- (-0.2,-0.35);
    \draw (1.2,-0.15)-- (1.2,0.15);
    \draw[dashed,black] (2,0) -- (3,0);
    \draw[black] (2,0)-- (2,0.15);
    \draw[black] (3,0)-- (3,0.15);
    \draw[black] (1.8,-.15)-- (1.8,0.15);
    \draw[black] (3.2,-0.15)-- (3.2,0.15);
    \draw[dashed,black] (5,0) -- (6,0);
    \draw[black] (5,0)-- (5,0.15);
    \draw[black] (6,0)-- (6,0.15);
    \draw[black] (4.8,0.15)-- (4.8,-0.15);
    \draw[black] (6.2,-0.15)-- (6.2,0.15);
    \draw[red,dashed] (7+0,0) -- (7+1,0);
    \draw[red] (7+0,0)-- (7+0,.15);
    \draw[red] (7+1,0)-- (7+1,.15);
    \draw[red] (7-0.2,0.15)-- (7-0.2,-0.15);
    \draw[red] (7+1.2,-0.15)-- (7+1.2,0.15);
    \draw[dashed,red] (7+2,0) -- (7+3,0);
    \draw[red] (7+2,0)-- (7+2,0.15);
    \draw[red] (7+3,0)-- (7+3,0.15);
    \draw[red] (7+1.8,-.15)-- (7+1.8,0.15);
    \draw[red] (7+3.2,-0.15)-- (7+3.2,0.15);
    \draw[dashed,red] (7+5,0) -- (7+6,0);
    \draw[red] (7+5,0)-- (7+5,0.15);
    \draw[red] (7+6,0)-- (7+6,0.15);
    \draw[red] (7+4.8,0.15)-- (7+4.8,-0.15);
    \draw[red] (7+6.2,-0.35)-- (7+6.2,0.15);
    \draw (1.2,-0.15)--(1.8,-0.15);
    \draw[black] (3.2,-0.15)--(3.45,-0.15);
    \draw (-0.2,-0.35)-- (6.2,-0.35);
    \draw[black] (4.5,-0.15)--(4.8,-0.15);
    \node[] at (4.,0.2) {$\cdots$};
    \draw[black] (7-2+1.2,-0.15)--(7-2+1.8,-0.15);
    \draw[black] (7-2+1.2,-0.35)--(7-2+1.8,-0.35);
    \draw[red] (7+1.2,-0.15)--(7+1.8,-0.15);
    \draw[red] (7+3.2,-0.15)--(7+3.45,-0.15);
    \draw[red] (7-0.2,-0.35)-- (7+6.2,-0.35);
    \draw[red] (7+4.5,-0.15)--(7+4.8,-0.15);
    \node[] at (7+4.,0.2) {$\cdots$};
    } \, ,
\end{align}
where there are $\alpha$ black density matrices and $m$ red density matrices. When ensemble averaging, we are only able to contract using the subgroup $S_{\alpha}\times S_m \subset S_{\alpha + m}$, leading to the sum over permutations 
\begin{align}
     \overline{\Tr\left[ \rho_A^{{\alpha}} \sigma_A^{m}\right]} = \frac{1}{(d_A d_B)^{{\alpha}+m}}\sum_{\tau \in S_{\alpha} \times S_m} d_A^{C(\eta^{-1}\circ \tau)}d_B^{C(\tau)}.
\end{align}
As can be seen by the diagram, even with the restricted sum, there are many ways to contract the lines that are non-crossing, hence maximizing the exponents. These are precisely the non-crossing permutations acting independently on the black and red indices, so the combinatorial factor will be given by the product of two Narayana numbers
\begin{align}
   \overline{\Tr\left[ \rho_A^{{\alpha}} \sigma_A^{m}\right]} = \frac{1}{(d_A d_B)^{{\alpha}+m}}\sum_{k = 1}^{\alpha}\sum_{j = 1}^m N_{\alpha,k}N_{m,j}d_A^{k+j-1} d_B^{2+\alpha + m-k-j}.
\end{align}
The reason why there is an additional ``$-1$'' in the exponent of $d_A$ is that the black and red lines are connected at the bottom of the diagram due to the matrix multiplication.
Note that this expression is a generalization of the replica trick used in the previous section for the relative entropy, \eqref{narayana_sum_rel}, if we set $\alpha = 1$ and $m = \alpha-1$. As before, the double sum can be expressed in terms of hypergeometric functions
\begin{align}
     \overline{\Tr\left[ \rho_A^{{\alpha}} \sigma_A^{m}\right]} =\begin{cases} d_A^{-m-{\alpha}+1} \, _2F_1\left(1-m,-m;2;\frac{d_A}{d_B}\right) \,
   _2F_1\left(1-{\alpha},-{\alpha};2;\frac{d_A}{d_B}\right), & d_A < d_B
   \\
 \frac{d_B^{-m-{\alpha}+2}}{d_A} \, _2F_1\left(1-m,-m;2;\frac{d_B}{d_A}\right) \,
   _2F_1\left(1-{\alpha},-{\alpha};2;\frac{d_B}{d_A}\right), & d_A > d_B
   \end{cases}.
\end{align}
Now that the sum that required $m$ to be an integer is complete, it is safe to take the $m \rightarrow 1-\alpha $ limit
\begin{align}
    \overline{\Tr\left[ \rho_A^{\alpha} \sigma_A^{1-{\alpha}}\right]} = \begin{cases}\, _2F_1\left(1-{\alpha},-{\alpha};2;\frac{d_A}{d_B}\right) \,
   _2F_1\left({\alpha}-1,{\alpha};2;\frac{d_A}{d_B}\right), & d_A < d_B
   \\
   \frac{d_B}{d_A} \, _2F_1\left(1-{\alpha},-{\alpha};2;\frac{d_B}{d_A}\right) \,
   _2F_1\left({\alpha}-1,{\alpha};2;\frac{d_B}{d_A}\right) , & d_A > d_B
   \end{cases}.
   \label{PRRE_pre_eq_replica}
\end{align}
When $d_A = d_B$, this precisely agrees with a formula from Ref.~\cite{2016PhRvA..93f2112P}.
Taking the logarithm leads to an exact closed-form expression for the PRRE in the large Hilbert space dimension limit
\begin{align}
    \overline{D_{\alpha}(\rho_A||\sigma_A)} = \frac{1}{{\alpha}-1}
    \begin{cases}\log\left[\, _2F_1\left(1-{\alpha},-{\alpha};2;\frac{d_A}{d_B}\right) \,
   _2F_1\left({\alpha}-1,{\alpha};2;\frac{d_A}{d_B}\right)\right], & d_A < d_B
   \\
   \log\left[
   \frac{d_B}{d_A} \, _2F_1\left(1-{\alpha},-{\alpha};2;\frac{d_B}{d_A}\right) \,
   _2F_1\left({\alpha}-1,{\alpha};2;\frac{d_B}{d_A}\right) \right], & d_A > d_B
   \end{cases}.
   \label{PRRE_eq}
\end{align}
This is a rare instance where we have an exact closed-form solution for relative entropies and can be thought of as the ``Page formula'' for PRRE. Importantly, this equation contains much more information about random quantum states than \eqref{relative_random}. A highlight is the finiteness of \eqref{PRRE_eq} for $\alpha < 1$ in the $d_A > d_B$ regime. This explains the approach of random quantum states to complete distinguishability. There are a few consistency checks that we can readily verify. Namely, we note that \eqref{PRRE_eq} reduces to \eqref{relative_random} if we send $\alpha \rightarrow 1$, $\eqref{PRRE_eq}$ is monotonically increasing in $d_A/d_B$ (data processing inequality), and monotonically increasing in $\alpha$.

An additional desirable property of \eqref{PRRE_eq} is that it is simple enough that we can perform the optimization needed to compute the quantum Chernoff distance
\begin{align}
    \overline{\xi(\rho_A || \sigma_A)} = \begin{cases}- 2\log \left[\,
   _2F_1\left(\frac{1}{2},-\frac{1}{2};2;\frac{d_A}{d_B}\right) \right], &  {d_A}<{d_B}
   \\
   -\log \left[\frac{d_B}{d_A}\,
   _2F_1\left(\frac{1}{2},-\frac{1}{2};2;\frac{d_B}{d_A}\right)^2 \right], & {d_A}>{d_B}
   \end{cases},
\end{align}
where the optimal value of $\alpha$ in \eqref{chernoff_bound} is found to be $1/2$.
This definitively establishes the error rate in quantum state discrimination for a measure one set of quantum states. Because $\alpha = 1/2$ is the optimal value, this adds to the usefulness of Holevo's just-as-good fidelity, which is given by
\begin{align}
    \overline{F_H(\rho_A || \sigma_A)} = \begin{cases}\, _2F_1\left(\frac{1}{2},-\frac{1}{2};2;\frac{d_A}{d_B}\right)^4,&d_A < d_B
    \\
    \frac{d_B^2}{d_A^2} \,_2F_1\left(\frac{1}{2},-\frac{1}{2};2;\frac{d_B}{d_A}\right)^4,&d_A > d_B
    \end{cases}.
    \label{holevo_fidelity}
\end{align}

In order to evaluate the quantum multiple Chernoff distance, we need to characterize the fluctuations in the PRRE. To compute the variance, we must compute
\begin{align}
    \Tr \left[ \rho_A^{\alpha} \sigma_A^{m}\right]^2 = 
    \tikz[scale=0.375,baseline=-0.5ex]{
    \draw[dashed] (0,0) -- (1,0);
    \draw (0,0)-- (0,.15);
    \draw (1,0)-- (1,.15);
    \draw (-0.2,0.15)-- (-0.2,-0.35);
    \draw (1.2,-0.15)-- (1.2,0.15);
    \draw[dashed,black] (2,0) -- (3,0);
    \draw[black] (2,0)-- (2,0.15);
    \draw[black] (3,0)-- (3,0.15);
    \draw[black] (1.8,-.15)-- (1.8,0.15);
    \draw[black] (3.2,-0.15)-- (3.2,0.15);
    \draw[dashed,black] (5,0) -- (6,0);
    \draw[black] (5,0)-- (5,0.15);
    \draw[black] (6,0)-- (6,0.15);
    \draw[black] (4.8,0.15)-- (4.8,-0.15);
    \draw[black] (6.2,-0.15)-- (6.2,0.15);
    \draw[red,dashed] (7+0,0) -- (7+1,0);
    \draw[red] (7+0,0)-- (7+0,.15);
    \draw[red] (7+1,0)-- (7+1,.15);
    \draw[red] (7-0.2,0.15)-- (7-0.2,-0.15);
    \draw[red] (7+1.2,-0.15)-- (7+1.2,0.15);
    \draw[dashed,red] (7+2,0) -- (7+3,0);
    \draw[red] (7+2,0)-- (7+2,0.15);
    \draw[red] (7+3,0)-- (7+3,0.15);
    \draw[red] (7+1.8,-.15)-- (7+1.8,0.15);
    \draw[red] (7+3.2,-0.15)-- (7+3.2,0.15);
    \draw[dashed,red] (7+5,0) -- (7+6,0);
    \draw[red] (7+5,0)-- (7+5,0.15);
    \draw[red] (7+6,0)-- (7+6,0.15);
    \draw[red] (7+4.8,0.15)-- (7+4.8,-0.15);
    \draw[red] (7+6.2,-0.35)-- (7+6.2,0.15);
    \draw (1.2,-0.15)--(1.8,-0.15);
    \draw[black] (3.2,-0.15)--(3.45,-0.15);
    \draw (-0.2,-0.35)-- (6.2,-0.35);
    \draw[black] (4.5,-0.15)--(4.8,-0.15);
    \node[] at (4.,0.2) {$\cdots$};
    \draw[black] (7-2+1.2,-0.15)--(7-2+1.8,-0.15);
    \draw[black] (7-2+1.2,-0.35)--(7-2+1.8,-0.35);
    \draw[red] (7+1.2,-0.15)--(7+1.8,-0.15);
    \draw[red] (7+3.2,-0.15)--(7+3.45,-0.15);
    \draw[red] (7-0.2,-0.35)-- (7+6.2,-0.35);
    \draw[red] (7+4.5,-0.15)--(7+4.8,-0.15);
    \node[] at (7+4.,0.2) {$\cdots$};
    \draw[dashed] (15,0) -- (15+1,0);
    \draw (15+0,0)-- (15+0,.15);
    \draw (15+1,0)-- (15+1,.15);
    \draw (15-0.2,0.15)-- (15-0.2,-0.35);
    \draw (15+1.2,-0.15)-- (15+1.2,0.15);
    \draw[dashed,black] (15+2,0) -- (15+3,0);
    \draw[black] (15+2,0)-- (15+2,0.15);
    \draw[black] (15+3,0)-- (15+3,0.15);
    \draw[black] (15+1.8,-.15)-- (15+1.8,0.15);
    \draw[black] (15+3.2,-0.15)-- (15+3.2,0.15);
    \draw[dashed,black] (15+5,0) -- (15+6,0);
    \draw[black] (15+5,0)-- (15+5,0.15);
    \draw[black] (15+6,0)-- (15+6,0.15);
    \draw[black] (15+4.8,0.15)-- (15+4.8,-0.15);
    \draw[black] (15+6.2,-0.15)-- (15+6.2,0.15);
    \draw[red,dashed] (15+7+0,0) -- (15+7+1,0);
    \draw[red] (15+7+0,0)-- (15+7+0,.15);
    \draw[red] (15+7+1,0)-- (15+7+1,.15);
    \draw[red] (15+7-0.2,0.15)-- (15+7-0.2,-0.15);
    \draw[red] (15+7+1.2,-0.15)-- (15+7+1.2,0.15);
    \draw[dashed,red] (15+7+2,0) -- (15+7+3,0);
    \draw[red] (15+7+2,0)-- (15+7+2,0.15);
    \draw[red] (15+7+3,0)-- (15+7+3,0.15);
    \draw[red] (15+7+1.8,-.15)-- (15+7+1.8,0.15);
    \draw[red] (15+7+3.2,-0.15)-- (15+7+3.2,0.15);
    \draw[dashed,red] (15+7+5,0) -- (15+7+6,0);
    \draw[red] (15+7+5,0)-- (15+7+5,0.15);
    \draw[red] (15+7+6,0)-- (15+7+6,0.15);
    \draw[red] (15+7+4.8,0.15)-- (15+7+4.8,-0.15);
    \draw[red] (15+7+6.2,-0.35)-- (15+7+6.2,0.15);
    \draw (15+1.2,-0.15)--(15+1.8,-0.15);
    \draw[black] (15+3.2,-0.15)--(15+3.45,-0.15);
    \draw (15-0.2,-0.35)-- (15+6.2,-0.35);
    \draw[black] (15+4.5,-0.15)--(15+4.8,-0.15);
    \node[] at (15+4.,0.2) {$\cdots$};
    \draw[black] (15+7-2+1.2,-0.15)--(15+7-2+1.8,-0.15);
    \draw[black] (15+7-2+1.2,-0.35)--(15+7-2+1.8,-0.35);
    \draw[red] (15+7+1.2,-0.15)--(15+7+1.8,-0.15);
    \draw[red] (15+7+3.2,-0.15)--(15+7+3.45,-0.15);
    \draw[red] (15+7-0.2,-0.35)-- (15+7+6.2,-0.35);
    \draw[red] (15+7+4.5,-0.15)--(15+7+4.8,-0.15);
    \node[] at (15+7+4.,0.2) {$\cdots$};
    }\, 
    .
\end{align}
Only the diagrams that connect the two blocks will contribute to the variance because the disconnected diagrams are subtracted. For small $d_A/d_B$, these contributions will be $O(d_A^{1-2\alpha-2m}d_B^{-1})$ or $O(d_A^{-1}d_B^{-1})$
after taking the relevant limit. We may use a Taylor expansion of the logarithm to determine that the variance of the PRRE, $\sigma^2$, will be the same order. The higher degree central moments, and therefore higher cumulants, will be subleading because in general, the $n^{th}$ central moment will be $O(d_A^{1-n\alpha-nm}d_B^{-1})$. Therefore the PRRE will follow a normal distribution at subleading order.


For a normal distribution, the probability of random variable $X$ being $r$ standard deviations, $\sigma$, below the mean, $\mu$, is
\begin{align}
    \mbox{Pr}(X-\mu \leq -r {\sigma}) = \int_{-\infty}^{-k{\sigma}}dx \frac{1}{\sigma\sqrt{2\pi}}e^{-\frac{x^2}{2\sigma^2}} = \frac{1}{2} \text{erfc}\left(\frac{r}{\sqrt{2}}\right).
\end{align}
Therefore, if we have $W$ independent samplings of $\rho_A$ and $\sigma_A$, 
the probability that the minimum relative entropy will be at most $r$ standard deviations from the mean is
\begin{align}
    \mbox{Pr}\left(\min\left[{D}_{\alpha}(\rho_A||\sigma_A)\right] \geq \mu - r\sigma\right) = \left(1- \frac{1}{2} \text{erfc}\left(\frac{r}{\sqrt{2}}\right)\right)^W
    \label{chernoff_prob}
\end{align}
If we are discriminating between $W$ states, the quantum multiple Chernoff distance will be
\begin{align}
    \xi_W(\rho_A || \sigma_A) \geq \frac{d_A}{4d_B} - {\sqrt{2} \text{erfc}^{-1}\left(2-2 (1-\epsilon_1 )^{\frac{1}{W}}\right)}\sigma
    \label{multiple_chernoff_distance}
\end{align}
with probability $1-\epsilon_1$. In order to be confident in the state discrimination ($P^*_{err} < \epsilon_2$), we need 
\begin{align}
    n \simeq  \frac{\log\left[ \epsilon_2^{-1}\right]}{\frac{d_A}{4d_B} - {\sqrt{2} \text{erfc}^{-1}\left(2-2 (1-\epsilon_1 )^{\frac{1}{W}}\right)}\sigma} 
\end{align}
copies of the state. Due to $\sigma$ being suppressed in the total Hilbert space dimension, this formula only mildly depends on $W$ even when $W$ is of order the Hilbert space dimension. Thus, the multiple Chernoff bound is essentially just as tight as the two-state Chernoff bound.

\subsection{Sandwiched R\'enyi relative entropy and Uhlmann fidelity}

Continuing our progression in difficulty, we now compute the SRRE using a new replica trick requiring two replica indices
\begin{align}
    \tilde{D}_{\alpha}(\rho_A ||
    \sigma_A) := \lim_{m\rightarrow \frac{1-{\alpha}}{2{\alpha}}}\frac{1}{{\alpha} - 1}\log\left[ \Tr\left[ \left(\sigma_A^{m}\rho_A \sigma_A^{m} \right)^{{\alpha}}\right]\right] .
\end{align}
The associated diagrams are more complicated because there the red and black lines are not cleanly partitioned
\begin{align}
    \Tr &\left[ \left(\sigma_A^{m}\rho_A \sigma_A^{m}\right)^{\alpha}\right] =
    \nonumber
    \\
    &\tikz[scale=0.5,baseline=-0.5ex]{
    \draw[red, dashed] (0,0) -- (1,0);
    \draw[red] (0,0)-- (0,.15);
    \draw[red] (1,0)-- (1,.15);
    \draw[red] (-0.2,0.15)-- (-0.2,-0.35);
    \draw[red] (1.2,-0.15)-- (1.2,0.15);
    \draw[dashed,red] (3,0) -- (4,0);
    \draw[red] (3,0)-- (3,0.15);
    \draw[red] (4,0)-- (4,0.15);
    \draw[red] (2.8,-.15)-- (2.8,0.15);
    \draw[red] (4.2,-0.15)-- (4.2,0.15);
    \draw[dashed,black] (5,0) -- (6,0);
    \draw[black] (5,0)-- (5,0.15);
    \draw[black] (6,0)-- (6,0.15);
    \draw[black] (4.8,0.15)-- (4.8,-0.15);
    \draw[black] (6.2,-0.15)-- (6.2,0.15);
    \draw[dashed,red] (7,0) -- (8,0);
    \draw[red] (7,0)-- (7,0.15);
    \draw[red] (8,0)-- (8,0.15);
    \draw[red] (6.8,0.15)-- (6.8,-0.15);
    \draw[red] (8.2,-0.15)-- (8.2,0.15);
    \draw[dashed,red] (7+3,0) -- (8+3,0);
    \draw[red] (7+3,0)-- (7+3,0.15);
    \draw[red] (8+3,0)-- (8+3,0.15);
    \draw[red] (6.8+3,0.15)-- (6.8+3,-0.15);
    \draw[red] (8.2+3,-0.15)-- (8.2+3,0.15);
    \node[] at (2.,0.2) {$\cdots$};
    \node[] at (7+2.,0.2) {$\cdots$};
    \draw[red] (7-2+1.2,-0.15)--(7-2+1.8,-0.15);
    \draw[red] (7-4+1.2,-0.15)--(7-4+1.8,-0.15);
    \draw[dashed,black] (5+7,0) -- (6+7,0);
    \draw[black] (5+7,0)-- (5+7,0.15);
    \draw[black] (6+7,0)-- (6+7,0.15);
    \draw[black] (4.8+7,0.15)-- (4.8+7,-0.15);
    \draw[black] (6.2+7,-0.15)-- (6.2+7,0.15);
    \draw[dashed,red] (7+7,0) -- (8+7,0);
    \draw[red] (7+7,0)-- (7+7,0.15);
    \draw[red] (8+7,0)-- (8+7,0.15);
    \draw[red] (6.8+7,0.15)-- (6.8+7,-0.15);
    \draw[red] (8.2+7,-0.15)-- (8.2+7,0.15);
    \draw[red] (7+7-2+1.2,-0.15)--(7+7-2+1.8,-0.15);
    \draw[red] (7+7-4+1.2,-0.15)--(7+7-4+1.8,-0.15);
    \node[] at (2+14.,0.2) {$\boldsymbol{\cdots}$};
    \draw[dashed,red] (7+3+7,0) -- (8+3+7,0);
    \draw[red] (7+3+7,0)-- (7+3+7,0.15);
    \draw[red] (8+3+7,0)-- (8+3+7,0.15);
    \draw[red] (6.8+3+7,0.15)-- (6.8+3+7,-0.15);
    \draw[red] (8.2+3+7,-0.15)-- (8.2+3+7,0.15);
    \draw[dashed,black] (5+7+7,0) -- (6+7+7,0);
    \draw[black] (5+7+7,0)-- (5+7+7,0.15);
    \draw[black] (6+7+7,0)-- (6+7+7,0.15);
    \draw[black] (4.8+7+7,0.15)-- (4.8+7+7,-0.15);
    \draw[black] (6.2+7+7,-0.15)-- (6.2+7+7,0.15);
    \draw[dashed,red] (7+7+7,0) -- (8+7+7,0);
    \draw[red] (7+7+7,0)-- (7+7+7,0.15);
    \draw[red] (8+7+7,0)-- (8+7+7,0.15);
    \draw[red] (6.8+7+7,0.15)-- (6.8+7+7,-0.15);
    \draw[red] (8.2+7+7,-0.15)-- (8.2+7+7,0.15);
    \draw[red] (7+7-2+1.2+7,-0.15)--(7+7-2+1.8+7,-0.15);
    \draw[red] (7+7-4+1.2+7,-0.15)--(7+7-4+1.8+7,-0.15);
    \node[] at (2.+21,0.2) {$\cdots$};
    \draw[red, dashed] (0+24,0) -- (1+24,0);
    \draw[red] (0+24,0)-- (0+24,.15);
    \draw[red] (1+24,0)-- (1+24,.15);
    \draw[red] (-0.2+24,0.15)-- (-0.2+24,-0.15);
    \draw[red] (1.2+24,-0.35)-- (1.2+24,0.15);
    \draw[red] (1.2+24,-0.35)-- (-0.2,-0.35);
    } \, .
\end{align}
There are still many ways to contract the above diagram without crossing lines. We take two steps. First, we need to have the black lines contract with themselves in a non-crossing manner. For example, we may have
\begin{align}
\tikz[scale=0.5,baseline=-0.5ex]{
    \draw[red, dashed] (0,0) -- (1,0);
    \draw[red] (0,0)-- (0,.15);
    \draw[red] (1,0)-- (1,.15);
    \draw[red] (-0.2,0.15)-- (-0.2,-0.35);
    \draw[red] (1.2,-0.15)-- (1.2,0.15);
    \draw[dashed,red] (3,0) -- (4,0);
    \draw[red] (3,0)-- (3,0.15);
    \draw[red] (4,0)-- (4,0.15);
    \draw[red] (2.8,-.15)-- (2.8,0.15);
    \draw[red] (4.2,-0.15)-- (4.2,0.15);
    \draw[dashed,black] (5,0) -- (6,0);
    \draw[black] (5,0)-- (5,0.15);
    \draw[black] (6,0)-- (6,0.15);
    \draw[black] (4.8,0.15)-- (4.8,-0.15);
    \draw[black] (6.2,-0.15)-- (6.2,0.15);
    \draw[dashed,red] (7,0) -- (8,0);
    \draw[red] (7,0)-- (7,0.15);
    \draw[red] (8,0)-- (8,0.15);
    \draw[red] (6.8,0.15)-- (6.8,-0.15);
    \draw[red] (8.2,-0.15)-- (8.2,0.15);
    \draw[dashed,red] (7+3,0) -- (8+3,0);
    \draw[red] (7+3,0)-- (7+3,0.15);
    \draw[red] (8+3,0)-- (8+3,0.15);
    \draw[red] (6.8+3,0.15)-- (6.8+3,-0.15);
    \draw[red] (8.2+3,-0.15)-- (8.2+3,0.15);
    \node[] at (2.,0.2) {$\cdots$};
    \node[] at (7+2.,0.2) {$\cdots$};
    \draw[red] (7-2+1.2,-0.15)--(7-2+1.8,-0.15);
    \draw[red] (7-4+1.2,-0.15)--(7-4+1.8,-0.15);
    \draw[dashed,black] (5+7,0) -- (6+7,0);
    \draw[black] (5+7,0)-- (5+7,0.15);
    \draw[black] (6+7,0)-- (6+7,0.15);
    \draw[black] (4.8+7,0.15)-- (4.8+7,-0.15);
    \draw[black] (6.2+7,-0.15)-- (6.2+7,0.15);
    \draw[dashed,red] (7+7,0) -- (8+7,0);
    \draw[red] (7+7,0)-- (7+7,0.15);
    \draw[red] (8+7,0)-- (8+7,0.15);
    \draw[red] (6.8+7,0.15)-- (6.8+7,-0.15);
    \draw[red] (8.2+7,-0.15)-- (8.2+7,0.15);
    \draw[red] (7+7-2+1.2,-0.15)--(7+7-2+1.8,-0.15);
    \draw[red] (7+7-4+1.2,-0.15)--(7+7-4+1.8,-0.15);
    \node[] at (2+14.,0.2) {$\boldsymbol{\cdots}$};
    \draw[dashed,red] (7+3+7,0) -- (8+3+7,0);
    \draw[red] (7+3+7,0)-- (7+3+7,0.15);
    \draw[red] (8+3+7,0)-- (8+3+7,0.15);
    \draw[red] (6.8+3+7,0.15)-- (6.8+3+7,-0.15);
    \draw[red] (8.2+3+7,-0.15)-- (8.2+3+7,0.15);
    \draw[dashed,black] (5+7+7,0) -- (6+7+7,0);
    \draw[black] (5+7+7,0)-- (5+7+7,0.15);
    \draw[black] (6+7+7,0)-- (6+7+7,0.15);
    \draw[black] (4.8+7+7,0.15)-- (4.8+7+7,-0.15);
    \draw[black] (6.2+7+7,-0.15)-- (6.2+7+7,0.15);
    \draw[dashed,red] (7+7+7,0) -- (8+7+7,0);
    \draw[red] (7+7+7,0)-- (7+7+7,0.15);
    \draw[red] (8+7+7,0)-- (8+7+7,0.15);
    \draw[red] (6.8+7+7,0.15)-- (6.8+7+7,-0.15);
    \draw[red] (8.2+7+7,-0.15)-- (8.2+7+7,0.15);
    \draw[red] (7+7-2+1.2+7,-0.15)--(7+7-2+1.8+7,-0.15);
    \draw[red] (7+7-4+1.2+7,-0.15)--(7+7-4+1.8+7,-0.15);
    \node[] at (2.+21,0.2) {$\cdots$};
    \draw[red, dashed] (0+24,0) -- (1+24,0);
    \draw[red] (0+24,0)-- (0+24,.15);
    \draw[red] (1+24,0)-- (1+24,.15);
    \draw[red] (-0.2+24,0.15)-- (-0.2+24,-0.15);
    \draw[red] (1.2+24,-0.35)-- (1.2+24,0.15);
    \draw[red] (1.2+24,-0.35)-- (-0.2,-0.35);
    \draw (1.2+19,0) arc (0:180:0.7);
    \draw[dashed] (1.0+19,0) arc (0:180:0.5);
    \draw (1.2+12,0) arc (0:180:4.2);
    \draw (1.2+12-1.4,0) arc (0:180:2.8);
    \draw[dashed] (1+12,0) arc (0:180:4);
    \draw[dashed] (1+11,0) arc (0:180:3);
    } \, .
\end{align}
This gives a factor of $d_B^{C(\tau)}$ where $\tau$ is the non-crossing permutation of the $\alpha $ black lines. We can see from this diagram that depending on how the black lines are contracted, this restricts the allowed permutations for the red lines. This is why this computation is more complicated than for the PREE
where the black and red permutations simply factorized as $NC_{\alpha}\times NC_m$. In order for the global permutation to be non-crossing, the red permutations must be non-crossing within each block partitioned off by the black permutations. In the above example diagram, the black ``rainbow'' restricts the red permutation to be of the form $NC_{2m} \times \dots$. The identity permutation on the black density matrix on the right places no additional restrictions. 
In terms of equations, the diagrams may be summed as
\begin{align}
    \overline{\Tr \left[ \left(\sigma_A^{m}\rho_A \sigma_A^{m}\right)^{\alpha}\right]}  &= \frac{1}{(d_A d_B)^{{\alpha}(2m+1)}}\sum_{\tau \in NC_{\alpha}} d_B^{C(\tau)}\prod_{\zeta \in cyc(\eta^{-1}\circ \tau)} \sum_{\gamma \in NC_{2m|\zeta|}}d_A^{C(\eta^{-1}\circ \gamma)}d_B^{C(\gamma)}
    \label{SRRE_init}
\end{align}
where the product is over the cycles of $\eta^{-1}\circ \tau$ and $|\cdot|$ represents the length of the cycle.

We first focus on the $d_A < d_B$ regime where the inner sum may be computed as before
\begin{align}
    &\overline{\Tr \left[ \left(\sigma_A^{m}\rho_A \sigma_A^{m}\right)^{\alpha}\right]}  
    =\frac{\sum_{\tau \in NC_{\alpha}} d_B^{C(\tau)}\prod_{\zeta \in cyc(\eta^{-1}\circ \tau)}d_A d_B^{2 |\zeta| m} \, _2F_1\left(-2 |\zeta| m,1-2 |\zeta|
  m;2;\frac{d_A}{d_B}\right)}{(d_A d_B)^{{\alpha}(2m+1)}}
  \nonumber
  \\
  &\hspace{1cm}=\frac{\sum_{\tau \in NC_{\alpha}} d_A^{\alpha+1-C(\tau)}d_B^{C(\tau)+2m\alpha}\prod_{\zeta \in cyc(\eta^{-1}\circ \tau)}\, _2F_1\left(-2 |\zeta| m,1-2 |\zeta|
  m;2;\frac{d_A}{d_B}\right)}{(d_A d_B)^{{\alpha}(2m+1)}},
\end{align}
where in the second line, we have pulled out the factors of $d_A$ and $d_B$ from the product by enforcing the global permutation to be noncrossing.
This formula does not need $m$ to be an integer, so it is now safe to take the $m \rightarrow \frac{1-\alpha}{2\alpha}$ limit
\begin{align}
    &\overline{\Tr\left[\left(
   {
    \sigma^{\frac{1-\alpha}{2\alpha}}_A
    \rho_A 
    \sigma^{\frac{1-\alpha}{2\alpha}}_A
    }
    \right)^{\alpha}\right]}\nonumber
    \\
    &\hspace{1cm}=\sum_{\tau \in NC_{\alpha}}  \left(\frac{d_A}{d_B}\right)^{{\alpha}-C(\tau)}\prod_{\zeta \in cyc(\eta^{-1}\circ \tau)} \, _2F_1\left(\frac{|\zeta|({\alpha}-1)}{{\alpha}},1+|\zeta|-\frac{|\zeta|}{{\alpha}}
   ;2;\frac{d_A}{d_B}\right).
\end{align}
The product over cycle structures makes this formula still very difficult. Fortunately, Kreweras solved exactly this combinatorial problem about cycle structure in his landmark paper on non-crossing partitions \cite{KREWERAS1972333}. He found that the number of non-crossing permutations of $\{1,2,\dots, \alpha\}$ with cycle structure\footnote{This notation means that there are $m_i$ cycles of length $i$.} $\{m_i\}$ is given by, what we will call, the Kreweras number \cite{KREWERAS1972333, SIMION2000367}
\begin{align}
    K_{\{m_i\}} := \frac{{\alpha}!}{(\alpha-b+1)!m_1!\dots m_{\alpha}!}, \quad b := \sum_im_i.
\end{align}
Therefore, we can reorganize 
the sum such that 
there are no more references to permutations, only natural numbers
\begin{align}
    &\overline{\Tr\left[\left( \sigma_A^{\frac{1-\alpha}{2\alpha}}
    {
    \rho^{\ }_A 
    }
    \sigma_A^{\frac{1-\alpha}{2\alpha}}\right)^{\alpha}\right]} \nonumber
    \\
    &\hspace{1cm}=\sum_{m_1, \dots, m_{\alpha} = 0}^{\alpha} K_{\{m_i\}} \left(\frac{d_A}{d_B}\right)^{\sum_i m_i- 1}\prod_{m_i \neq 0} \, _2F_1\left(\frac{i({\alpha}-1)}{{\alpha}},1+i-\frac{i}{{\alpha}}
   ;2;\frac{d_A}{d_B}\right)^{m_i }.
   \label{exact_SRRE_moments}
\end{align}
This formula still presents a daunting task to evaluate in terms of elementary functions for generic $\alpha$, though it provides a tractable, controlled expansion in $d_A/d_B$. This is because, for small $d_A/d_B$, the hypergeometric function is close to one. We then must consider the smallest values of $\sum_i m_i$. First, we take only the leading term with $\sum_i m_i = 1$ (cyclic permutation)
\begin{align}
    \overline{\Tr\left[\left( \sigma_A^{\frac{1-\alpha}{2\alpha}}
    {
    \rho^{\ }_A
    }
    \sigma_A^{\frac{1-\alpha}{2\alpha}}\right)^{\alpha}\right]} &=   \, _2F_1\left({{\alpha}-1},\alpha
   ;2;\frac{d_A}{d_B}\right)+O\left(\frac{d_A}{d_B}\right).
\end{align}
This is not terribly useful because, as explained above, to this order, the RHS is exactly one, which would lead to the SRRE 
being identically zero. To find a nontrivial result, we need the next term where $\sum_i m_i = 2$ which can be achieved in many ways. These are the all the ways to sum to integers between $1$ and $\alpha-1$ to $\alpha$
\begin{align}
    &\overline{\Tr\left[\left( \sigma_A^{\frac{1-\alpha}{2\alpha}}
    {
    \rho^{\ }_A 
    }\sigma_A^{\frac{1-\alpha}{2\alpha}}\right)^{\alpha}\right]} =\frac{d_A}{d_B}\sum_{j = 1}^{\lfloor \frac{\alpha}{2} \rfloor} K_{\{m_j , m_{\alpha - j} = 1\}} \, _2F_1\left(\frac{j({\alpha}-1)}{{\alpha}},1+j-\frac{j}{{\alpha}}
   ;2;\frac{d_A}{d_B}\right)
   \nonumber
   \\
   &\times\, _2F_1\left(\frac{(\alpha -j)({\alpha}-1)}{{\alpha}},1+(\alpha -j)-\frac{\alpha -j}{{\alpha}}
   ;2;\frac{d_A}{d_B}\right)+\, _2F_1\left({{\alpha}-1},\alpha
   ;2;\frac{d_A}{d_B}\right)\nonumber
   \\
   &\hspace{10cm}+O\left(\frac{d_A}{d_B}\right)^2,
   \label{SRRE_first_order}
\end{align}
where the floor function in the sum ensures that we do not double count. The Kreweras number is
\begin{align}
    K_{\{m_j , m_{\alpha - j} = 1\}} = \begin{cases}
        \alpha, & j \neq \frac{\alpha}{2}
        \\
        \frac{\alpha}{2}, & j = \frac{\alpha}{2}
    \end{cases}.
\end{align}
$ j =\frac{\alpha}{2}$ will only occur when $\alpha$ is even. The exact form of the hypergeometric functions in the sum are not important at this order because for small $d_A/d_B$, they are all close to one. Therefore, only the Kreweras number is important. We can easily compute the sum at this order for any integer $\alpha$ and find that this parity effect disappears,
\begin{align}
    \overline{\Tr\left[\left( \sigma_A^{\frac{1-\alpha}{2\alpha}}
    {\rho^{\ }_A} \sigma_A^{\frac{1-\alpha}{2\alpha}}\right)^{\alpha}\right]} =
        1+ \alpha (\alpha - 1)\frac{d_A}{d_B}+O\left(\frac{d_A}{d_B}\right)^2,  
\end{align}
leading to an SRRE of
\begin{align}
    \overline{\tilde{D}_{\alpha}(\rho_A || \sigma_A)} = \alpha \frac{d_A}{d_B}+O\left(\frac{d_A}{d_B}\right)^2.
    \label{SRRE_smalldA}
\end{align}
Note that this agrees with the previously derived von Neumann relative entropy in the relevant $\alpha \rightarrow 1$ limit. Moreover, it obeys the data processing inequality for all positive $\alpha $ if we take the quantum channel to be the partial trace.

The Uhlmann fidelity is found\footnote{We note that an exact expression was recently found for the fidelity of two random density matrices, consistent with our large-$N$ results \cite{2021arXiv210502743L}. Our result is complementary as the exact expression is very complicated and not tractable at large-$N$.} by setting $\alpha = 1/2$
\begin{align}
    \overline{F(\rho_A||\sigma_A)} = 1-\frac{d_A}{2d_B} + O\left(\frac{d_A}{d_B}\right)^2.
\end{align}
At this order, the Uhlmann fidelity is identical to Holevo's just-as-good fidelity \eqref{holevo_fidelity}. The value of the fidelity exactly at $d_A = d_B$ was found in Ref.~\cite{2016PhRvA..93f2112P} to be $\frac{9}{16}$ and additional results may be found in Ref.~\cite{2005PhRvA..71c2313Z}.

We can also evaluate the SRRE exactly for any integer moment using \eqref{exact_SRRE_moments}. Here, we work out the least tedious case of $\alpha = 2$ which is also known as the \textit{collision relative entropy} \cite{2005PhDT.......176R}. In this case, \eqref{SRRE_first_order} is actually exact and does not contain $O\left(\frac{d_A}{d_B}\right)^2$ corrections. We only sum over $j = 1$, so
\begin{align}
    &\overline{\Tr\left[ \left(\sigma_A^{-\frac{1}{4}}\rho_A \sigma_A^{-\frac{1}{4}}\right)^2\right]} =\frac{d_A}{d_B}\, _2F_1\left(\frac{1}{{2}},\frac{3}{{2}}
   ;2;\frac{d_A}{d_B}\right)^2+\frac{1}{1-\frac{d_A}{d_B}},
\end{align}
leading to an SRRE of
\begin{align}
    \overline{\tilde{D}_2(\rho_A || \sigma_A)} = \begin{cases}\log\left[\frac{d_A}{d_B}\, _2F_1\left(\frac{1}{{2}},\frac{3}{{2}}
   ;2;\frac{d_A}{d_B}\right)^2+\frac{1}{1-\frac{d_A}{d_B}}\right], & d_A < d_B
   \\
   \infty, & d_A > d_B
   \end{cases},
   \label{SRRE2_eq}
\end{align}
where we have set the SRRE to infinity when $d_A > d_B$ because the von Neumann relative entropy is infinite in this regime and the SRRE's are monotonically increasing with $\alpha$.
It is straightforward to evaluate the higher integer SRRE's if desired.

It is equally important to investigate the opposite regime where $d_A/d_B$ is large. In this case, the inner sum in \eqref{SRRE_init} gives
\begin{align}
    &\overline{\Tr\left[\left( \sigma_A^m\rho_A \sigma_A^m\right)^{\alpha} \right]}
    =\frac{\sum_{\tau \in NC_{\alpha} } d_B^{C(\tau)}\prod_{\zeta \in cyc(\eta^{-1}\circ \tau)}d_A^{2 |\zeta| m} d_B \, _2F_1\left(-2 |\zeta| m,1-2 |\zeta|
   m;2;\frac{d_B}{d_A}\right)}{(d_A d_B)^{{\alpha} (2m+1)}}
   \nonumber
   \\
   &\hspace{1cm}=\frac{\sum_{\tau \in NC_{\alpha}} \left(\frac{d_B}{d_A}\right)^{C(\tau)+C(\eta^{-1}\circ\tau)} d_A^{{\alpha}(2m+1)+1} \prod_{\zeta \in cyc(\eta^{-1}\circ \tau)}  \, _2F_1\left(-2 |\zeta| m,1-2 |\zeta|
   m;2;\frac{d_B}{d_A}\right)}{(d_A d_B)^{{\alpha}(2m+1)}}.
\end{align}
Now that we have done the sums over the $m$ permutations, we can safely take $m \rightarrow \frac{1-\alpha}{2\alpha}$ and rewrite the sum in terms of Kreweras numbers 
\begin{align}
    &\overline{\Tr\left[\left( \sigma_A^{\frac{1-{\alpha}}{2{\alpha}}}\rho_A \sigma_A^{\frac{1-{\alpha}}{2{\alpha}}}\right)^{\alpha}\right]}
   \nonumber
   \\
   &\hspace{1cm}=\left(\frac{d_B}{d_A}\right)^{\alpha }\sum_{m_1,\dots, m_{\alpha} = 0}^{\alpha}K_{\{m_i\}}\prod_{m_i \neq 0} \, _2F_1\left(\frac{i(\alpha-1)}{\alpha},1+\frac{i(\alpha-1)}{\alpha};2;\frac{d_B}{d_A}\right)^{m_i}.
\end{align}
This is an exact formula, but is difficult to evaluate away from limits. For large $d_A/d_B$, all of the hypergeometric functions are close to one so all that matters is the total number of noncrossing permutations, which is given by the Catalan number
\begin{align}
    C_{\alpha}:= \frac{1}{\alpha + 1}\binom{2\alpha}{\alpha}.
\end{align}
Therefore, at leading order, we have
\begin{align}
    \overline{\Tr\left[\left( \sigma_A^{\frac{1-{\alpha}}{2{\alpha}}}\rho^{\ }_A \sigma_A^{\frac{1-{\alpha}}{2{\alpha}}}\right)^{\alpha}
    \right]}
   &=C_{\alpha}\left(\frac{d_B}{d_A}\right)^{\alpha } + O\left(\frac{d_B}{d_A}\right)^{\alpha + 1}.
\end{align}
Unlike for small $d_A/d_B$, there are no additional terms at leading order. The SRRE is thus
\begin{align}
    \overline{\tilde{D}_{\alpha}(\rho_A|| \sigma_A)} = \frac{\alpha}{{\alpha}-1}\log \left[\frac{d_B}{d_A} \right]+ \frac{1}{{\alpha}-1}\log\left[ C_{\alpha}\right] + O\left(\frac{d_B}{d_A}\right).
    \label{SRRE_largedA}
\end{align}
Important to note is that this is only well-defined for ${\alpha} < 1$. This is to be expected because of rank deficiency. In the well-defined regime, the SRRE is monotonic in $\alpha$ and manifestly obeys the data processing inequality.

We can evaluate the asymptotic expression at $\alpha = 1/2$ to find the Uhlmann fidelity
\begin{align}
    \overline{F(\rho_A || \sigma_A)} = \frac{64d_B}{9\pi^2d_A} + O\left( \frac{d_B}{d_A}\right)^{3/2}.
\end{align}
The prefactor comes from the Catalan number which is nonintegral for noninteger $\alpha$. We see that the fidelity is inversely proportional to $d_A/d_B$, decaying to zero when subsystem $A$ occupies most of the Hilbert space. The full spectrum of $\sigma_A^{\frac{1-\alpha}{2\alpha}}\rho_A\sigma_A^{\frac{1-\alpha}{2\alpha}}$, and hence the fidelity, may be evaluated using techniques of free probability theory. This is completed in Appendix \ref{free_prob_app}. The answer is the free multiplicative convolution of two Marchenko-Pastur distributions.

\subsection{Trace distance}

The final distinguishability measure we discuss is the trace distance. This is the ideal measure when discussing one-shot state discrimination \eqref{trace_distance_operational}. More general than \eqref{trace_norm_def}, we can define an $\alpha$-norm version of the trace distance
\begin{align}
    T_{\alpha}(\rho_A|| \sigma_A) := \frac{1}{2^{1/{\alpha}}}|\rho_A -\sigma_A|_{\alpha},
\end{align}
where the $\alpha$-norm of an operator, $A$, is defined as
\begin{align}
    |A |_{\alpha} := \left(\Tr \left[ \sqrt{A^{\dagger} A}^{\alpha}\right]\right)^{1/\alpha}.
\end{align}
For even $\alpha$ and Hermitian $A$, we can dispose of the square root
\begin{align}
    |A |_{\tilde{\alpha}} = \left(\Tr \left[  A^{2\tilde{\alpha}}\right]\right)^{1/{2\tilde{\alpha}}}, \quad 2\tilde{\alpha} :=  \alpha.
\end{align}
The trace norm is then the $\tilde{\alpha} \rightarrow 1/2$ limit of this expression.

The replica trick, while only requiring a single replica parameter, is quite difficult as we must compute all even powers of $\rho_A - \sigma_A$ which involves arbitrary mixing of $\rho_A$ and $\sigma_A$ \cite{2019PhRvL.122n1602Z}
\begin{align}
    \Tr\left[ (\rho_A- \sigma_A)^{\alpha}\right] = \sum_{\mathcal{S} \in \mathcal{P}(\{1,2,\dots,  \alpha\})}(-1)^{|\mathcal{S}|}\Tr (\tau_{\mathcal{S}_1}\dots \tau_{\mathcal{S}_{\alpha}} ),
\end{align}
where the sum runs over the power set of $\{ 1,2,\dots, \alpha\}$ and $|\cdot|$ is the cardinality of the subset. We have $\tau_{\mathcal{S}_i} = \rho_A$ 
if $i \in \mathcal{S}$ and $\tau_{\mathcal{S}_i} = \sigma_A$ 
if $i \notin \mathcal{S}$. Each term in the sum can be expressed as an appropriate summation over the symmetric group, though this is far from straightforward.

Consider the small $d_A/d_B$ limit. In this case, the terms that maximize $C(\tau)$ will dominate the sum. This is when $\tau$ is the identity. This permutation is always present, regardless of $\mathcal{S}$ and universally contributes as $d_A^{1-\alpha}$ which in the $\alpha \rightarrow 1$ limit contributes at $O(1)$. To see if this contributes to the overall sum, we need to understand the cardinalities. The number of subsets with cardinality $k$ is given by the binomial coefficient, so the identity contributes as
\begin{align}
    \overline{\Tr\left[ (\rho_A- \sigma_A)^{\alpha}\right]} \supset \sum_{k = 0}^{{\alpha}}\binom{{\alpha}}{k}(-1)^{k}d_A^{1-{\alpha}} = 0.
\end{align}
To get a nontrivial answer, we must therefore move beyond the identity permutation. This is to be expected because the trace distance will be small for small $d_A/d_B$ and should not be $O(1)$. The next leading term is when $C(\tau) = \alpha-1$ which corresponds to the identity on all sites except for two which are swapped; this is always non-crossing. This contributes universally as $d_A^{2-\alpha} d_B^{-1}$, which will lead to the $O(d_A/d_B)$ contribution. The combinatorics are slightly more complicated. If $|\mathcal{S}| = k$, then there are $
    \binom{k}{2}+\binom{\alpha-k}{2}
$
ways to have a single pairing because we can only choose pairings within the block of $k$ $\rho_A$'s or $n-k$ $\sigma_A$'s. Therefore, the contribution at this order is
\begin{align}
    \overline{\Tr\left[ (\rho_A- \sigma_A)^{\alpha}\right]} \supset \sum_{k = 0}^{\alpha}\binom{{\alpha}}{k}\left(\binom{k}{2}+\binom{{\alpha}-k}{2}\right)(-1)^{k}d_A^{2-{\alpha}} d_B^{-1}= 2\delta_{\alpha, 2}d_A^{2-{\alpha}} d_B^{-1}.
\end{align}
This is not an analytic function, so the $\tilde{\alpha} \rightarrow 1/2$ limit is quite ambiguous. We are free to work to higher orders, though we will argue that this will not help our cause. 

In the small $d_B/d_A$ regime, 
the expansion is
more involved. The leading terms come from maximizing $C(\eta^{-1}\circ\tau)$. We can only have $C(\eta^{-1}\circ\tau) = 
\alpha$ in the case that $\mathcal{S} = \{ 1,2,\dots, \alpha\}$ or is empty because otherwise, $\tau = \eta$ will not be an allowed permutation. At this order, we therefore have
\begin{align}
    \overline{\Tr\left[ (\rho_A- \sigma_A)^{\alpha}\right]} = \begin{cases}
       2 d_B^{1-\alpha}, & \alpha \in 2\mathbb{Z}
       \\
       0 & \alpha \in 2\mathbb{Z} + 1
    \end{cases}.
\end{align}
The parity effect arises from the exponent of the sign in the sum. Analytically continuing the even integers to one, we find
\begin{align}
    \overline{T(\rho_A || \sigma_A)} = 1 + O\left( \frac{d_B}{d_A}\right),
\end{align}
meaning that the states are nearly maximally distant. To understand how the trace distance approaches one, we need to work at the next order. The noncrossing permutations that give $C(\eta^{-1}\circ\tau) = 
\alpha - 1$ are those that are of the form $\eta_{\alpha_1}\times \eta_{\alpha_2}$. This means that the $\rho_A$'s and $\sigma_A$'s must be in disjoint blocks i.e.~$\mathcal{S}$ is a set only containing consecutive integers. There are $(\alpha-1)$ ways to partition $\alpha$ into nonzero integers $\alpha_1$ and $\alpha_2$ if we define the tuple ($\alpha_1, \alpha_2$) to be distinct from ($\alpha_2, \alpha_1$). There is an additional factor of $\alpha$ coming from the rotations of $\mathcal{S}$ to $\mathcal{S}+1$, leading to  
\begin{align}
    \overline{\Tr\left[ (\rho_A- \sigma_A)^{\alpha}\right]} \supset \sum_{k = 1}^{\alpha -1}\alpha(-1)^{k}d_A^{-1}d_B^{2-{\alpha}} = \begin{cases}
       -\alpha d_A^{-1}d_B^{2-{\alpha}}, & \alpha \in 2\mathbb{Z}
       \\
       0 & \alpha \in 2\mathbb{Z} + 1
    \end{cases}.
\end{align}
Finally, when $\mathcal{S} = \{ 1,2,\dots, \alpha\}$ or $\mathcal{S} = \emptyset$, we again can partition the elements into $\alpha_1$ and $\alpha_2$ size blocks but this time ($\alpha_1, \alpha_2$) and ($\alpha_2, \alpha_1$) are indistinguishable. Therefore, for even $\alpha$, there are $\alpha/2$ possibilities while for odd $\alpha$, there are only $(\alpha-1)/2$ possibilities plus the rotation factors\footnote{The rotation factor for even $\alpha$ when $\alpha_1 =\alpha_2$ is only $\alpha/2$ because of indistinguishability.}, leading to
\begin{align}
    \overline{\Tr\left[ (\rho_A- \sigma_A)^{\alpha}\right]} \supset  \begin{cases}
       \alpha(\alpha -1 ) d_A^{-1}d_B^{2-{\alpha}}, & \alpha \in 2\mathbb{Z}
       \\
       0 & \alpha \in 2\mathbb{Z} + 1
    \end{cases},
\end{align}
where the odd terms are trivial because the  $\mathcal{S} = \{ 1,2,\dots, \alpha\}$ and $\mathcal{S} = \emptyset$ terms exactly cancel in the sum due to the power of the sign.
Taking the $\alpha \rightarrow 1$ limit of even $\alpha$, we find the trace norm at this order to be
\begin{align}
    \overline{T(\rho_A || \sigma_A)} = 1 -\frac{d_B}{2d_A}+  O\left( \frac{d_B}{d_A}\right)^2.
    \label{trace_distance_largedA}
\end{align}
The trace distance may be evaluated away from limits using free probability techniques as we review in Appendix \ref{free_prob_app} \cite{2015arXiv151107278M}
\begin{align}
    \overline{T(\rho_A || \sigma_A)} = \begin{cases}
        \frac{\sqrt{{d_A (2d_B-d_A)}}
   (d_A+d_B)+d_B(4d_A-2d_B) \sin
   ^{-1}\left(\sqrt{\frac{{d_A}}{{2d_B}}}\right)}{2 \pi 
  d_A d_B}, & d_A < 2d_B
   \\
   1 -\frac{d_B}{2d_A}, & d_A > 2d_B
    \end{cases}.
    \label{tracedistance_exact}
\end{align}
Interestingly, our asymptotic formula was exact. The trace distance at $d_A = d_B$ was found in Ref.~\cite{2016PhRvA..93f2112P} to be $\frac{1}{2}+\frac{1}{\pi}$.


Without free probability, we can still use bounds from Section \ref{disting_sec} to place strong constraints on the trace distance. In particular, this is helpful for the small $d_A/d_B$ where we were unable to find an analytic answer. First, we use Pinsker's inequality \eqref{pinsker_eq}, using the relative entropy \eqref{relative_random} as an upper bound for $d_A < d_B$
\begin{align}
    \overline{T({\rho_A} ||
    { \sigma_A})}
    \leq \sqrt{2+\frac{d_A}{ d_B}+2\left(\frac{d_B}{d_A}-1\right) \log
   \left(1-\frac{d_A}{d_B}\right)}.
\end{align}
This is only a useful bound when the RHS is less than one. Recall that when $d_A = d_B$, the RHS will be $\sqrt{3} > 1$. For small $d_A/d_B$, we have
\begin{align}
    \overline{T({\rho_A} ||
    {\sigma_A})}\leq \sqrt{\frac{2d_A}{d_B}} + O\left(\frac{d_A}{d_B} \right)^{3/2}.
\end{align}
This means that the trace distance is very small, though without a lower bound, we cannot yet say that we could not find the leading order trace distance from the above expansion. We can determine a lower bound using Holevo's just-as-good fidelity which is, in general, stronger than the lower bound from Uhlmann fidelity
\begin{align}
    1- {}_2F_1\left(\frac{1}{2},-\frac{1}{2};2;\frac{d_A}{d_B}\right)^2 \leq 
    \overline{T({\rho_A} || 
    {\sigma_A} )}\leq  \sqrt{1- {}_2F_1\left(\frac{1}{2},-\frac{1}{2};2;\frac{d_A}{d_B}\right)^4},
    \label{T_bound_smalldA}
\end{align}
where we have also included the upper bound that is, in general, weaker than the upper bound from Uhlmann fidelity.
For small $d_A/d_B$, this gives
\begin{align}
    \frac{d_A}{4d_B} + O\left(\frac{d_A}{d_B} \right)^2 \leq \overline{T(
    {
    \rho_A} 
    || 
    {
    \sigma_A})} \leq  \sqrt{\frac{d_A}{2d_B}}+ O\left(\frac{d_A}{d_B} \right)^{3/2}.
\end{align}
This is a stronger upper bound than from Pinsker's inequality, though the scaling is still not nailed down, only constrained to between linear and square root with $d_A/d_B$. Because the scaling is at most linear, we cannot hope to find the leading order behavior from the expansion at the beginning of this subsection because the linear term was zero if we are to trust the ``continuation.'' 
The Uhlmann fidelity for small $d_A/d_B$ does not strengthen the upper bound at leading order.

We also want to characterize the $d_A > d_B$ regime. While Pinsker's inequality does not help here because the relative entropy is infinite, Holevo's just-as-good fidelity places nontrivial upper and lower bounds
\begin{align}
    1 - \frac{d_B}{d_A} \,_2F_1\left(\frac{1}{2},-\frac{1}{2};2;\frac{d_B}{d_A}\right)^2 \leq 
    \overline{T(
    {
    \rho_A} 
    || 
    {
    \sigma_A})}  \leq \sqrt{1-\frac{d_B^2}{d_A^2} \,_2F_1\left(\frac{1}{2},-\frac{1}{2};2;\frac{d_B}{d_A}\right)^4}.
    \label{T_bound_largedA}
\end{align}
For small $d_B/d_A$, this is
\begin{align}
    1 - \frac{d_B}{d_A} + O\left(\frac{d_B}{d_A}\right)^2 \leq \overline{T(
    {
    \rho_A} 
    || 
    {
    \sigma_A})}  \leq  1- \frac{d_B^2}{ 2 d_A^2}+ O\left(\frac{d_B}{d_A}\right)^3 ,
\end{align}
meaning the states are almost as far away from each other as possible, approaching one exponentially in $N_A-N_B$. The upper bound can be improved by the Uhlmann fidelity at leading order such that the scaling behavior is completely fixed
\begin{align}
    1 - \frac{d_B}{d_A} + O\left(\frac{d_B}{d_A}\right)^2 \leq \overline{T(
    {
    \rho_A} 
    || 
    {
    \sigma_A})}  \leq  1- \frac{32 d_B}{9\pi^2 d_A}+ O\left(\frac{d_B}{d_A}\right)^2 .
\end{align}
These bounds are consistent with the analytic expressions \eqref{trace_distance_largedA} and \eqref{tracedistance_exact}.

\subsection{Small-$N$ numerics}

All of our computations thus far have been in the limit where both $d_A$ and $d_B$ are large. It is important to ask whether these asymptotic results are accurate when $d_A$ and $d_B$ are finite. One motivation is if these predictions can be observed in experiments and Noisy Intermediate-Scale Quantum (NISQ) technology \cite{2018arXiv180100862P}. Of course, the Hilbert space dimensions are exponentially large in the number of qubits, so there is hope that our results are predictive for small-scale experiments. In this section, we numerically compute the various distance measures and compare to the asymptotic formulas. This serves as a further consistency check of our results, which we find to be extraordinarily accurate. 

\begin{figure}
    \centering
    \includegraphics[width = .32\textwidth]{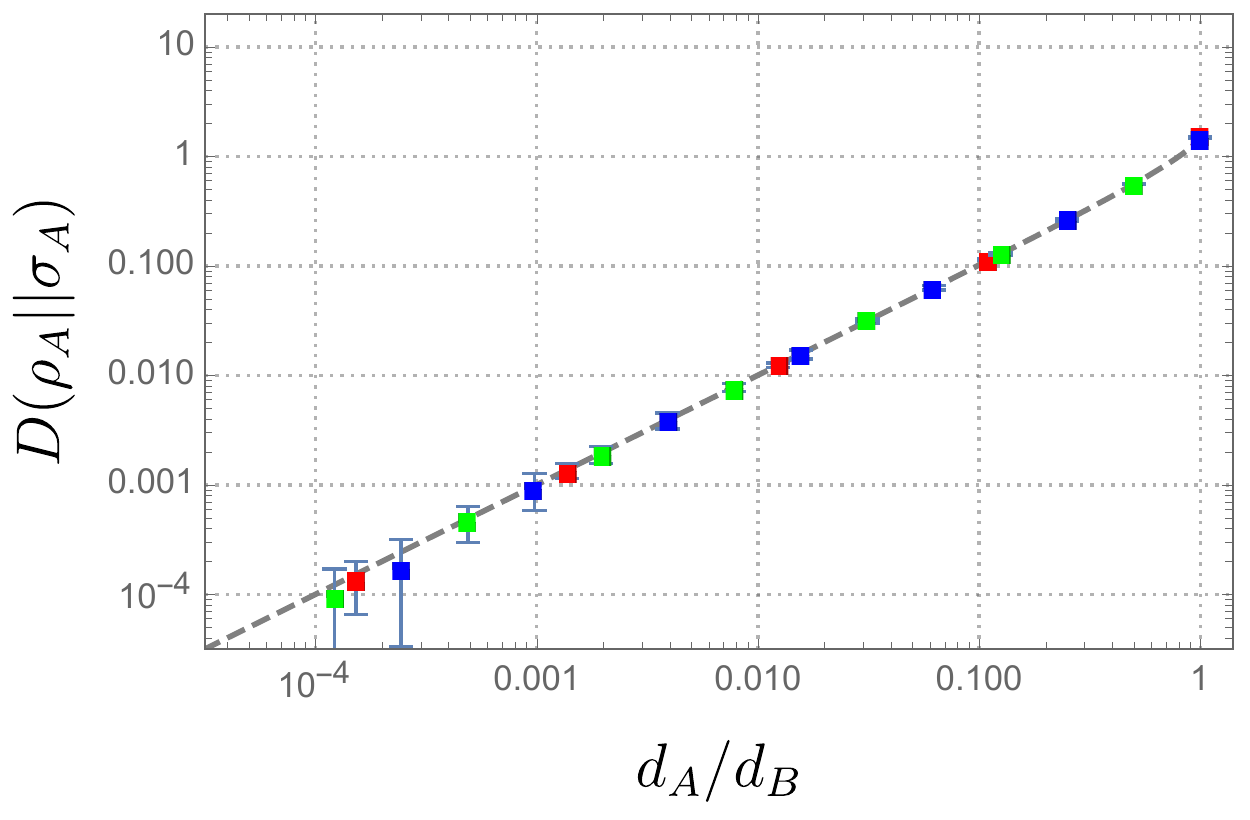}
    \includegraphics[width = .32\textwidth]{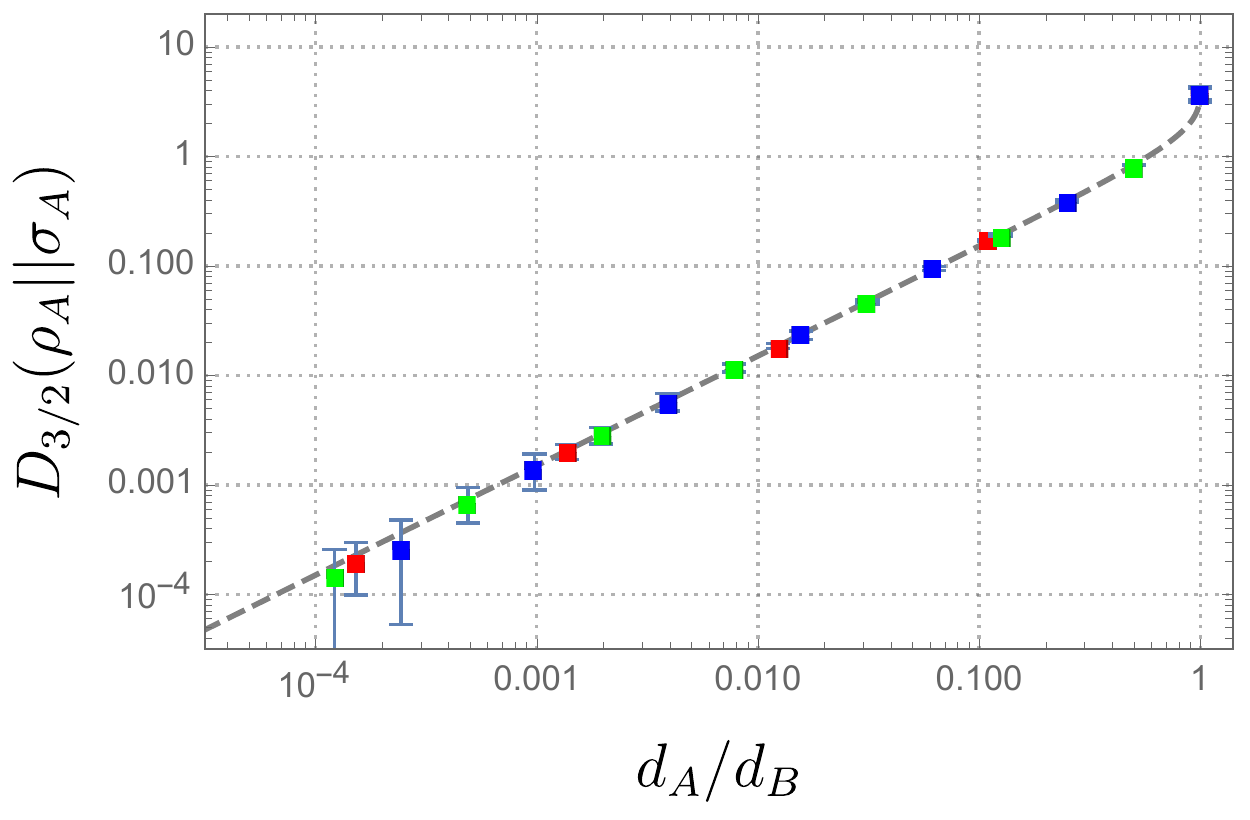}
    \includegraphics[width = .32\textwidth]{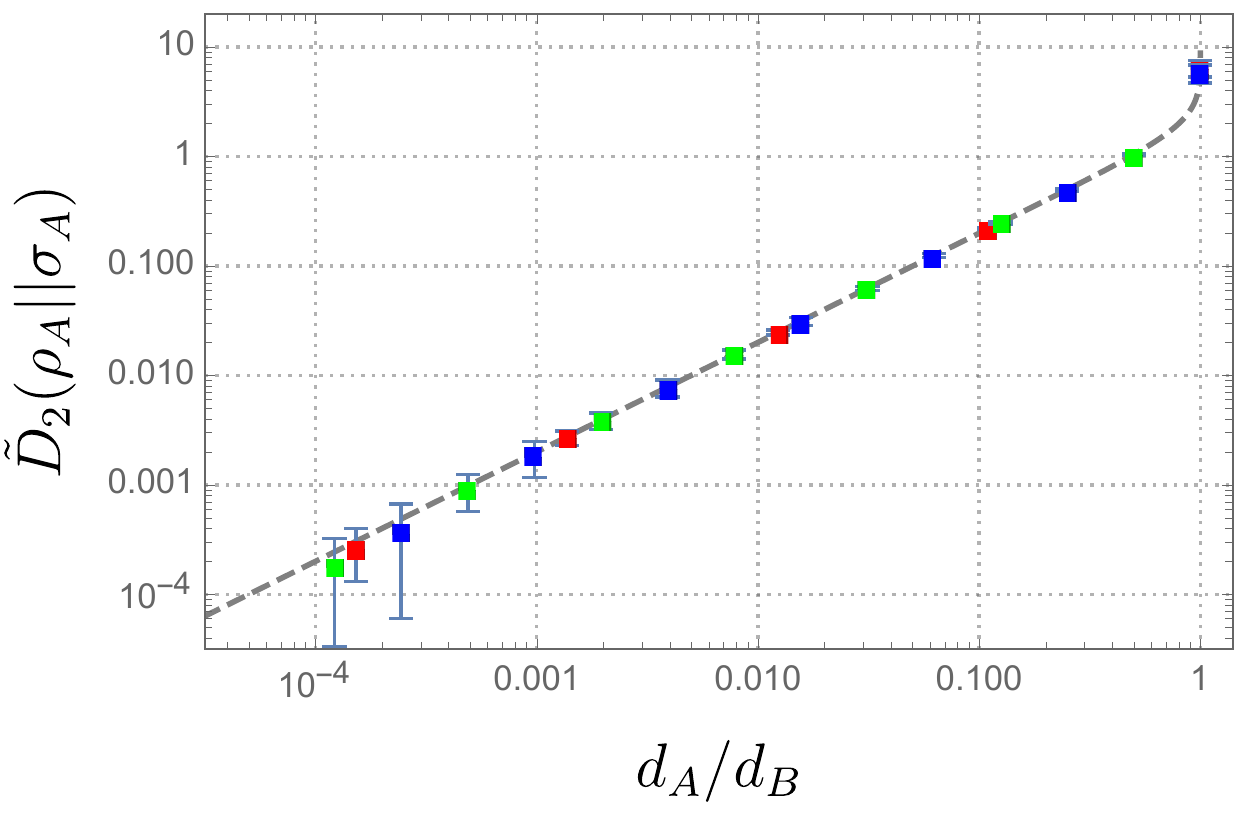}
    \caption{Three representative examples of relative entropies are shown, all of which with $\alpha \geq 1$, so they are infinite for $d_A/d_B > 1$. Left: von Neumann relative entropy with dashed line given by \eqref{relative_random}. Center: PRRE with $\alpha = 3/2$ with dashed line given by \eqref{PRRE_eq}. Right SRRE with $\alpha = 2$ with dashed line given by \eqref{SRRE2_eq}. The data are given for total Hilbert space dimensions of $2^{14}$ (blue), $2^{15}$ (green), and $3^{10}$ (red). The error bars represent the statistical fluctuations in the $10^3$ disorder realizations which decay for large Hilbert spaces.}
    \label{relative_entropy_numerics}
\end{figure}

In Fig.~\ref{relative_entropy_numerics}, we plot the von Neumann relative entropy, $D_{3/2}$, and $\tilde{D}_2$. All of these quantities are infinite for $d_A > d_B$ due to the rank deficiencies in the reduced density matrix. For this reason, we are able to sample very large Hilbert space dimensions because the bottleneck on classical computers is $d_A$ and not the total system size. We find very accurate agreement between the exact large-$N$ predictions and the small-$N$ numerics. The fluctuations in the entropies are noticeably larger for small $d_A$ because of the subleading corrections that we have thus far ignored.

\begin{figure}
    \centering
    \includegraphics[width = .48\textwidth]{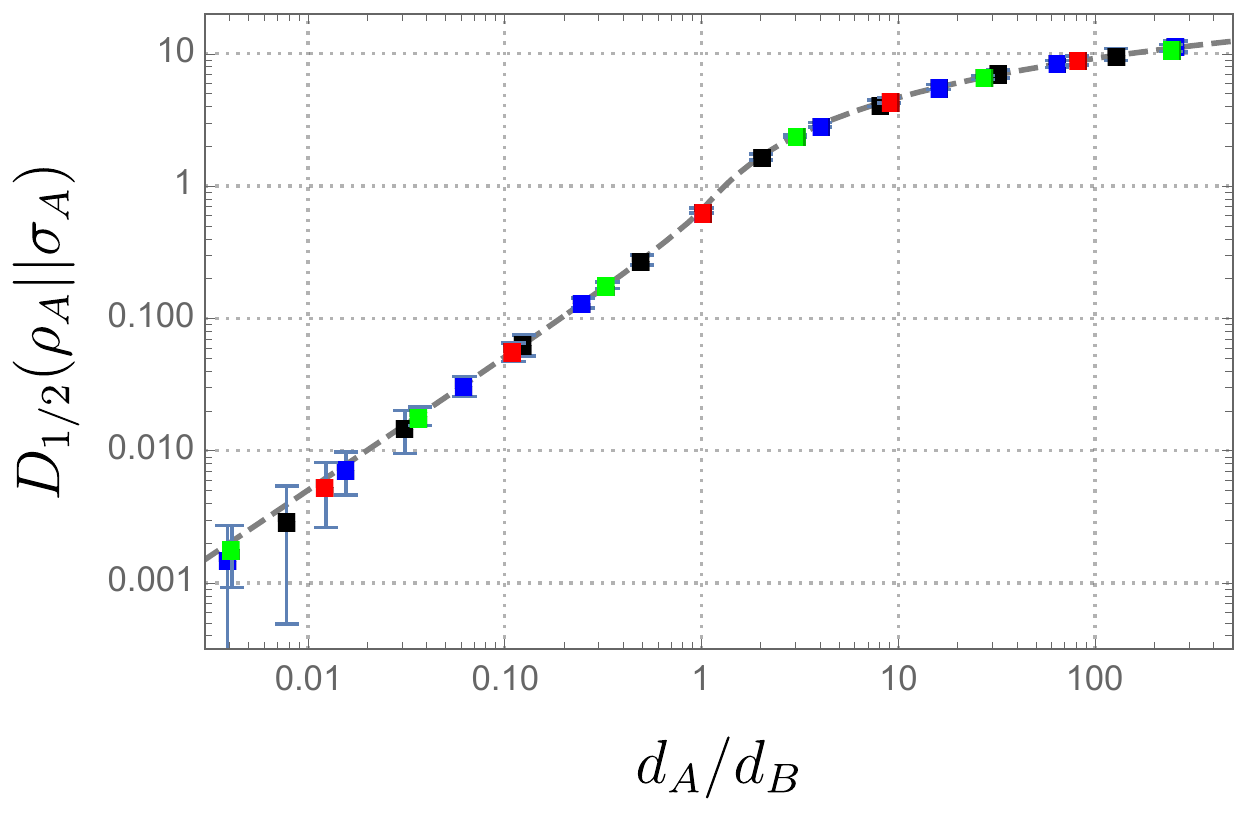}
    \includegraphics[width = .48\textwidth]{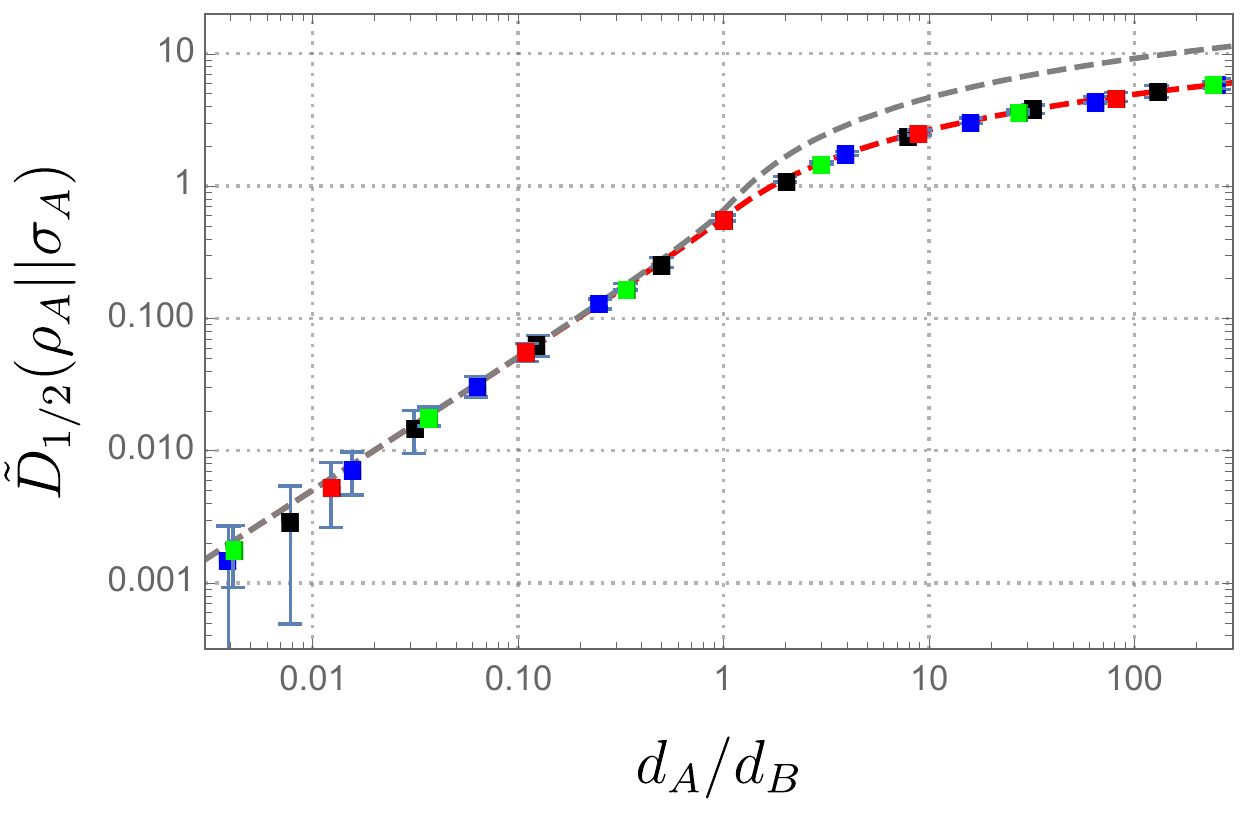}
    \caption{Representative examples of relative entropies are shown for $\alpha < 1$ such that there are no divergences. Left: PRRE for $\alpha = 1/2$ (related to Holevo's just-as-good fidelity) with dashed line given by \eqref{PRRE_eq}. Right: SRRE for $\alpha = 1/2$ (related to Uhlmann fidelity) with dashed red lines given by the answer from the spectrum \eqref{MP_squared_eq}. The grey line is the upper bound from the PRRE. The data are given for total Hilbert space dimensions $2^9$ (black), $2^{10}$ (blue), $3^6$ (red), and $3^7$ (green). The error bars represent the statistical fluctuations in the $10^3$ disorder realizations which decay for large Hilbert spaces. }
    \label{REhalf_numerics}
\end{figure}

In Fig.~\ref{REhalf_numerics}, we investigate the other regime by plotting $D_{1/2}$ and $\tilde{D}_{1/2}$. These quantities are related to Holevo's just-as-good and Uhlmann fidelities respectively and are therefore well-defined in the $d_A > d_B$ regime. This limits the Hilbert space sizes we can probe, though we still find very accurate agreement with the large-$N$ analysis.

\begin{figure}
    \centering
    \includegraphics[width = .48\textwidth]{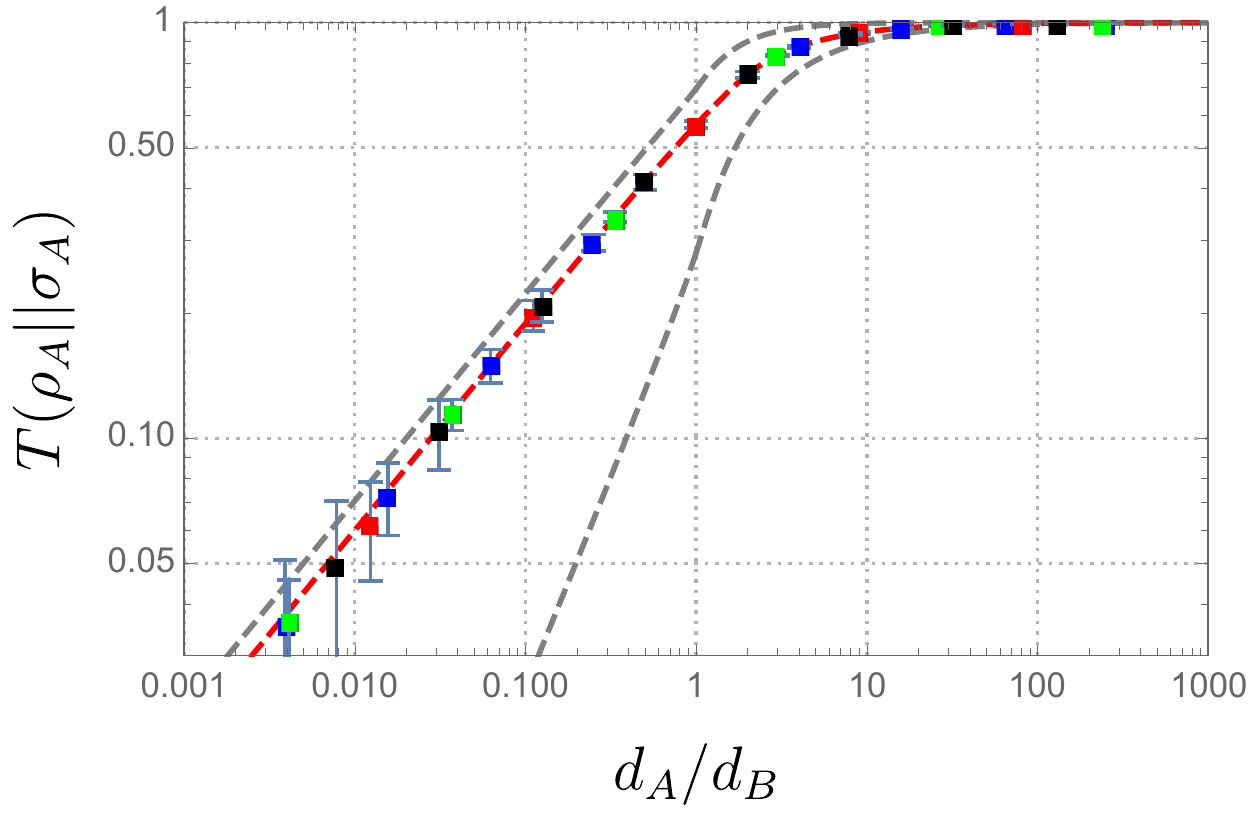}
    \includegraphics[width = .48\textwidth]{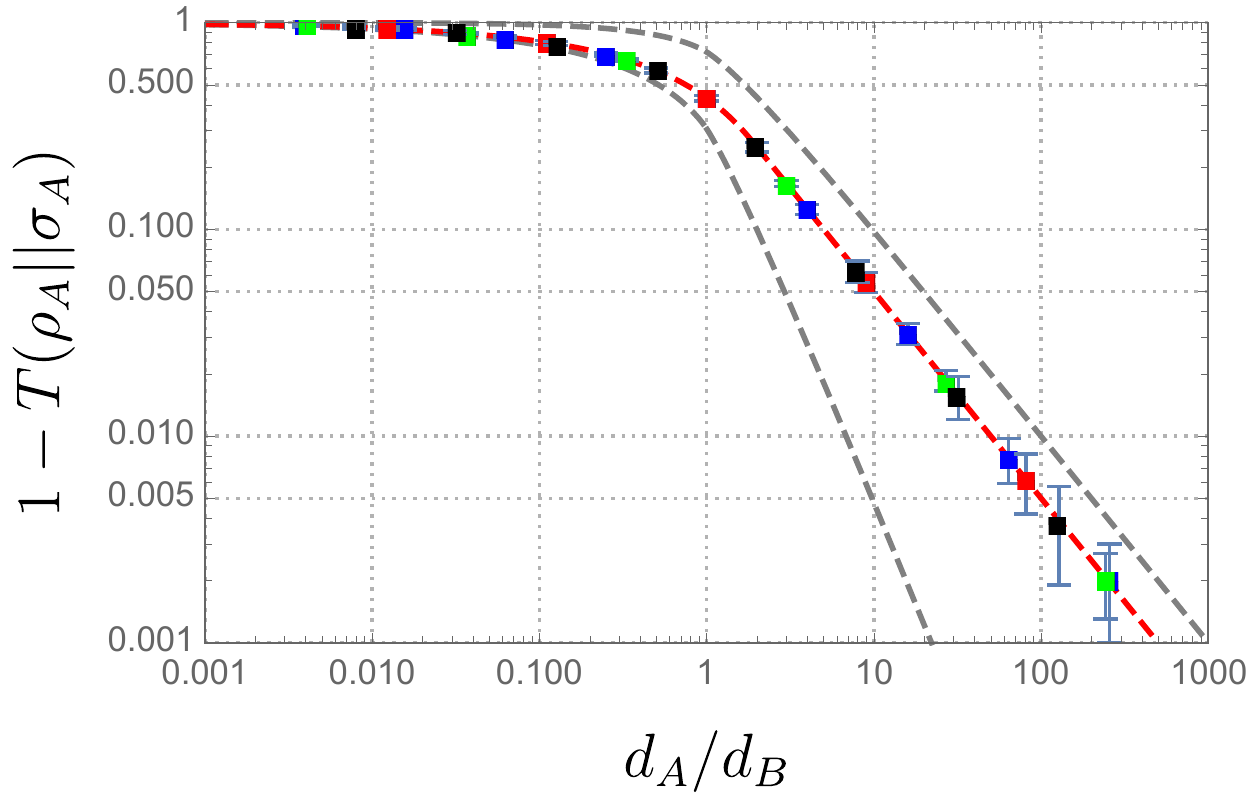}
    \caption{Left: The trace distance is shown with dashed red line given by \eqref{tracedistance_exact} and grey lines given by the bounds from the fidelity \eqref{T_bound_smalldA} and \eqref{T_bound_largedA}. Right: One minus the trace distance is shown to display the approach to one. The data are given for total Hilbert space dimensions $2^9$ (black), $2^{10}$ (blue), $3^6$ (red), and $3^7$ (green). The error bars represent the statistical fluctuations in the $10^3$ disorder realizations which decay for large Hilbert spaces.}
    \label{trace_distance_numerics}
\end{figure}

Finally, in Fig.~\ref{trace_distance_numerics}, we plot the trace distance, examining both the small $d_A/d_B$ and large $d_A/d_B$ regimes. The large-$N$ expressions precisely agree with numerics and are bounded within the fidelities.


\section{Distinguishing black holes}
\label{black_hole_sec}

While studying random states is interesting in its own right, the physical implications of our results becomes significantly richer when we apply them to gravitational systems. We will explain how the connection between random states and gravitational systems is more than an analogy and in some ways, quantitatively identical. 

\subsection{Fixed-area states in holography}

In quantum field theory, we compute the moments of reduced density matrices by evaluating the partition function on certain replica manifolds \cite{1994NuPhB.424..443H,2004JSMTE..06..002C}. These are glued according to the relevant trace structure. If the quantum field theory is holographic, we may map the calculation to an evaluation of the gravitational path integral with boundary conditions prescribed by this trace structure. In the gravitational path integral, we are instructed to sum over all geometries with the given boundary conditions. In the derivation of the Ryu-Takayanagi formula \cite{2013JHEP...08..090L}, only replica symmetric geometries were considered. In contrast, we find that replica symmetry breaking saddles are important for the evaluation of relative entropies.

In general, it is very difficult to evaluate the gravitational path integral for multiple replicas. This is because the nontrivial coupling between the replicas leads to backreaction, changing the bulk geometry \cite{2016NatCo...712472D}. A great simplification in the gravitational path integral can be made if we focus on ``fixed area states'' \cite{2019JHEP...05..052A,2019JHEP...10..240D}. These are states where the area of one or more surface is fixed and not integrated over. In general, the different replicas will not backreact among themselves, so we are left with copies of the original bulk geometry except for potential conical singularities appearing at the locations of the fixed surfaces. 

\begin{figure}
    \centering
    \includegraphics[height = 6cm]{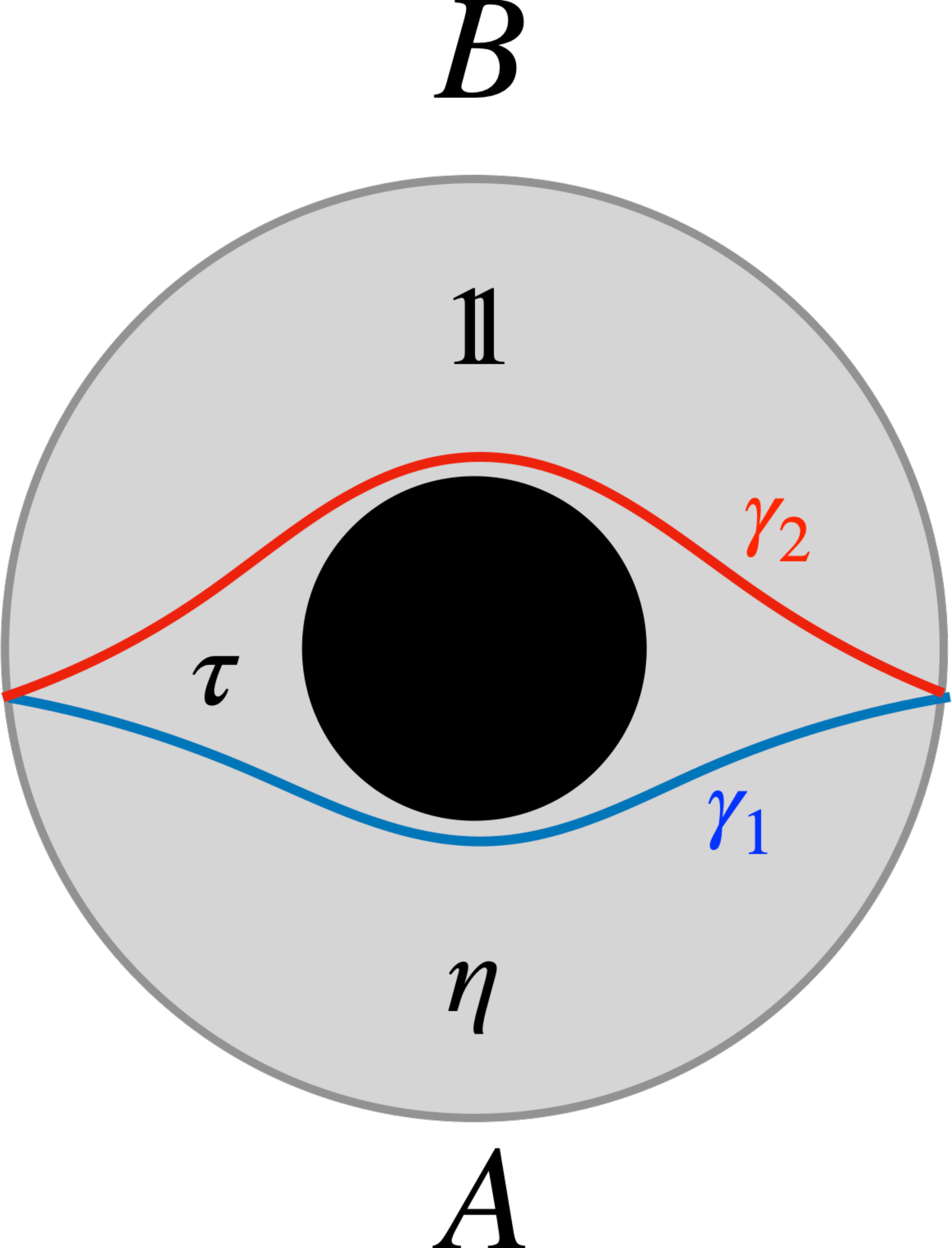}
    \caption{An AdS black hole is shown with the asymptotic boundary partitioned into two regions $A$ and $B$. The two candidate RT surfaces are shown in blue and red respectively. In the gravitational replica trick, the region bounded by $\gamma_1$ and $A$ is glued cyclically ($\eta$) among the replicas. The region bounded by $\gamma_2$ and $B$ is glued according to the identity ($\mathbbm{1}$). The region between $\gamma_1$ and $\gamma_2$ is not fixed by the asymptotic boundary conditions and may be glued among the replicas according to arbitrary $S_{\alpha}$ permutations ($\tau$).}
    \label{fixed_area_cartoon} 
\end{figure}

As an example, consider the R\'enyi entropies of a region on the boundary of a pure state black hole background\footnote{On the CFT side of the duality, these states should be thought of as highly-energy pure states.}. There exist two extremal surfaces that are candidate Ryu-Takayanagi surfaces, $\gamma_1$ and $\gamma_2$, each wrapping the black hole horizon in topologically distinct manners. Denote the areas of these two surfaces $A_1$ and $A_2$ respectively. The moments of the reduced density matrix are
\begin{align}
    \Tr \left[ \rho_A^{\alpha} \right]= \frac{\mathcal{Z}(\rho_A^{{\alpha}})}{\mathcal{Z}(\rho_A)^{\alpha}},
\end{align}
where the numerator is the gravitational path integral on the replicated geometry and the denominator is the path integral on a single copy, necessary for normalization. Because the geometry is identical in both geometries away from the conical singularities, the numerator and denominator will almost completely cancel. The nontrivial terms come from the actions of the conical singularities which are determined by their opening angles
\begin{align}
    \Tr \left[ \rho_A^{\alpha} \right]= \sum_{\tau \in S_{\alpha}} e^{(C(\eta^{-1}\circ \tau)-\alpha) \frac{A_1}{4G_N} + (C(\tau)-\alpha) \frac{A_2}{4G_N}}\Tr\left[\rho_b^{n_1}\right]\Tr\left[\rho_b^{n_2}\right]\dots\Tr\left[\rho_b^{n_{C(\tau)}}\right],
\end{align}
where $G_N$ is Newton's
constant and
the $n_k$'s are the lengths of the cycles in $\tau$ and $\rho_b$ is the bulk state labeling the black hole microstate. These account for the bulk entropy term in the FLM formula \cite{2013JHEP...11..074F}.
The sum over the permutation group arises from all of the ways the replicas may be glued together in the codimension-one region bounded by the two fixed surfaces (see Fig.~\ref{fixed_area_cartoon}). We have chosen the bulk state to be pure such that all of the bulk traces are one
\begin{align}
    \Tr \left[ \rho_A^{\alpha} \right]= \sum_{\tau \in S_{\alpha}} e^{(C(\eta^{-1}\circ \tau)-\alpha)\frac{A_1}{4G_N} + (C(\tau)-\alpha) \frac{A_2}{4G_N}}.
\end{align}
This sum should now look familiar as it is identical to the sum needed for the R\'enyi entropies of Haar random states, \eqref{renyi_sum}, once identifying $d_A \leftrightarrow e^{A_1/4G_N}$ and $d_B \leftrightarrow e^{A_2/4G_N}$. In this way, entropies in fixed-area states in holography are identical to entropies in Haar random states\footnote{This was pointed out in Ref.~\cite{2019arXiv191111977P}.}. 

This connection becomes even richer when we consider more than one gravitational state to compute the relative entropies. Consider the following moments needed for the von Neumann relative entropy
\begin{align}
    \Tr \left[ \rho_A \sigma_A^{\alpha-1} \right]= \frac{\mathcal{Z}(\rho_A \sigma_A^{\alpha-1})}{\mathcal{Z}(\rho_A)\mathcal{Z}(\sigma_A)^{\alpha-1}}.
\end{align}
Both states have fixed areas and the same semiclassical geometry, but come from different black hole microstates, $\rho_b$ and $\sigma_b$. In the language of Refs.~\cite{2015JHEP...04..163A,2017CMaPh.354..865H}, they are orthogonal states in the same code subspace. 
Just as before, the gravitational path integral instructs us to sum over all topologies, meaning that the region between the two fixed-area surfaces can be glued according to \textit{any} $S_{\alpha}$ permutation. This seems different than the calculation in Haar random states which only contained a sum over a subgroup $\mathbbm{1} \times S_{\alpha -1}$, \eqref{vn_RE_sum}
\begin{align}
    \Tr \left[ \rho_A \sigma_A^{\alpha-1} \right]= \sum_{\tau \in  S_{\alpha}}{e^{(C(\eta^{-1} \circ \tau)-\alpha) \frac{A_1}{4G_N} + \left(C(\tau)-\alpha\right) \frac{A_2}{4G_N}}}\Tr\left[\rho_b \sigma_b^{n_1-1}\right]\Tr\left[\sigma_b^{n_2}\right]\dots\Tr\left[\sigma_b^{n_{C(\tau)}}\right].
\end{align}
However, because $\rho_b$ and $\sigma_b$ are orthogonal, $\Tr\left[\rho_b \sigma_b^{n_1-1}\right]$ is only non-zero if $n_1 = 1$. This reduces the sum to 
\begin{align}
    \Tr \left[ \rho_A \sigma_A^{\alpha-1} \right]= \sum_{\tau \in  \mathbbm{1}\times S_{\alpha-1}}{e^{(C(\eta^{-1} \circ \tau)-\alpha) \frac{A_1}{4G_N} + \left(C(\tau)-\alpha\right) \frac{A_2}{4G_N}}},
\end{align}
which is identical to \eqref{vn_RE_sum} under the same identification. A nearly identical argument holds for the PRRE, SRRE, and trace distance. Therefore we conclude that not only do fixed-area states have the same entropies as Haar random states, but they also have identical Hilbert space geometries.

These results have interesting implications for the distinguishability of black hole microstates. Namely, the asymptotic observer with arbitrarily small information about the state (small $A$), is able to distinguish between any black hole microstates. This is surprising because we usually consider all black holes to look the same from outside the horizon to any observer, especially local observers. The catch is that the microstates are only distinguishable nonperturbatively in Newton's constant, $O(e^{-1/G_N})$. This is because all distinguishability measures are linear in $d_A/d_B$ for small region $A$ which translates to proportional to $e^{(A_1-A_2)/4G_N}$. This means that while distinguishability is in principle possible, the error rates in state discrimination will be very high unless the observer has an exponentially large number of copies of the system. The distinguishability is nonperturbatively small up until region $A$ is roughly one qubit less than half the boundary system, at which point it becomes $O(1)$. When the observer has access to more than half of the boundary, the black hole microstates become completely distinguishable up to nonperturbatively small corrections.

We also note that these results represent nonperturbative corrections to the JLMS formula which asserts that the boundary relative entropy equals the bulk relative entropy within the entanglement wedge \cite{2016JHEP...06..004J}. We have considered bulk states that are pure, orthogonal, and localized between the two extremal surfaces; the bulk states are identical outside of the black hole\footnote{Recently, the SRRE between a state with a single fixed-area surface and a state with two fixed-area surfaces was evaluated \cite{2021arXiv210700009H}.}. When $A$ is sufficiently small, the black hole is not within its entanglement wedge so the bulk states are identical i.e.~the bulk relative entropy is zero. Therefore, the JLMS formula asserts that boundary relative entropy is zero. We have shown that there are nonperturbative corrections to this statement.

\subsection{The PSSY model and replica wormholes}

In a landmark achievement, the Page curve \cite{1993PhRvL..71.3743P} for an evaporating black hole was computed for the first time in two independent papers \cite{2020JHEP...09..002P,2019JHEP...12..063A}. The key mechanism that ``fixed'' Hawking's calculation was the inclusion of certain wormhole saddles in the gravitational path integral, referred to as ``replica wormholes'' \cite{2019arXiv191111977P,2020JHEP...05..013A}. Using the toy model of black hole evaporation presented in Ref.~\cite{2019arXiv191111977P} (PSSY), we now show the role of replica wormholes in calculations of relative entropy. This elucidates how the assumptions of Hawking fail. We call this a violation of the no-hair theorem, which is a non-perturbative effect and therefore not present in Hawking's calculation.

The PSSY model consists of two-dimension Jackiw-Teitelboim gravity decorated with end of the world (EOW) branes with $k$ flavors. The Euclidean action is given by
\begin{align}
    I = -\frac{S_0}{2\pi}\left[\frac{1}{2}\int_{\mathcal{M}}\sqrt{g}R+ \int_{\partial \mathcal{M}}\sqrt{h}K \right]-\left[\frac{1}{2}\int_{\mathcal{M}}\sqrt{g}\phi(R+2)+\int_{\partial \mathcal{M}}\sqrt{h}\phi K \right] +\mu \int_{brane}ds,
\end{align}
where $S_0$ is the large ground state entropy, $g$ ($h$) is the bulk (asymptotic boundary) metric with curvature $R$ ($K$), $\phi$ is the dilaton, and $\mu$ is the tension of the EOW brane.

The EOW brane has $k \gg 1$ internal microstates. The global states on the black hole and radiation that we consider are of a maximally entangled form
\begin{align}
    \ket{\Psi} = \frac{1}{\sqrt{k}}\sum_{i = 1}^k \ket{\psi_i}_B\ket{i}_R,
    \label{PSSYstate1}
\end{align}
where $\ket{i}_R$ represents an orthonormal basis of the states of the radiation. Consider a second microstate
\begin{align}
    \ket{\Psi'} = \frac{1}{\sqrt{k}}\sum_{i = 1}^k \ket{\psi_i}_B\ket{i+1}_R,
    \label{PSSYstate2}
\end{align}
where $i+k \sim i$ is implied. The definitions of these states are not microscopic in the sense that the $\ket{\psi_i}_B$'s are defined by a gravitational path integral and are not exactly orthogonal. As for the fixed-area state calculation, they may also be thought of as being orthogonal in the code subspace. 
the von Neumann entropy \cite{2019arXiv191111977P}.
Because of the non-orthogonality, the reduced states of \eqref{PSSYstate1} and \eqref{PSSYstate2} on the radiation are not a priori identical even though the states appear to be related only by a local unitary transformation on $B$\footnote{We thank Daniel Harlow for comments on this point.}. 

The overlap between these two states is
\begin{align}
    \bra{\Psi'}\Psi\rangle = \frac{1}{k} \sum_{il = 1}^k\bra{\psi_l}\psi_i\rangle_B
    \bra{l}i+1\rangle_R = \frac{1}{k}\sum_{i = 1}^k \bra{\psi_{i+1}}\psi_i\rangle_B.
\end{align}
The overlap on the right hand side is given by a gravity amplitude
\begin{align}
    {
    \bra{\psi_i}\psi_j\rangle_B} =\includegraphics[width = .2\textwidth]{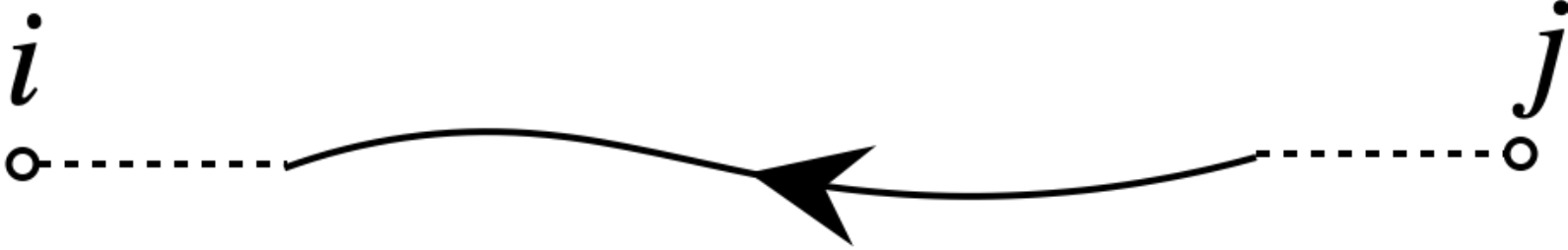}.
    \label{westcoast_overlap_eq}
\end{align}
Because $i \neq i + 1$, connecting the brane in the gravity diagram is incompatible and the amplitude is zero. This means that $\ket{\Psi}$ and $\ket{\Psi'}$ are roughly orthogonal but there are important caveats to this statement because the overlap should be thought of as an ensemble averaged statement. In particular, $|\bra{\psi_i}\psi_{i+1}\rangle_B|^2$ is non-zero. This is completely analogous to the Haar random story where, on average, two independently chosen vectors will have zero overlap, but the variance is non-zero. The analog of \eqref{westcoast_overlap_eq} for random matrices is
\begin{align}
{    \bra{\psi_i}\psi_{j\neq i}\rangle_B} =
    \,
    \tikz[baseline=0ex]{
    \draw[dashed] (0,0.4) -- (0,0) ;
    \draw[dashed] (0,0) --(.5,0);
    \draw[dashed, red] (0.5,0.0) -- (1,0.0) -- (1,0.4);
    \draw (-0.2,0.4)-- (-0.2,-0.15)--(0.5,-0.15);
    \draw[red] (0.5,-0.15)--(1.2,-0.15)-- (1.2,0.4);
    }\ .
\end{align}
Because, when ensemble averaging, we cannot contract black and red indices, the average, $\overline{\bra{\psi_i}\psi_{j\neq i}\rangle_B}$, equals zero. In complete analogy, the ensemble average of
\begin{align}
{
    |\bra{\psi_i}\psi_{j\neq i}\rangle_B|^2
    }=
    \,
    \tikz[baseline=0ex]{
    \draw[dashed] (0,0.4) -- (0,0) ;
    \draw[dashed] (0,0) --(0.5,0);
    \draw[dashed, red] (0.5,0.0) --(1,0.0) -- (1,0.4);
    \draw (-0.2,0.4)-- (-0.2,-0.15)--(0.5,-0.15);
    \draw[red] (0.5,-0.15)--(1.2,-0.15)-- (1.2,0.4);
    \draw[red] (2.3,-0.15)--(1.6,-0.15)-- (1.6,0.4);
    \draw[dashed, red] (2.3,0.0) --(1.8,0.0) -- (1.8,0.4);
    \draw[dashed] (2.3,0.0)--(2.8,0.0)--(2.8,0.4);
    \draw (2.3,-0.15)--(3.0,-0.15)--(3.0,0.4);
    }\ 
\end{align}
is non-zero (though very small) because we may now contract red with red and black with black.

We are interested in the relative entropy of the radiation for two different microstates. Hawking and even the island formula papers assumed that the radiation is seen as purely thermal\footnote{Known greybody factors do not qualitatively change this statement in any meaningful way.} before the Page time in accordance with the no-hair theorem i.e.~all black holes of the same mass, charge, and angular momentum look the same from the outside. After the Page time, while the island formula papers did not assume the radiation to be purely thermal, there was no difference between the calculations for different microstates of the black hole. From one perspective, this is great because unitarity can be realized without knowing the microscopic theory. On the other hand, it is disappointing because it bypasses the question of why all initial states appear to lead to the same final state. We resolve this part of the information problem within the PSSY model and believe analogous results should hold in more realistic models of black hole evaporation.

The reduced density matrix on the radiation for the first state is given by
\begin{align}
    \rho_R := \Tr_B \ket{\Psi} \bra{\Psi} = \frac{1}{k}\sum_{i,j = 1} \bra{\psi_j}\psi_i\rangle_B \ket{i}\bra{j}_R
    \label{PSSYrho1}
\end{align}
and similarly for the second state 
\begin{align}
    \rho_R' := \Tr_B \ket{\Psi'} \bra{\Psi'} = \frac{1}{k}\sum_{i,j = 1} \bra{\psi_j}\psi_i\rangle_B \ket{i+1}\bra{j+1}_R.
    \label{PSSYrho2}
\end{align}
From here on out, we will drop the subscripts labeling the Hilbert spaces as it should be clear.

\begin{figure}
    \centering
    \includegraphics[width = .32\textwidth]{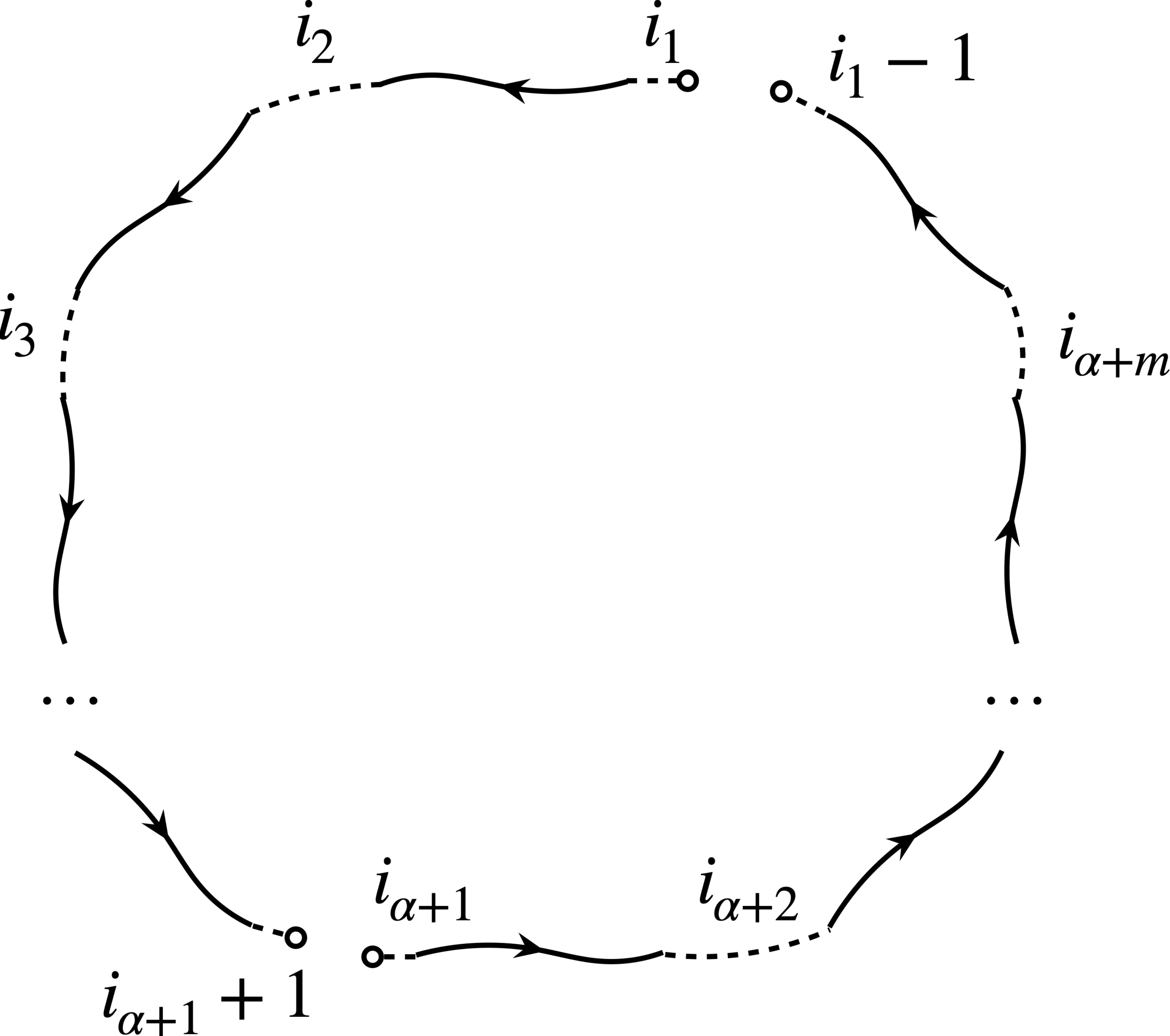}
    \includegraphics[width = .32\textwidth]{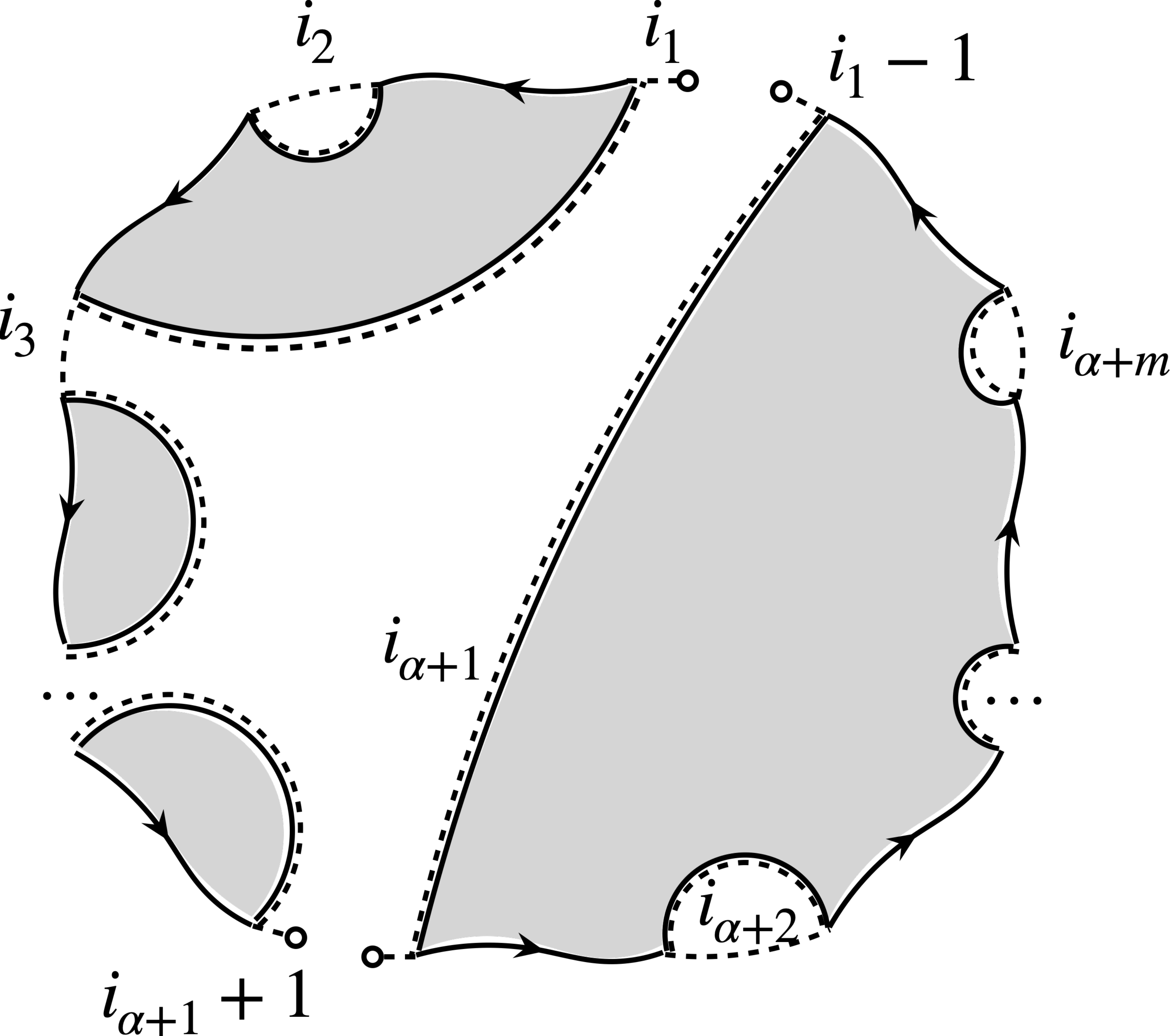}
    \includegraphics[width = .32\textwidth]{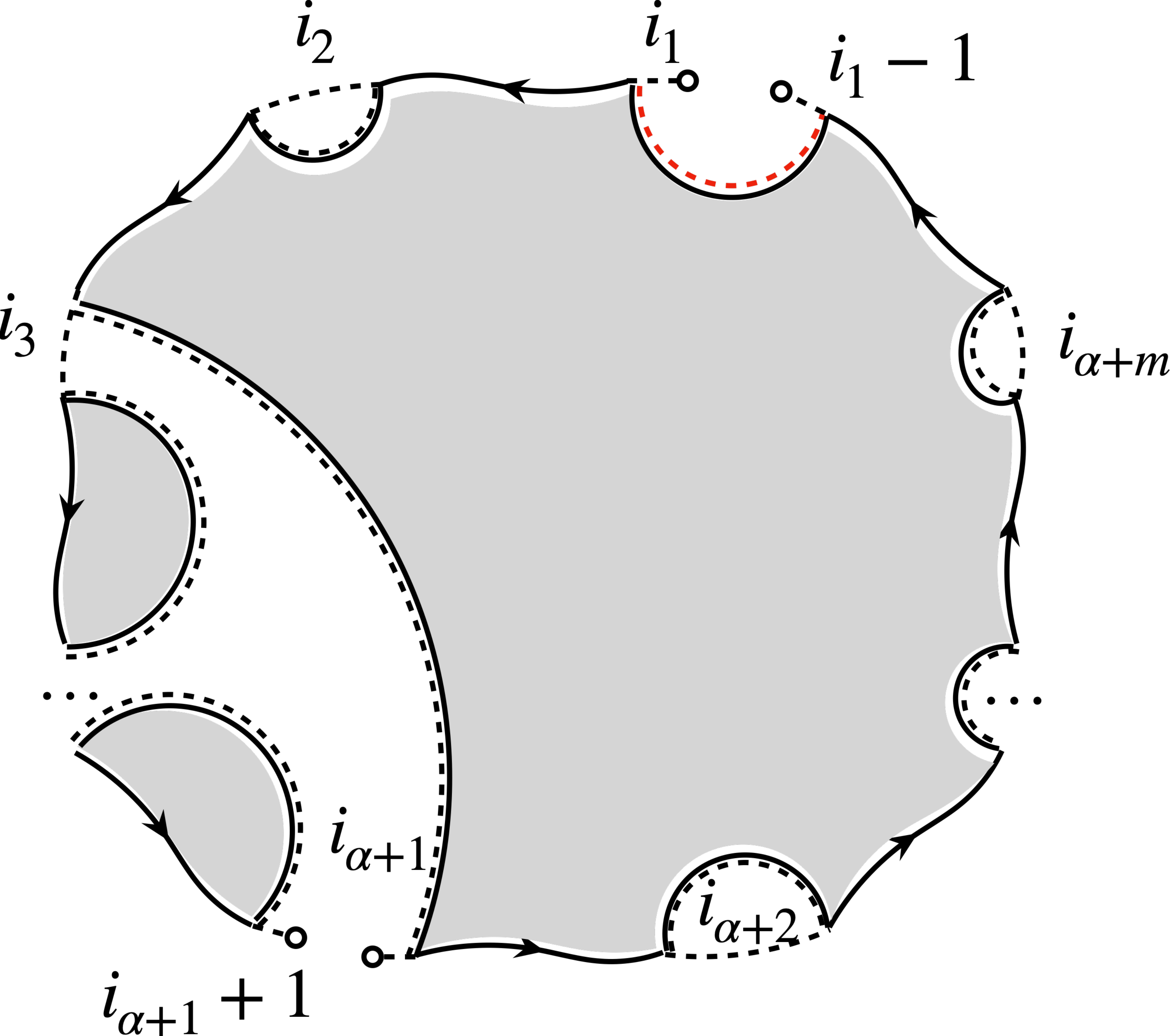}
    \caption{Left: The boundary conditions for the path integral in \eqref{PRRE_pathintegral_eq}. Center: An example of a legal way of filling in the geometry. This geometry is also planar, so it contributes at leading order. Right: An example of an illegal way of filling in the geometry because the EOW brane with label $i_1$ cannot be connected to the EOW brane with label $i_1 -1$. It is not hard to convince oneself that every diagram that connects the left $n$ and right $m$ boundaries will always lead to an inconsistency as in the right diagram.}
    \label{PREE_replica_wormhole}
\end{figure}

We now compute the PRRE between two states of the radiation using the replica trick.
\begin{align}
    \Tr \left[\rho_R^{\alpha} \rho_R'^m\right] &= \frac{1}{k^{{\alpha}+m}}
    \sum_{i_1,\dots, i_{{\alpha}+m},j_1, \dots, j_{{\alpha}+m} = 1}^k\bra{\psi_{j_1}}\psi_{i_1}\rangle \dots \bra{\psi_{j_{{\alpha}+m}}}\psi_{i_{{\alpha}+m}}\rangle
    \nonumber
    \\
    &\times \delta_{j_1 i_2} \delta_{j_2 i_3} \dots \delta_{j_{{\alpha}-1}i_{\alpha}}\delta_{j_{\alpha} i_{{\alpha}+1}+1}\delta_{j_{{\alpha}+1},i_{{\alpha}+2}}\dots \delta_{j_{{\alpha}+m-1},i_{{\alpha}+m}}\delta_{j_{{\alpha}+m}+1,i_1}
    \nonumber
    \\
    &= \frac{1}{k^{{\alpha}+m}}\sum_{i_1,\dots, i_{{\alpha}+m}= 1}^k\bra{\psi_{i_2}}\psi_{i_1}\rangle\bra{\psi_{i_3}}\psi_{i_2}\rangle \dots\bra{\psi_{i_{{\alpha}}}}\psi_{i_{{\alpha}-1}}\rangle\bra{\psi_{i_{{\alpha}+1}+1}}
    \psi_{i_{{\alpha}}}\rangle
    \nonumber
    \\
    &\times \bra{\psi_{i_{{\alpha}+2}}}\psi_{i_{{\alpha}+1}}\rangle \bra{\psi_{i_{{\alpha}+3}}}\psi_{i_{{\alpha}+2}}\rangle\dots  \bra{\psi_{i_{{\alpha}+m}}}\psi_{i_{{\alpha}+m-1}}\rangle
     \bra{\psi_{i_{1}-1}}\psi_{i_{{\alpha}+m}}\rangle.
     \label{PRRE_pathintegral_eq}
\end{align}
This is a more complicated but still tractable gravitational path integral. As shown in Fig.~\ref{PREE_replica_wormhole}, the sum is over only the non-crossing permutations in the $S_{\alpha} \times S_m$ subgroup of $S_{{\alpha}+m}$ due to the boundary conditions on the EOW brane $i_1$ and $i_1 -1$ being incompatible as well as $i_{{\alpha}+1}$ and $i_{{\alpha}+1}+1$ being incompatible. There are crossing permutations in $S_{\alpha +m}/S_{\alpha}\times S_m$ that are compatible with the boundary conditions, but these are subleading. The only other difference in diagrams from the random matrix theory calculation is that the geometries are allowed to have additional handles. This, however, is unimportant because each handle will contribute a factor of $e^{-2 S_0}$ and we have assumed the ground state entropy to be large. 

\begin{figure}
    \centering
    \includegraphics[width = .6 \textwidth]{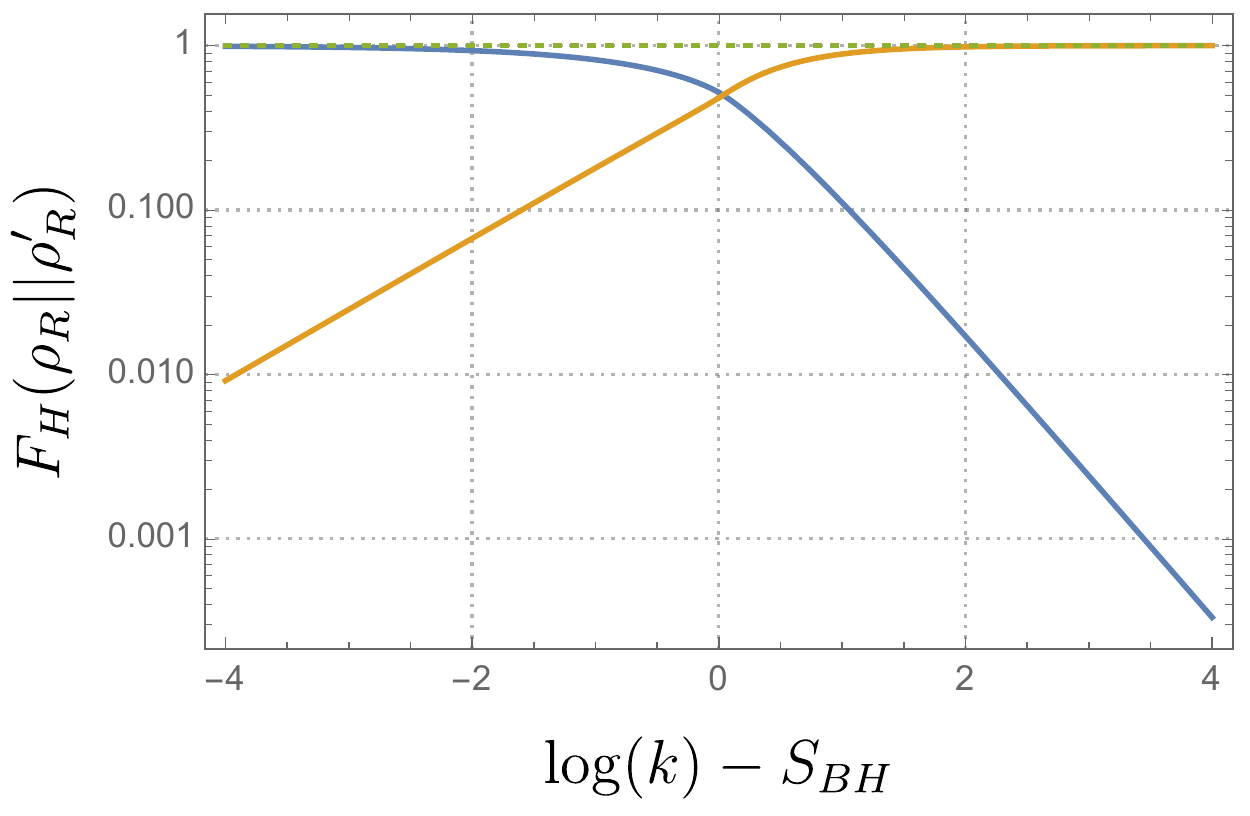}
    \caption{Holevo's just-as-good fidelity (blue) and one minus the fidelity (orange) are shown for the PSSY model following \eqref{FH_PSSY}. Before the Page time ($\log \left[k\right] = S_{BH}$), the fidelity is exponentially close to one. After the Page, time, the fidelity exponentially decays to zero.
    }
    \label{holevo_fidelity_fig}
\end{figure}

To compute the PREE, we must evaluate the gravitational path integral on these replica geometries. For simplicity, we consider the case where the black hole is in the microcanonical ensemble i.e.~instead of fixing the lengths of the boundary, we fix the energy, $E$. The path integral of an $n$-boundary wormhole is \cite{2019arXiv191111977P}
\begin{align}
    \textbf{Z}_n = e^{\textbf{S}}y\left(\sqrt{2E}\right)^n,
\end{align}
where $\textbf{S}$ is the microcanonical entropy at energy $E$. Because of the simply power of $n$, after normalizing the density matrix, the function $y$ will drop out of the final answer. All calculations are then identical to random matrix theory with the identification of $k \leftrightarrow d_A$ and $e^\textbf{S} \leftrightarrow d_B$\footnote{For analogous reasons, the exact same analysis as in Section \ref{randomstate_sec} can be made for the SRRE 
and trace distance but we do not write these out explicitly to avoid repetition.}. We choose to only write the PRRE
\begin{align}
    D_{\alpha}(\rho_R|| \rho'_R) 
    = 
    \frac{1}{\alpha-1}
    \begin{cases}
    \log \left(\, _2F_1\left(1-\alpha,-\alpha;2;\frac{k}{e^\textbf{S}}\right) \,
   _2F_1\left(\alpha-1,\alpha;2;\frac{k}{e^\textbf{S}}\right)\right), & k < e^{\textbf{S}}
   \\
   \log \left(
   \frac{e^{\textbf{S}}}{k} \, _2F_1\left(1-\alpha,-\alpha;2;
   \frac{e^{\textbf{S}}}{k}\right) \,
   _2F_1\left(\alpha-1,\alpha;2;
   \frac{e^{\textbf{S}}}{k}\right) \right), & k > e^{\textbf{S}}
   \end{cases}.
   \label{PRRE_PSSY}
\end{align}
The quantum Chernoff bound asserts
\begin{align}
    \lim_{n\rightarrow \infty}-& \frac{\log\left[\min_{A}\left[ \alpha_n(A) + \beta_n(A) \right]\right] }{n} \nonumber
   \\
   &=\max_{\alpha \in (0, 1)}\begin{cases}- \log \left[\, _2F_1\left(1-\alpha,-\alpha;2;\frac{k}{e^\textbf{S}}\right) \,
   _2F_1\left(\alpha-1,\alpha;2;\frac{k}{e^\textbf{S}}\right)\right], & k < e^{\textbf{S}}
   \\
   - \log \left[\frac{e^{\textbf{S}}\, _2F_1\left(1-\alpha,-\alpha;2;\frac{e^\textbf{S}}{k}\right) \,
   _2F_1\left(\alpha-1,\alpha;2;\frac{e^\textbf{S}}{k}\right)}{k}\right], & k > e^{\textbf{S}}
   \end{cases}
    \nonumber
   \\
   &=  \begin{cases}- 2\log \left[\,
   _2F_1\left(\frac{1}{2},-\frac{1}{2};2;\frac{k}{e^\textbf{S}}\right) \right], & k < e^{\textbf{S}}
   \\
   -\log \left[\frac{e^\textbf{S}}{k}\,
   _2F_1\left(\frac{1}{2},-\frac{1}{2};2;\frac{e^\textbf{S}}{k}\right)^2 \right], & k > e^{\textbf{S}}
   \end{cases},
   \label{chernoff_PSSY}
\end{align}
where we found the maximum to be at $\alpha = 1/2$ i.e.~Holevo's just-as-good fidelity
\begin{align}
    \min_{A}\left[ \alpha_n(A) + \beta_n(A) \right] \leq 
    {
    F_H(\rho_R || \rho'_R)^{n/2}.
    }
    \label{PSSY_fidel}
\end{align}
While \eqref{PSSY_fidel} saturates at large $n$, the RHS holds as an upper bound for all integer $n$.
Holevo's just-as-good fidelity, plotted in Fig.~\ref{holevo_fidelity_fig}, is
\begin{align}
        {
    F_H(\rho_R || \rho'_R)
    }
    = \begin{cases}
     \, _2F_1\left(-\frac{1}{2},\frac{1}{2};2;\frac{k}{e^\textbf{S}}\right){}^4, & k < e^{\textbf{S}}
   \\
   \frac{e^{2\textbf{S}} \,
   _2F_1\left(-\frac{1}{2},\frac{1}{2};2;\frac{e^\textbf{S}}{k}\right){}^4}{k^2}, & k > e^{\textbf{S}}
    \end{cases}.
    \label{FH_PSSY}
\end{align}

When observing a black hole from the outside, our task is not as simple as distinguishing \textit{two} states. Rather, we need to distinguish between all $e^\textbf{S}\gg 2$ states of the black hole. On the face of it, this seems like an insurmountable task. However, using the multiple quantum Chernoff bound \eqref{mult_chernoff} and the normal distribution for relative entropies leading to \eqref{multiple_chernoff_distance}, we determine that our asymptotic error in the multistate discrimination is identical, at leading order, to that of the two state discrimination.

This has important implications on the nature of black hole evaporation that have not been addressed in the calculations of the entropy. The island formula (or quantum Ryu-Takayanagi formula), stated below, was the main tool in recent calculations of entropy of Hawking radiation
\cite{2015JHEP...01..073E, 2020arXiv200606872A}
\begin{align}
    S_{vN}(\rho_R) = \min_{\chi}\left[\frac{A_{\chi}}{4G_N}+S_{
    \textit{semi-cl}}(\Sigma_{\chi}) \right],
\end{align}
where $A_{\chi}$ is the area of the codimension-two quantum extremal surface $\chi$ and $S_{\textit{semi-cl}}(\Sigma_{\chi})$ is the von Neumann entropy of the bulk quantum fields in the codimension-one region, $\Sigma_{\chi}$, bounded by $\chi$ in the bulk i.e.~the entanglement wedge of the radiation, $R$. While this formula accurately computes the von Neumann entropy of the radiation, restoring \textit{consistency} with unitarity, it leaves more to be desired. In particular, the bulk entropy term is completely semi-classical and given by a quantum field theory calculation in curved space. The calculation is agnostic to the details of the black hole microstate. One of the central pieces of the apparent paradox was that Hawking radiation always looked the same on the outside regardless of the dynamics in the black hole interior, the phenomenon of \textit{no hair}. \eqref{FH_PSSY} instead tells us that there is \textit{detectable} hair in the radiation even before the Page time. In fact, information about the particular microstate is present even in the first Hawking quantum. Because the fidelity of the radiation for any two different microstates is strictly less than one, we can always tell the difference between Hawking radiation coming from black holes that are in different microstates, even if they have the same macroscopic parameters mass, charge, and angular momentum. 

The caveat is that the difference between states of the radiation coming from distinct black hole microstates is exponentially small in the black hole entropy i.e.~the deviation of the fidelity from one before the Page time is $O(e^{-\textbf{S}})$. This means that while in principle possible, any reasonable observer will be hard-pressed to observe this difference. If we want the probability of error in distinguishing the black hole microstates to be less than $\epsilon$, we need an $O(e^{\textbf{S}}\log \epsilon)$ number of copies of the state of the radiation, more precisely
\begin{align}
    n =  \frac{\log (\epsilon )}{2 \log \left(\,
   _2F_1\left(-\frac{1}{2},\frac{1}{2};2;\frac{k}{e^{\textbf{S}}
   }\right)\right)}.
\end{align}

After the Page time, there is a different caveat. The fidelity is exponentially close to zero, so the states are essentially fully distinguishable. Precisely, with just one copy of the state, the error probability is bounded above as
\begin{align}
    \min_{A}\left[ \alpha_n(A) + \beta_n(A) \right] \leq \frac{e^{\textbf{S}} \,
   _2F_1\left(-\frac{1}{2},\frac{1}{2};2;\frac{e^\textbf{S}}{k}\right){}^2}{k} = O(k^{-1}).
\end{align}
The issue is that the amount of radiation needed to perform this discrimination is of order the size of the black hole. This means that the observer will have to perform a very complex computation, which again is not so feasible in practice.

Now, consider what we would have concluded if we did not include replica wormholes in the gravitational path integral. This is the analog of Hawking's calculation of the state of the radiation that led to information loss. Removing replica wormholes corresponds to only including the identity permutation in the sum. This means
\begin{align}
    \Tr \left[\rho_R^{\alpha}\rho_R'^{m} \right] =  k^{1-\alpha-m},
\end{align}
leading to all PRRE's being identically zero, regardless of how much radiation is collected. This is consistent with the initial paradox where the radiation was thought to be in the same state regardless of the black hole microstate. In fact, this was clear from \eqref{PSSYrho1} and \eqref{PSSYrho2} because, if the states of the black hole are orthogonal, the reduced density matrices on $R$ would be identical.

Finally, we note that the computation of the relative entropy between two states in the PSSY model was recently studied as a way to detect the violation of global symmetries in theories of quantum gravity \cite{2020arXiv201106005C}. The simpler quantity $\Tr\left[\rho_A \sigma_A \right]$ was evaluated as a proxy with the full relative entropy computation left as an open question. \eqref{PRRE_PSSY} is the (generalized) solution to this question. While an $O(1)$ answer was anticipated for the relative entropy after the Page time in Ref.~\cite{2020arXiv201106005C}, we conclude that the relative entropy is indeed infinite. It is only $O(1)$ slightly prior to the Page time and exponentially small but finite at earlier times.

\section{Tensor networks}
\label{sec_tensor}

Tensor networks represent a generalization of the states we have considered, adding in the ingredient of locality. As such, tensor networks have been particularly useful as toy models of holographic duality \cite{2015JHEP...06..149P,2016JHEP...11..009H}. They are also independently interesting as presenting new classes of ensembles of random states with novel spectral properties \cite{2010JPhA...43A5303C}. In this section, we generalize the computations of Section \ref{randomstate_sec} to generic random tensor networks, finding qualitatively new phenomena. A specific application of these results is for the random tensor networks used for modeling holography. We clarify which random states faithfully represent holographic states and which do not.


We begin with the warm-up example of a random tensor network with two tensors, $T_1$ and $T_2$, contracted together 
\begin{align}
\begin{tikzpicture}
    \node[draw, shape=rectangle] (T1) at (0,0) {$T_1$};
    \node[draw, shape=rectangle] (T2) at (1,0) {$T_2$};
    \draw [thick] (T1) -- (T2);
    \draw [thick] (T1) -- (-1,0);
    \draw [thick] (T2) -- (2,0);
    \end{tikzpicture}.
\end{align}
This is the simplest generalization of the single-tensor network i.e.~Haar random state
\begin{align}
\begin{tikzpicture}
    \node[draw, shape=rectangle] (T1) at (0,0) {$T$};
    \draw [thick] (T1) -- (-1,0);
    \draw [thick] (T1) -- (1,0);
    \end{tikzpicture}.
\end{align}
The two-tensor network has one additional degree of freedom, the dimension of the internal bond, $d_b$. For $T_1$ and $T_2$ independently Gaussian, it is straightforward to generalize the diagrammatic approach. The state is now
\begin{align}
\label{eq:tr_r2}
    \ket{\Psi}:=  
    \,
    \tikz[scale=1.5,baseline=0ex,decoration={
    markings,
    mark=at position 0.5 with {\arrow{>}}}]{
    \draw[dotted] (0,0.2)-- (0,0);
    \draw[dotted,postaction={decorate}] (0,0) -- (.5,0);
    \draw[dotted]  (.5,0.2) -- (.5,0);
    \draw (-0.2,0.2)-- (-0.2,-0.25);
    \draw[dashed] (.7,-0.25)-- (.7,0.2);
    }\ ,
\end{align}
where the dotted line is for $d_b$ and is always contracted. The arrow indicates that the dotted lines must be connected in a way that has all arrows with the same orientation. The reduced density matrix is
\begin{align}
\label{eq:tr_r2}
    \rho_A:=  
    \,
    \tikz[scale=1.5,baseline=0ex,decoration={
    markings,
    mark=at position 0.5 with {\arrow{>}}}]
    {
    \draw[dotted,postaction={decorate}] (0,0.2)-- (0,0)--(.5,0)-- (.5,0.2);    
    \draw (-0.2,0.2)-- (-0.2,-0.25);
    \draw[dotted,postaction={decorate}]  (1.5,0.2) --(1.5,0)-- (2,0)-- (2,0.2) ;    
    \draw (2.2,-0.25)-- (2.2,0.2);
    \draw[dashed] (1.5-0.2,0.2)-- (1.5-0.2,0)-- (.7,0)-- (.7,0.2);
    }\ .
\end{align}
Note the directions of the arrows. We can see that the normalization associated with each density matrix is $(d_A d_Bd_b)^{-1}$.
When taking the average of the moments, we now have a double sum over the permutation group, corresponding to the two random tensors. For example, the purity moments will be
\begin{align}
    \overline{\Tr\left[ \rho_A^{\alpha}\right]} = \frac{1}{(d_A d_B d_b)^{\alpha}} \sum_{\tau_1, \tau_2 \in S_{\alpha}} d_A^{C(\eta^{-1}\circ \tau_1)} d_B^{C(\tau_2)}d_b^{C(\tau_1^{-1} \circ \tau_2)}.
\end{align}
To solve this equation at leading order, we need to maximize the exponents. That is, we must find the set of permutations, $\{\tau_1,\tau_2\}$ that maximize $C(\eta^{-1}\circ \tau_1) + C(\tau_2) + C(\tau_1^{-1} \circ \tau_2)$. This is already a significantly harder problem than the single-tensor network where the answer is that $\tau$ must be a non-crossing permutation. Interestingly, this maximization may be rephrased as a classical network flow problem \cite{2010JPhA...43A5303C}. We attach a ``source'' to $B$ and a ``sink'' to $A$ and determine the maximal flow, $w_{\text{max-flow}}$, of the network where each edge has a weight corresponding to the logarithm of the Hilbert space dimension. We apply the Ford-Fulkerson method in which, one at a time, we take a path from the source to the sink through the tensors, subtracting the weight of the edges by one as we go along the path \cite{ford_fulkerson_1956}. Each one of these paths is called an \textit{augmenting path}. We repeat this process until there are no more paths from the source to the sink such that we are left with a \textit{residual network}. The rules for each permutation are that
\begin{enumerate}
    \item All $\tau_i$'s are non-crossing.
    \item $\tau_i$'s are non-decreasing along each augmenting path in the network i.e.~each permutation is contained within all permutations further along the path. 
    \item All $\tau_i$'s in the connected component of the source in the residual network are set to the identity.
    \item All $\tau_i$'s in the connected component of the sink in the residual network are set to $\eta$.
    \item All $\tau_i$'s in the same connected component are identical. 
\end{enumerate}
At leading order, the moments will then be
\begin{align}
    \overline{\Tr\left[ \rho_A^{\alpha}\right]} =N^{ - (\alpha-1) w_{\text{max-flow}} } \sum_{\{\tau_1,\tau_2\}} \tilde{d}_A^{C(\eta^{-1}\circ \tau_1)-\alpha} \tilde{d}_B^{C(\tau_2)-\alpha}\tilde{d}_b^{C(\tau_1^{-1} \circ \tau_2)-\alpha},
    \label{network_eq}
\end{align}
where $\{ \tau_1,\tau_2\}$ is the set of permutations obeying the constraints and the dimensions with tildes are $O(1)$ due to multiples of $N$, a large parameter, being pulled out. For example, if $d_A = O(N^2)$, then $\tilde{d}_A := d_A N^{-2}$. In the special case that all $\tilde{d}_i$'s are one, we have
\begin{align}
    \overline{\Tr\left[ \rho_A^{\alpha}\right]} =F_{\alpha}N^{ - (\alpha-1) w_{\text{max-flow}} } ,
    \label{network_eq_special}
\end{align}
where $F_{\alpha}$ represents the number of paths satisfying the constraints. 

For example, consider the case where $d_A = d_B = d_b = N$. There will be a single augmenting path ($w_{\text{max-flow}} = 1$) 
\begin{align}
\begin{tikzpicture}
    \node[draw, shape=rectangle] (T1) at (2,0) {$T_1$};
    \node[draw, shape=rectangle] (T2) at (0,0) {$T_2$};
    \node[draw, shape=rectangle] (so) at (-2,0) {source};
    \node[draw, shape=rectangle] (si) at (4,0) {sink};
    \draw [ultra thick,-latex,preaction={draw,green,-,double=green,double distance=2\pgflinewidth,}] (T2) -- (T1);
    \draw [ultra thick,-latex,preaction={draw,green,-,double=green,double distance=2\pgflinewidth,}] (T1) -- (si);
    \draw [ultra thick,-latex,preaction={draw,green,-,double=green,double distance=2\pgflinewidth,}] (so) -- (T2);
    \end{tikzpicture}
\end{align}
such that the resulting network will consist of disconnected tensors with the constraint that $\tau_1 \leq \tau_2 \in NC_{\alpha}$
\begin{align}
\begin{tikzpicture}
    \node[draw, shape=rectangle] (T1) at (2,0) {$T_1$};
    \node[draw, shape=rectangle] (T2) at (0,0) {$T_2$};
    \node[draw, shape=rectangle] (so) at (-2,0) {source};
    \node[draw, shape=rectangle] (si) at (4,0) {sink};
    \end{tikzpicture}.
\end{align}
The number of such permutations is given by the second Fuss-Catalan number
\begin{align}
    FC^{(2)}_{\alpha} :=  \frac{1}{2\alpha +1}\binom{3\alpha }{\alpha},
\end{align}
so the moments will be given by
\begin{align}
    \overline{\Tr\left[ \rho_A^{\alpha}\right]} =\frac{1}{2\alpha +1}\binom{3\alpha }{\alpha}N^{ 1- \alpha } .
\end{align}
The associated von Neumann entropy is
\begin{align}
    S_{vN}(A) = \log\left[N\right] -\frac{5}{6}.
\end{align}
This generalizes Page's formula.

Had we instead taken, for example, $\sqrt{d_A} = d_B = d_b = N$, the same augmenting path would have led to the following residual network
\begin{align}
\begin{tikzpicture}
    \node[draw, shape=rectangle] (T1) at (2,0) {$T_1$};
    \node[draw, shape=rectangle] (T2) at (0,0) {$T_2$};
    \node[draw, shape=rectangle] (so) at (-2,0) {source};
    \node[draw, shape=rectangle] (si) at (4,0) {sink};
    \draw [thick] (T1) -- (si);
    \end{tikzpicture}.
\end{align}
Because $T_1$ is still connected to the sink $\tau_1$ will be set to $\eta$ while $\tau_2$ can be any non-crossing permutation of which there are a Catalan number's worth, leading to 
\begin{align}
    S_{vN}(A) = \log\left[N\right] -\frac{1}{2}.
\end{align}

More generally, we can have a tensor network with $n$ tensors, $\{T_1, T_2, \dots, T_n \}$. A set of indices of these tensors will be contracted. We refer to the dimensions of these indices by the tensors they connect e.g.~$d_{ij}$. There is also a set of uncontracted indices which correspond to systems $A$ and $B$. We refer to the dimensions of these indices as $d_{Ai}$ and $d_{Bi}$ which label the subsystem they belong to and the tensors that they are indices of. The purity moments can then be expressed as a sum over $n$ permutation elements
\begin{align}
    \overline{\Tr\left[ \rho_A^{\alpha}\right]} = \frac{1}{(\prod_{i}d_{Ai} d_{Bi}\prod_j d_{ij})^{\alpha}} \sum_{\tau_1, \dots, \tau_n \in S_{\alpha}}\prod_{i=1}^n d_{Ai}^{C(\eta^{-1}\circ \tau_i)} d_{Bi}^{C(\tau_i)}\prod_{j=1}^n d_{ij}^{C(\tau_i^{-1} \circ \tau_j)}.
\end{align}
Here, we must maximize the more complicated exponent which can also be formulated as a network flow problem. \eqref{network_eq} is generalized to 
\begin{align}
    \overline{\Tr\left[ \rho_A^{\alpha}\right]} =N^{ - (\alpha-1) w_{\text{max-flow}} } \sum_{\{\tau_i\}} \prod_{i}^n\tilde{d}_{Ai}^{C(\eta^{-1}\circ \tau_i)-\alpha} \tilde{d}_{Bi}^{C(\tau_i)-\alpha}\prod_{j=1}^n \tilde{d}_{ij}^{C(\tau_i^{-1} \circ \tau_j)-\alpha},
\end{align}
where $\{ \tau_i\}$ is the set of permutations obeying the updated rules. 
\eqref{network_eq_special} still applies, though the combinatorics may become significantly more difficult. If we are not concerned with the $O(1)$ constant, we only need to determine the maximal flow. By the max-flow min-cut theorem, the maximal flow from the source to the sink will always be equal to the minimal cut, $\gamma_A$, in the network needed to separate the source and sink into disconnected components \cite{ford_fulkerson_1956,1056816}. The von Neumann entropy is then
\begin{align}
    S_{vN}(A) = \gamma_A \log\left[ N\right] + O(1).
\end{align}

We can now generalize this, as before, to relative entropy. We will explicitly compute the PRRE. This only changes the permutation allowed in the sum 
\begin{align}
    \overline{\Tr\left[ \rho_A^{\alpha}\sigma_A^m\right]} = \frac{1}{(\prod_{i}d_{Ai} d_{Bi}\prod_j d_{ij})^{\alpha+m}} \sum_{\tau_1, \dots, \tau_n \in S_{\alpha}\times S_m}\prod_{i=1}^n d_{Ai}^{C(\eta^{-1}\circ \tau_i)} d_{Bi}^{C(\tau_i)}\prod_{j=1}^n d_{ij}^{C(\tau_i^{-1} \circ \tau_j)}.
\end{align}
The key difference between this replica trick and the one for R\'enyi entropies is that $\eta$ is not an allowed permutation in the sum. This effects all of the $C(\eta^{-1}\circ \tau_i)$ terms because they are maximized not by $\alpha + m$, but $\alpha + m -1$ which occurs with $\tau_i = \eta_{\alpha} \times \eta_{m} \in S_{\alpha}\times S_m$. This changes rule (1) of the Ford-Fulkerson algorithm to ``All $\tau_i$'s are in $NC_{\alpha}\times NC_m$'' and rule (4) to  ``All $\tau_i$'s in the connected component of the sink in the residual network are set to $\eta_{\alpha} \times \eta_{m}$.'' The moments are then
\begin{align}
    \overline{\Tr\left[ \rho_A^{\alpha}\sigma_A^m\right]} =N^{ - (\alpha + m-1) \gamma_A }N^{-(E_A-\gamma_A)} \sum_{\{\tau_i\}}\prod_{i=1}^n \tilde{d}_{Ai}^{C(\eta^{-1}\circ \tau_i)-\alpha-m} \tilde{d}_{Bi}^{C(\tau_i)-\alpha-m}\prod_{j=1}^n \tilde{d}_{ij}^{C(\tau_i^{-1} \circ \tau_j)-\alpha-m},
\end{align}
where $E_A$ is the weight of the external $A$ edges before applying the Ford-Fulkerson algorithm, which, in the single-tensor case simply equaled the maximum flow. In the special case where all $\tilde{d}_i$'s are one,
\begin{align}
    \overline{\Tr\left[ \rho_A^{\alpha}\sigma_A^m\right]} = F_{\alpha,m} N^{ - (\alpha + m-1) \gamma_A }N^{\gamma_A-E_A} .
\end{align}
First, consider the two-tensor network when all dimensions equal $N$. There is a single augmenting path and $E_A = \gamma_A$. However, due to the restriction to $S_{\alpha}\times S_m$, $\tau_1$ and $\tau_2$ are restricted to non-crossing within the subgroup, such that
\begin{align}
    F_{\alpha,m} = FC^{(2)}_{\alpha}FC^{(2)}_{m},
\end{align}
which is much smaller than $FC_{\alpha +m }$. The von Neumann relative entropy is then given by
\begin{align}
    D(\rho_A||\sigma_A) = \frac{17}{6},
\end{align}
which should be compared with $3/2$ which was found for the single-tensor network. Apparently, adding a random tensor makes the state more distinguishable. Generalizing this conclusion, if the tensor network is a string of $n$ tensors
\begin{align}
\begin{tikzpicture}
    \node[draw, shape=rectangle] (T1) at (0,0) {$T_1$};
    \node[draw, shape=rectangle] (T2) at (1,0) {$T_2$};
    \node[draw, shape=rectangle] (T3) at (2,0) {$T_3$};
    \node[draw, shape=rectangle] (T4) at (3.5,0) {$T_n$};
    \draw [thick] (T1) -- (T2);
    \draw [thick] (T1) -- (-1,0);
    \draw [thick] (T2) -- (T3);
    \draw [thick,dotted] (T3) -- (T4);
    \draw [thick] (T4) -- (4.5,0);
    \end{tikzpicture},
\end{align}
the combinatorial factor is given by a product of the $n^{th}$ Fuss-Catalan number
\begin{align}
    F_{\alpha,m} = FC^{(n)}_{\alpha}FC^{(n)}_{m} = \frac{1}{(n\alpha + 1)(nm+1)}\binom{(n+1)\alpha }{\alpha}\binom{(n+1)m}{m}.
\end{align}
The relative entropy is then 
\begin{align}
    D(\rho_A||\sigma_A) = H_{n+1}+n-1,
\end{align}
where $H_n:=\sum_{k=1}^n\frac{1}{k}$ is the harmonic number. 
This function is monotonically increasing in $n$.

If we are not concerned with the $O(1)$ contribution, the PRRE will generally be given by
\begin{align}
    D_{\alpha}(\rho_A || \sigma_A) = \frac{\gamma_A -E_A}{\alpha-1}\log \left[N\right] +O(1).
\end{align}
This implies that the quantum relative entropy is always divergent if the $A$ edges do not coincide with the minimal cut. We will come back to this point shortly. Similarly, note that Holevo's just-as-good fidelity is exponentially small in this case
\begin{align}
    F_H(\rho_A ||\sigma_A) = N^{-2(E_A-\gamma_A)}.
\end{align}
This means that for many tensor networks, two independent states will be easily distinguishable, no matter the relative size of $A$ and $B$. Note that when $E_A=\gamma_A$, the $O(1)$ and subleading terms are very interesting and were the main topic of this paper.

\begin{figure}
    \centering
    \includegraphics[height = 6cm]{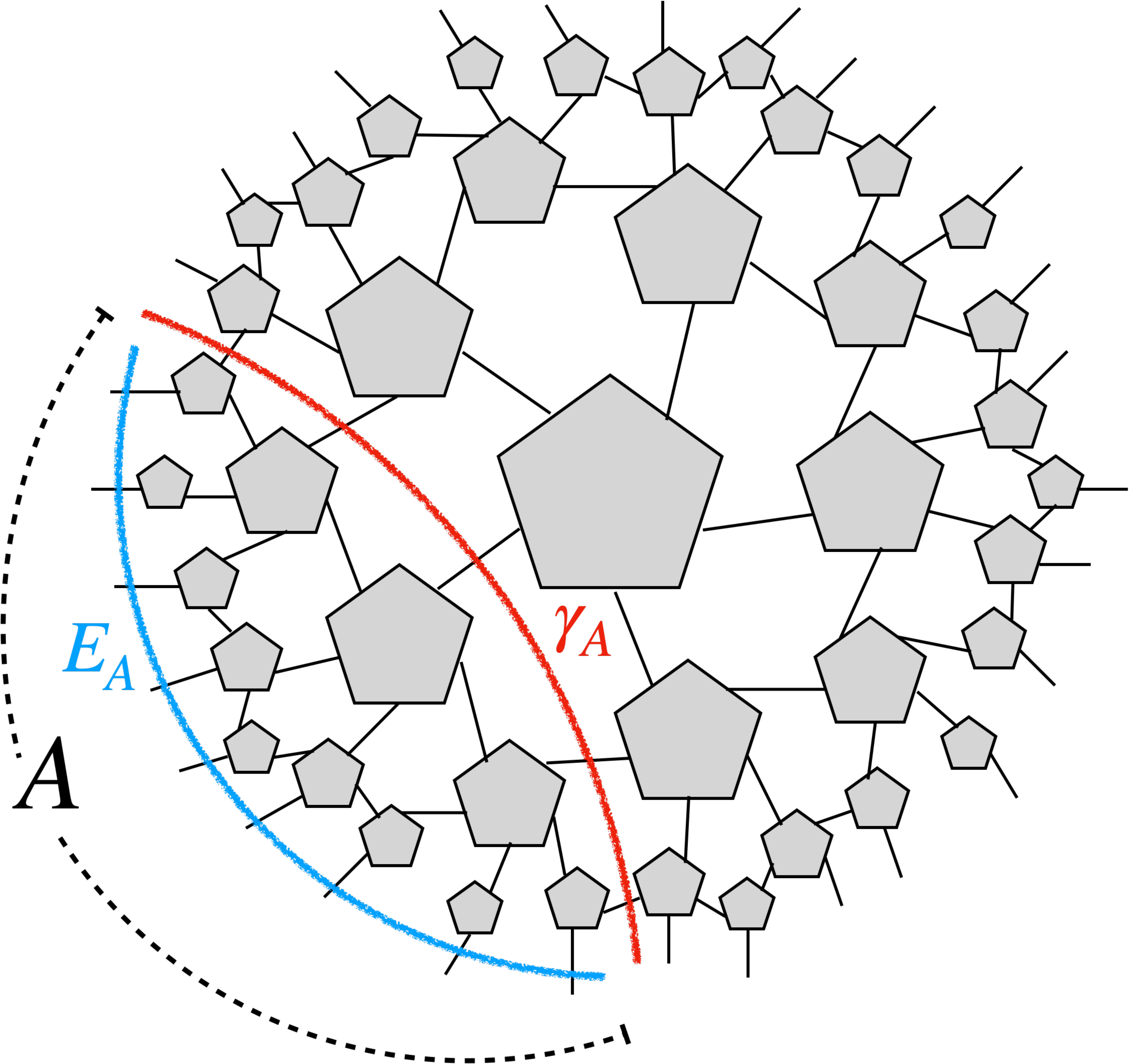}
    \caption{A discretization of hyperbolic space is shown at a tensor network. For boundary subregion $A$, the minimal cut through the network, $\gamma_A$, always dips into the bulk and is smaller than the boundary cut $E_A$}.
    \label{holographic_TN_cartoon}
\end{figure}

Recall that holographic random tensor networks are tensor networks composed of random tensors that are arranged geometrically as discretized hyperbolic space (see Fig.~\ref{holographic_TN_cartoon}) \cite{2016JHEP...11..009H}. Due to the negative curvature of this space, the minimal surfaces for boundary regions always lie in the bulk. This means that we always have $E_A > \gamma_A$, so independent states will always be completely distinguishable. This seems to be in tension with the holographic results of Section \ref{black_hole_sec}. As it seems, single-tensor networks, which have no built in locality, exactly match holographic states while the tensor networks that naively look like Anti-de Sitter space do not share any information theoretic properties with holography except for the entropy.

At face value, the above conclusions are a bit unsettling. Fortunately, this can be remedied by more carefully stating how a tensor network should model holographic states. Tensor network models represent the holographic map as a quantum error correcting code where the bulk degrees of freedom play the role of ``logical qubits'' that are protected by being embedded in the larger boundary Hilbert space. The logical qubits live in a code subspace. In the random tensor networks we have been considering, the code subspace (the ensemble of states we are sampling from) is identical to the Hilbert space 
on the boundary. This equality between bulk effective field theory and boundary Hilbert space dimensions only occurs in AdS/CFT when one has a large black hole whose horizon approaches the asymptotic boundary of the space. This is the reason for the requirement that $E_A = \gamma_A$; all minimal surfaces in the large black hole geometry hug the asymptotic boundary. In order to model other holographic states using tensor networks, we must make the code subspace significantly smaller than the total Hilbert space. Additionally, the bulk density matrices should not be orthogonal. For example, when considering perturbations about vacuum AdS, the total Hilbert space dimension is $O(e^{-1/G_N})$ while the code subspace is $O(e^{G_N^{0}})$. In practice, this means that for the two states, $\rho$ and $\sigma$, we must take the random tensors to be correlated with each other i.e.~the measure for each random tensor only has support on a proper subset of the Hilbert space.

Another important class of random tensor networks is random unitary circuits. In these tensor networks, all tensors are random unitary operators drawn from the Haar measure. Such networks have been the focus of intense study because they present an exactly solvable minimal model of chaotic many-body dynamics, only preserving locality and unitarity. Using the replica trick and Weingarten calculus, many measures of entanglement and operator growth have been computed using geometric quantities in these circuits \cite{2017PhRvX...7c1016N,2018PhRvX...8b1013V,2018PhRvX...8b1014N,2018arXiv180300089J,2019arXiv190512053H,2020JHEP...01..031K,2019JHEP...12..020W,2019arXiv191208918L,2020JHEP...04..074K,2020arXiv200514243K}. Analogously, the distinguishibility measures discussed in this paper will be computable. For the dynamics of evolving a state from a product state with a random unitary circuit, twice the time (depth of the circuit) plays the role of $\gamma_A$ when it is smaller than the length of the region $A$, $l_A$, which plays the role of $E_A$. We therefore will find that states are easily distinguishable for times $t < l_A/2$ and very hard to distinguish afterwards. This describes process of thermalization where different initial states become indistinguishable at late times. The details of this calculation, including the precise approach to equilibrium, are left to future work.

\section{Subsystem eigenstate thermalization}
\label{sETH_sec}

The eigenstate thermalization hypothesis (ETH) was a major development in understanding the emergence of thermal physics from isolated quantum many-body systems in pure states \cite{1991PhRvA..43.2046D, 1994PhRvE..50..888S,2018RPPh...81h2001D}. The statement of eigenstate thermalization is that given two energy eigenstates, $\ket{E_i}$ and $\ket{E_j}$, and a ``simple'' few-body operator $\mathcal{O}$, the expectation value varies smoothly with the macroscopic, thermodynamic quantities such as energy
\begin{align}
    \bra{E_i}\mathcal{O}\ket{E_j} = f_{\mathcal{O}}(E)\delta_{ij} + e^{-S(E)/2}R_{ij}, \quad E := \frac{E_i + E_j}{2},
\end{align}
where $f_{\mathcal{O}}(E)$ is a smooth function of the energy, $S(E)$ is the thermodynamic entropy, and $R_{ij}$ is an $O(1)$ pseudorandom matrix. The ETH is expected to hold for generic nonintegrable systems and violated in integrable systems. In words, it states that expectation values of simple observables appear thermal, up to exponentially small corrections. 

Note that the standard, \textit{local} ETH 
is a statement only about local or few-body operators. A significant strengthening of the ETH 
can be made by asserting that the entire reduced density matrix supported on a finite spatial region appears thermal. More precisely, the subsystem eigenstate thermalization hypothesis states that the reduced density matrices of eigenstates, $\rho_A(\psi)$, are exponentially close in trace distance to a universal thermal density matrix, $\rho_{\mbox{\tiny univ}}(E)$, that only depends on the total energy \cite{2018PhRvE..97a2140D,2018PhRvX...8b1026G}
\begin{align}
    \left|\rho_A(\psi) - \rho_{\mbox{\tiny univ}}(E) \right|_1 = O\left(e^{-S(E)/2}\right).
    \label{SETH1}
\end{align} 
In addition, ``off-diagonal'' matrices are exponentially suppressed
\begin{align}
    \left|\Tr_B\left[\ket{E_i}\bra{E_j} \right] \right|_1 = \delta_{ij}+  O\left(e^{-S(E)/2}\right).
    \label{SETH2}
\end{align}
These conditions imply the local ETH for \textit{all} operators in region $A$ and are significantly stronger \cite{2016arXiv161000302L}. 

It is important to understand which systems obey the subsystem ETH. Of course, for all systems, when $A$ is the entire system, 
the subsystem ETH 
completely fails because the distance between a pure state and a thermal state is $O(1)$. It is then nontrivial to determine at which point the subsystem ETH breaks down and thermal physics no longer applies. In the following, we show that holographic CFTs obey the subsystem ETH 
whenever $A$ is smaller than half the total system size. More generally, we find generic chaotic Hamiltonian systems, whose eigenstate ansatzes were put forward in Refs.~\cite{2010NJPh...12g5021D,2017arXiv170908784L,2019PhRvE.100b2131M}, obey the subsystem ETH.

\subsection{Generic chaotic Hamiltonians}

We use the following ansatz for the tensor product decomposition of energy eigenstates of energy $E$ for generic chaotic quantum many-body systems \cite{2010NJPh...12g5021D}
\begin{align}
    \ket{E} = \mathcal{N}^{-1/2}\sum_{E -\Delta < E_i + E_J <E +\Delta}c_{iJ}\ket{E_i}_A\ket{E_J}_B,
    \label{ketE_ansatz}
\end{align}
where $\Delta \ll E$, $\mathcal{N}$ is the normalization, $\ket{E_i}_A$ and $\ket{E_J}_B$ are subsystem energy eigenstates\footnote{In fact, it is not quite correct to consider these subsystem energy eigenstates due to the interaction terms in the Hamiltonian coupling $A$ and $B$ that lead to correlations near the boundary, as emphasized in Ref.~\cite{2017arXiv170908784L}. It is more accurate to consider very similar states referred to as ``many-body Berry (MBB) states'' in Ref.~\cite{2017arXiv170908784L}. These are constructed from perturbing an integrable Hamiltonian by an integrability breaking term. In MBB states, the subsystem eigenstates are not \textit{energy} eigenstates, but local product states. The following calculation is unchanged.}, and the coefficients are complex Gaussian random variables
\begin{align}
    \overline{c_{iJ}c^*_{i'J'}} = \delta_{ii'}\delta_{JJ'},
\end{align}
where, 
with the proper normalization,
the variance is set to one.
The reduced density matrix is
\begin{align}
    \rho_A = \mathcal{N}^{-1}\sum_{E_i -2\Delta < E_j <E_i +2\Delta}\sum_{E-E_i-\Delta<E_J<E-E_i+\Delta}c_{iJ}c^*_{jJ}\ket{E_i}\bra{E_j}_A.
\end{align}
Using this ansatz for the reduced density matrix, we perform the replica trick for the PRRE. In analogy with Refs.~\cite{2010NJPh...12g5021D,2017arXiv170908784L,2019PhRvE.100b2131M,2020JHEP...11..007D} where the  R\'enyi and von Neumann entropies were evaluated for this ansatz, we find, in analogy with Section \ref{PRRE_sec}, that in the thermodynamic limit, after ensemble averaging
\begin{align}
    \overline{\Tr \left[\rho_A^{\alpha}\sigma_A^{1-\alpha}\right]}= \mathcal{N}^{-1} \int d\mathcal{E}e^{S_A(\mathcal{E})+S_B(E-\mathcal{E})}G_{\alpha}(\mathcal{E}),
    \label{PRRE_int_cha}
\end{align}
where
\begin{align}
    G_{\alpha}(\mathcal{E}) := \begin{cases}\, _2F_1\left(1-{\alpha},-{\alpha};2;e^{S_A(\mathcal{E})-S_B(E-\mathcal{E})}\right)
    \,
   _2F_1\left({\alpha}-1,{\alpha};2;e^{S_A(\mathcal{E})-S_B(E-\mathcal{E})}\right), 
       \\& \hspace{-5cm}S_A(\mathcal{E}) < S_B(E-\mathcal{E})
   \\
   \\
   e^{S_B(E-\mathcal{E})-S_A(\mathcal{E})}{ \, _2F_1\left(1-{\alpha},-{\alpha};2;e^{S_B(E-\mathcal{E})-S_A(\mathcal{E})}\right)
     \,
   _2F_1\left({\alpha}-1,{\alpha};2;e^{S_B(E-\mathcal{E})-S_A(\mathcal{E})}\right)} , 
       \\&\hspace{-5cm} S_A(\mathcal{E}) > S_B(E-\mathcal{E})
   \end{cases},
\end{align}
and
\begin{align}
    \mathcal{N} = \int d\mathcal{E}e^{S_A(\mathcal{E})+S_B(E-\mathcal{E})} .
\end{align}
We use the following ansatz for the thermodynamic entropies
\begin{align}
    S_A(\mathcal{E})=fVs\left( \frac{\mathcal{E}}{Vf}\right), \quad  S_B(E-\mathcal{E})=(1-f)Vs\left( \frac{E-\mathcal{E}}{V(1-f)}\right),
\end{align}
where $s(u)$ is the entropy density, $V$ is the volume of the total system and $f$ is the fractional volume of subsystem $A$.
The saddle point equation for the main integral is 
\begin{align}
    s'\left( \frac{\mathcal{E}_1}{Vf}\right) = s'\left( \frac{E-\mathcal{E}_1}{V(1-f)}\right) - \frac{\partial_{\mathcal{E}}G_{\alpha}(\mathcal{E}_1)}{G_{\alpha}(\mathcal{E}_1)}
\end{align}
and for the normalization
\begin{align}
    s'\left( \frac{\mathcal{E}_2}{Vf}\right) = s'\left( \frac{E-\mathcal{E}_2}{V(1-f)}\right).
\end{align}

We can now evaluate the PRRE in various regimes. The saddle point equation for the normalization is simple to solve because $s'(u)$ is single-valued
\begin{align}
    \mathcal{E}_2 = f E.
\end{align}
This is not so surprising as it implies a constant energy density. The normalization is then evaluated to
\begin{align}
    \mathcal{N} = \sqrt{ \frac{2\pi V f(1-f)}{s''\left( \frac{E}{V}\right)}}e^{V s\left( \frac{E}{V}\right)}
\end{align}

We now specify to $f < 1/2$ where we claim that $\mathcal{E}_1 < E/2$. In this regime, we can expand the hypergeometric function
\begin{align}
    G_{\alpha}(\mathcal{E}) \simeq 1 + \alpha(\alpha-1)e^{S_A(\mathcal{E})- S_B(E-\mathcal{E})}.
\end{align}
In this approximation,
\begin{align}
    \frac{\partial_{\mathcal{E}}G_{\alpha}(\mathcal{E}_1)}{G_{\alpha}(\mathcal{E}_1)} = \frac{(\alpha -1) \alpha  e^{f V s\left(\frac{\mathcal{E}_1}{f V}\right)}
   \left(s'\left(\frac{E-\mathcal{E}_1}{(1-f) V}\right)+s'\left(\frac{\mathcal{E}_1}{f
   V}\right)\right)}{e^{(1-f) V s\left(\frac{E-\mathcal{E}_1}{(1-f)
   V}\right)}+(\alpha -1) \alpha  e^{f V s\left(\frac{\mathcal{E}_1}{f V}\right)}}.
\end{align}
This term is exponentially small for $f < 1/2$. Therefore, the saddle point equation for $\mathcal{E}_1$ can be treated as an expansion around $\mathcal{E}_2$
\begin{align}
    \mathcal{E}_1 = \mathcal{E}_2(1+\delta), \quad \delta \ll 1.
\end{align}
To leading order, the saddle point equation is 
\begin{align}
    \frac{E s''\left(\frac{E}{V}\right)}{V}\delta &= -\frac{E s''\left(\frac{E}{V}\right)f}{V(1-f)}\delta -\frac{2 (\alpha -1) \alpha  e^{(2 f-1) V s\left(\frac{E}{V}\right)}
   s'\left(\frac{E}{V}\right)}{(\alpha -1) \alpha  e^{(2 f-1) V
   s\left(\frac{E}{V}\right)}+1}
   \nonumber
   \\
   &\simeq 
   -\frac{E s''\left(\frac{E}{V}\right)f}{V(1-f)}\delta -{2 (\alpha -1) \alpha  e^{(2 f-1) V s\left(\frac{E}{V}\right)}
   s'\left(\frac{E}{V}\right)}.
\end{align}
Solving for $\delta$, we find self-consistency with the claim that $\mathcal{E}_1$ is very close to $\mathcal{E}_2$
\begin{align}
    \delta \simeq \frac{2 (\alpha -1) \alpha  (f-1) V e^{(2 f-1) V
   s\left(\frac{E}{V}\right)} s'\left(\frac{E}{V}\right)}{E
   s''\left(\frac{E}{V}\right)}.
\end{align}
The saddle point solution is then
\begin{align}
     \int d\mathcal{E}e^{S_A(\mathcal{E})+S_B(E-\mathcal{E})}G_{\alpha}(\mathcal{E}) \simeq \sqrt{ \frac{2\pi V f(1-f)}{s''\left( \frac{E}{V}\right)}} e^{Vs\left(\frac{E}{V}\right)+\alpha(\alpha -1)e^{(2 f-1) V s\left(\frac{E}{V}\right)}+ O(\delta^2)}.
\end{align}
Therefore, the moments are
\begin{align}
    \overline{\Tr \left[\rho_A^{\alpha}\sigma_A^{1-\alpha}\right]}\simeq  e^{\alpha(\alpha -1)e^{(2 f-1) V s\left(\frac{E}{V}\right)}},
\end{align}
so the PRRE is exponentially suppressed in the entropy for all values of $\alpha$
\begin{align}
    \overline{D_{\alpha}(\rho_A || \sigma_A)} \simeq \alpha e^{(2 f-1) V s\left(\frac{E}{V}\right)}.
\end{align}
This places strict bounds on the trace distance
\begin{align}
    \frac{e^{(2 f-1) V s\left(\frac{E}{V}\right)}}{4}\leq \overline{T(\rho_A || \sigma_A)}\leq\frac{e^{( f-1/2) V s\left(\frac{E}{V}\right)}}{\sqrt{2}}.
\end{align}
This provides further evidence for the refined subsystem eigenstate thermalization hypothesis in Ref.~\cite{2018PhRvE..97a2140D} that postulated the scaling of the trace distance on the size of subregion $A$ because the upper bound precisely matches that scaling when replacing $N_A$, the number of qubits in region $A$, with the subsystem thermodynamic entropy.

Note also that these results generalize our result for random matrix theory because at infinite temperature, we can identify
\begin{align}
    f V s\left(\frac{E}{V} \right) = \log d_A ,\quad  (1-f) V s\left(\frac{E}{V} \right) = \log d_B.
\end{align}

Next, consider the $f > 1/2$. Unfortunately, the maximum of the integral occurs right near the transition $S_A(\mathcal{E}_1)=S_B(E-\mathcal{E}_1)$.
Because of this, we cannot simply make a saddle point approximation. However, it is straightforward to argue that the PRRE will be large for $\alpha < 1$ and infinite for $\alpha \geq 1$. Note that the integrands of the numerator and denominator of ${\Tr \left[\rho_A^{\alpha}\sigma_A^{1-\alpha}\right]}$ are exponentially close for $S_A(\mathcal{E}_1)<S_B(E-\mathcal{E}_1)$ while the numerator is exponentially suppressed in relation to the denominator for $S_A(\mathcal{E}_1)>S_B(E-\mathcal{E}_1)$. Because the saddle point for the denominator occurs when $S_A(\mathcal{E}_1)>S_B(E-\mathcal{E}_1)$, ${\Tr \left[\rho_A^{\alpha}\sigma_A^{1-\alpha}\right]}$ will be exponentially small i.e.~$\log\left[{\Tr \left[\rho_A^{\alpha}\sigma_A^{1-\alpha}\right]}\right]$ will be negative and of order the entropy. Due to the factor of $(\alpha-1)^{-1}$ in the PRRE, this implies that the PRRE for $\alpha < 1$ is of order the entropy and ill-defined (infinite) for $\alpha \geq 1$, in analogy with the random matrix theory result \eqref{PRRE_eq}.

We have found that whenever $A$ is less than half the total system size, the PRRE and therefore the subsystem trace distance between any two eigenstates of the same energy is exponentially suppressed in the entropy. The trace distance is a metric on the space of density matrices, so these eigenstates lie within a ball with radius $O(e^{-S(E)})$. The universal density matrix then must also lie within this ball such that \eqref{SETH1} is satisfied.

For \eqref{SETH2}, we need to perform an additional computation. The off diagonal ($i\neq j$) matrix for two random states is represented as
\begin{align}
    \label{eq:rho_offdiag}
    \Tr_B\left[\ket{\psi_i}\bra{\psi_j} \right]
:= 
\,
    \tikz[scale=1.0,baseline=-0.5ex]{
    \draw[dashed] (0,0.2)  -- (0,0);
    \draw[dashed] (0,0)  -- (0.5,0);
    \draw[dashed,red] (0.5,0)  -- (1,0);
    \draw[dashed,red]  (1,0.2)  -- (1,0);
    \draw[] (-0.2,0.2)-- (-0.2,-0.15);
    \draw[red] (1.2,0.2) -- (1.2,-0.15);
    }\ .
\end{align}
To compute the trace norm, we need the integer powers of 
\begin{align}
    \label{eq:rho_offdiag_square}
    \Tr_B\left[\ket{\psi_i}\bra{\psi_j} \right]\Tr_B\left[\ket{\psi_j}\bra{\psi_i} \right]
= 
\,
    \tikz[scale=1.0,baseline=-0.5ex]{
    \draw[dashed] (0,0.2)  -- (0,0);
    \draw[dashed] (0,0)  -- (0.5,0);
    \draw[dashed,red] (0.5,0)  -- (1,0);
    \draw[dashed,red]  (1,0.2)  -- (1,0);
    \draw[] (-0.2,0.2)-- (-0.2,-0.15);
    \draw[red] (1.2,0.2) -- (1.2,-0.15);
     \draw[dashed,red] (2,0.2)  -- (2,0);
    \draw[dashed,red] (2,0)  -- (2.5,0);
    \draw[dashed,] (2.5,0)  -- (3,0);
    \draw[dashed,]  (3,0.2)  -- (3,0);
    \draw[red] (2-0.2,0.2)-- (2-0.2,-0.15);
    \draw[] (3.2,0.2) -- (3.2,-0.15);
    \draw[red]  (1.2,-0.15)-- (2-0.2,-0.15);
    }\ .
\end{align}
The moments are given by a new sum over permutations
\begin{align}
    \overline{\Tr\left[ \left(\Tr_B\left[\ket{\psi_i}\bra{\psi_j} \right]\Tr_B\left[\ket{\psi_j}\bra{\psi_i} \right]\right)^{\alpha}\right]}= \frac{1}{(d_A d_B)^{2\alpha}}\sum_{\tau \in S_{odd}} d_A^{C(\eta^{-1}\circ \tau)}d_B^{C( \tau) }.
\end{align}
Here, $S_{odd}$ represents the set of permutations where, within each cycle, the difference between consecutive numbers is always odd. For example, the cycle $(1,2,5,6)$ is allowed, but $(2,4,5)$ is not.
Crucially, the identity permutation is not an allowed permutation. If we want to maximize the number of dashed loops for small $d_A/d_B$, $\tau$ must be composed of $\alpha$ noncrossing two-cycles. The degeneracy is given by the Catalan number so that
\begin{align}
    \overline{\Tr\left[ \left(\Tr_B\left[\ket{\psi_i}\bra{\psi_j} \right]\Tr_B\left[\ket{\psi_j}\bra{\psi_i} \right]\right)^{\alpha}\right]} = C_{\alpha} d_A^{1-\alpha}d_B^{-\alpha} + O\left(d_B^{-\alpha -1} \right).
\end{align}
The trace norm is the $\alpha \rightarrow 1/2$ limit, such that
\begin{align}
    \overline{\left|\Tr_B\left[\ket{\psi_i}\bra{\psi_j} \right] \right|_1} = \frac{8}{3\pi}\sqrt{\frac{d_A}{d_B}} + O\left(\frac{d_A}{d_B} \right)^{3/2}.
\end{align}
When $d_A/d_B$ is large, the cyclic permutation is an allowed permutation and will dominate, leading to 
\begin{align}
    \overline{\Tr\left[ \left(\Tr_B\left[\ket{\psi_i}\bra{\psi_j} \right]\Tr_B\left[\ket{\psi_j}\bra{\psi_i} \right]\right)^{\alpha}\right]}= d_B^{1-2\alpha} + O\left(d_A^{-1} \right).
\end{align}
Taking the $\alpha \rightarrow 1/2$ limit tells us that $\left|\Tr_B\left[\ket{\psi_i}\bra{\psi_j} \right] \right|_1 $ is exponentially close to one. 

For finite energy eigenstates, this translates to 
\begin{align}
    \overline{\Tr\left[ \left(\Tr_B\left[\ket{\psi_i}\bra{\psi_j} \right]\Tr_B\left[\ket{\psi_j}\bra{\psi_i} \right]\right)^{\alpha}\right]} = \mathcal{N}^{-2\alpha} \int d\mathcal{E}e^{S_A(\mathcal{E}) +S_B(E-\mathcal{E})}G_{\alpha}(\mathcal{E}),
\end{align}
where $G_{\alpha}(\mathcal{E})$ is now given by
\begin{align}
    G_{\alpha}(\mathcal{E}) = \begin{cases}
       C_{\alpha}e^{\alpha S_A(\mathcal{E}) +(\alpha-1)S_B(E-\mathcal{E})}, & e^{S_A(\mathcal{E})-S_B(E-\mathcal{E})} \ll 1
       \\
       e^{ (2\alpha-1)S_A(\mathcal{E})}, &  e^{S_A(\mathcal{E})-S_B(E-\mathcal{E})} \gg 1
    \end{cases}.
\end{align}
At $\alpha = 1/2$, for sufficiently small $f$, the saddle point will fall in the $e^{S_A(\mathcal{E})-S_B(E-\mathcal{E})} \ll 1$ regime leading to a saddle point equation for the numerator of 
\begin{align}
    3s'\left( \frac{\mathcal{E}_1}{Vf}\right) = s'\left( \frac{E-\mathcal{E}_1}{V(1-f)}\right)
\end{align}
Due to the factor of $3$, $\mathcal{E}_1$ will be larger than $\mathcal{E}_2$. We then find
\begin{align}
    \overline{\left|\Tr_B\left[\ket{\psi_i}\bra{\psi_j} \right] \right|_1} \simeq \frac{8 e^{\frac{3}{2}S_A(\mathcal{E}_1) +\frac{1}{2}S_B(E-\mathcal{E}_1)}}{3\pi e^{S_A(\mathcal{E}_2) +S_B(E-\mathcal{E}_2)}}.
\end{align}
This is exponentially small because $ e^{S_A(\mathcal{E}_1) +S_B(E-\mathcal{E}_1)}\ll e^{S_A(\mathcal{E}_2) +S_B(E-\mathcal{E}_2)}$ and $S_A(\mathcal{E}_1) \ll S_B(E-\mathcal{E}_1)$. For sufficiently large $f$, the saddle point will fall in the other regime such that the numerator and denominator are identical at $\alpha = 1/2$, leading to $\overline{\left|\Tr_B\left[\ket{\psi_i}\bra{\psi_j} \right] \right|_1} \simeq 1$ at leading order.

\subsection{Holographic states}

We could now simply posit that because holographic systems are believed to be chaotic, their eigenstates will also have a spatial decomposition according to \eqref{ketE_ansatz} and thus, will obey the subsystem ETH 
for $f < 1/2$. However, this line of reasoning is somewhat unsatisfying because it is not constructive. 
Instead, we implement a gravitational calculation, using the fixed-area state analysis from Section \ref{black_hole_sec}, to evaluate the PREE in normal states without any areas fixed. Our strategy follows Ref.~\cite{2020JHEP...11..007D} in manipulating the gravitational path integral into a form identical to \eqref{PRRE_int_cha}, deriving the validity of using \eqref{ketE_ansatz} for holographic eigenstates when computing the PRRE.
Due to the similarities with Ref.~\cite{2020JHEP...11..007D}, we keep the derivation brief, referring the interested reader to the original literature.

To begin, we make the assumption that black hole microstates can be represented as a random superposition of energy eigenstates in a microcanonical energy window $I_E = [E- \Delta , E+ \Delta ]$
\begin{align}
    \ket{E,\hat{c}} \propto \sum_{E_i \in I_E }c^ie^{-\frac{\beta E_i}{2}} \ket{E_i},
\end{align}
where $\beta$ is an effective temperature. These states are believed to be holographically dual to black hole geometries with end-of-world branes specifying the microstate lying behind the horizon \cite{2013JHEP...05..014H,2017arXiv170702325K,2018arXiv180304434A,2019JHEP...07..065C,2021arXiv210306893M}, similar to the PSSY model. 

The corresponding density matrix is represented as the path integral on a strip of width $\beta$ with boundary conditions determined by $\hat{c}$. We will consider two microstates in the same energy window, corresponding to two independent sets of Gaussian random variables $\hat{c}$ and $\hat{d}$. As argued in Ref.~\cite{2020JHEP...11..007D}, after disorder averaging, the random variables match up the boundary conditions of the strips according to the same Wick contractions previously discussed for Haar random states. Therefore, the path integral is given by a sum over all allowed Wick contractions
\begin{align}
    \Tr \left[ \rho_A^{\alpha }\sigma_A^{m}\right] = \frac{\mathcal{Z}_{\alpha, m}}{\mathcal{Z}_{1,0}^{\alpha }\mathcal{Z}_{0,1}^{ m}}   ,\quad  \mathcal{Z}_{\alpha, m} \simeq \sum_{\mathcal{M}_i}\int \mathcal{D}\phi e^{-I_E(\mathcal{M}_i,\phi)},
\end{align}
where $\mathcal{M}_i$ are the replica manifolds. Using the holographic dictionary, this is a sum over bulk geometries with asymptotic boundary conditions $\mathcal{M}_i$. In these bulk geometries, the EOW branes have disappeared due to the disorder averaging. This is an alternative way to see the reduction of bulk saddles from $S_{\alpha + m}$ to $S_{\alpha }\times S_m$. In general, solving the bulk equations of motion to evaluate the path integrals on shell is very difficult. We use the ``double-defect'' construction of Ref.~\cite{2020JHEP...11..007D} to separate the action into a bulk contribution $I_{bulk}(g,\phi, E)$ and actions, $I_{brane}(\Sigma_1)$ and $I_{brane}(\Sigma_2)$, for cosmic branes, $\Sigma_1$ and $\Sigma_2$, that are located at the two extremal surfaces
\begin{align}
    \mathcal{Z}_{\alpha, m} = \sum_{\mathcal{M}_i}\int \mathcal{D}\phi \mathcal{D}g \mathcal{D}\Sigma_1 \mathcal{D}\Sigma_2 e^{-(\alpha + m)I_{bulk}(g,\phi, E) -\frac{\alpha + m -k_i}{4G_N}I_{brane}(\Sigma_1)-\frac{k_i-1}{4G_N}I_{brane}(\Sigma_2)},
\end{align}
where $k_i$ plays the role of the number of cycles in the Wick contraction corresponding to $\mathcal{M}_i$. The sum over $\mathcal{M}_i$ may be done prior to the path integral and we also take $m\rightarrow 1- \alpha$ to arrive at
\begin{align}
    \mathcal{Z}_{\alpha, m} =\int \mathcal{D}\phi \mathcal{D}g \mathcal{D}\Sigma_1 \mathcal{D}\Sigma_2 e^{-I_{bulk}(g,\phi, E) }G_{\alpha}(\Sigma_1,\Sigma_2),
\end{align}
where
\begin{align}
    G_{\alpha}(\Sigma_1,\Sigma_2) = \begin{cases}\, _2F_1\left(1-{\alpha},-{\alpha};2;e^{\frac{\Delta I_{brane}}{4G_N}}\right)
    \,
   _2F_1\left({\alpha}-1,{\alpha};2;e^{\frac{\Delta I_{brane}}{4G_N}}\right), 
       \\
       &\hspace{-2cm}\Delta I_{brane} < 0
   \\
   \\
   e^{-\frac{\Delta I_{brane}}{4G_N}}{ \, _2F_1\left(1-{\alpha},-{\alpha};2;e^{-\frac{\Delta I_{brane}}{4G_N}}\right)
     \,
   _2F_1\left({\alpha}-1,{\alpha};2;e^{-\frac{\Delta I_{brane}}{4G_N}}\right)} , 
     \\
       &\hspace{-2cm}  \Delta I_{brane} > 0
   \end{cases},
\end{align}
and $\Delta I_{brane}:=I_{brane}(\Sigma_1)-I_{brane}(\Sigma_2) $. Next, we make use of the fixed-area basis by postponing the integrals over the areas of the branes until the very end, such that the integral is rewritten as 
\begin{align}
    \mathcal{Z}_{\alpha, 1-\alpha} =\int dA_1dA_2 P(A_1, A_2) G_{\alpha}(A_1,A_2),
\end{align}
where $G_{\alpha}(A_1,A_2)$ is identical to $G_{\alpha}(\Sigma_1,\Sigma_2)$ with $I_{brane}(\Sigma_1)\leftrightarrow A_1$ and $I_{brane}(\Sigma_2)\leftrightarrow A_2$ and $ P(A_1, A_2)$ is the (unnormalized) probability of being in the state with areas $A_1$ and $A_2$
\begin{align}
    P(A_1, A_2) := \int \mathcal{D}g\mathcal{D}\phi\Big|_{A_{\Sigma_1} = A_1,A_{\Sigma_2} = A_2}e^{-I_{bulk}(g,\phi, E) }.
\end{align}
For high-energy eigenstates, $P(A_1,A_2)$ will localize to a trajectory where $A_2$ is a function of $A_1$ \cite{2020JHEP...11..007D}
\begin{align}
    P(A_1, A_2) \simeq \delta_{A_2, A_2(A_1)} \int \mathcal{D}g\mathcal{D}\phi\Big|_{A_{\Sigma_1} = A_1}e^{-I_{bulk}(g,\phi, E) }.
\end{align}
Finally, to compare with \eqref{PRRE_int_cha}, we want to change the integration variable from $A_1$ to the energy density in region $A$, $\mathcal{E}$. Using the Bekenstein-Hawking formula \cite{cmp/1103899181}, we write the areas in terms of entropy densities
\begin{align}
    \frac{A_1}{4G_N} = f Vs\left(\frac{\mathcal{E}(A_1, A_2)}{f V} \right) + A_{\infty},\quad \frac{A_2}{4G_N} = (1-f) Vs\left(E-\frac{\mathcal{E}(A_1, A_2)}{(1-f) V} \right)+ A_{\infty},
\end{align}
where $A_{\infty}$ is the divergent piece of the Ryu-Takayanagi surface which approximately cancels in all expressions because we are in the high-energy limit where the surfaces are approximately purely radial until they reach the horizon and subsequently tightly wrap the horizon. $\mathcal{E}(A_1, A_2)$ is the ADM energy which is a function of the horizon area $A_{BH} \simeq A_1 + A_2 -2 A_{\infty}$. In a saddle point approximation, the probability then becomes \cite{2020JHEP...11..007D}
\begin{align}
    P(\mathcal{E}) \simeq e^{S_A(\mathcal{E}) + S_{B}(E- \mathcal{E})}.
\end{align}
In total, we find 
\begin{align}
    \frac{\mathcal{Z}_{\alpha, 1-\alpha}}{\mathcal{Z}_{1,0}^{\alpha}\mathcal{Z}_{0,1}^{1-\alpha}} \simeq \frac{\int d\mathcal{E}  e^{S_A(\mathcal{E}) + S_{B}(E- \mathcal{E})} G_{\alpha}(\mathcal{E})}{\int d\mathcal{E}  e^{S_A(\mathcal{E}) + S_{B}(E- \mathcal{E})} }.
\end{align}
It should now be evident that this formula is identical to \eqref{PRRE_int_cha}, so we conclude that the PRRE for holographic theories is exponentially small in the entropy when $f < 1/2$ and subsystem eigenstate thermalization will hold.

\section{Discussion}

We take the opportunity to now comment on future directions that are out of the scope of this work but deserve attention.

On the formal end, one may be worried that the basis of our calculations have assumed finite Hilbert space dimensions and tensor factorization of the Hilbert space into $\mathcal{H}_A \otimes \mathcal{H}_B$. This is not true in quantum field theory and is the reason why reduced density matrices and von Neumann entropies are not well-defined. However, the relative entropy and related quantities are well-defined quantities in the continuum using modular theory (see, for example, Refs.~\cite{2017arXiv170204924H, 2018arXiv180304993W}). 
For this reason, we expect that our calculations that assumed tensor factorization are accurate even though we ignored this subtlety. However, this expectation is not guaranteed. In particular, the rank deficiency in the reduced density matrices that led to infinite relative entropy has no analog in the continuum because subregions are described by Type III von Neumann algebras which are roughly ``full rank.''

More practically, we were only able to compute the relative entropies of the simple PSSY toy model of black hole evaporation. We expect that this model captures the essential features of the distinguishability of evaporating black holes, like the wormhole contributions that restore unitarity. It is an important future direction to complete analogous calculations in more realistic models of black hole evaporation. Though a complete calculation including all saddles like in JT gravity is most likely out of reach, one may be able to identify the saddles that lead to $O(e^{-1/G_N})$ fidelity prior to the Page time and $O(1)$ fidelity after the Page time.

For non-evaporating black holes, our calculation was limited to large black holes away from phase transitions. It is clearly of interest to generalize these holographic calculations to smaller, normal black holes. Moreover, there may be interesting cross-over behavior in relative entropies near the phase transitions. In the Page states, the transitions were $O(1)$ (e.g.~for relative entropy, the $D(\rho_A || \sigma_A) = 3/2$ at the transition). However, for finite energy states, there may be enhanced corrections analogous to the $O(\sqrt{V})$ corrections in the von Neumann entropy \cite{2017PhRvL.119v0603V,2019PhRvE.100b2131M,2020JHEP...11..007D,2020JHEP...12..084M}. These should be visible in chaotic, non-holographic systems as well. Preliminary numerical results were given in Ref.~\cite{2021PhRvL.126q1603K}, though a more systematic study of both integrable and chaotic Hamiltonians is warranted.

Finally, it would be interesting to study random tensor networks that act as quantum error correcting codes from ``bulk'' Hilbert spaces to ``boundary'' Hilbert spaces. These quantum error correcting codes are more closely related to the semiclassical holographic states than the tensor network states studied in Section \ref{sec_tensor}. Using these tensor networks, various interesting corrections to the JLMS formula \cite{2016JHEP...06..004J} may be explored. We expect this to have important implications for approximate quantum error correction and entanglement wedge reconstruction in AdS/CFT.

\acknowledgments

We thank Anatoly Dymarsky, Santosh Kumar, Pratik Rath, Jonathan Sorce, and Karol \.{Z}yczkowski for helpful discussions and comments. S.R.~is supported
by the National Science Foundation under Award No.
DMR-2001181, and by a Simons Investigator Grant from
the Simons Foundation (Award No.~566116).

\appendix

\section{Distinguishibility from free probability theory}
\label{free_prob_app}

We have tried to avoid the use of free probability theory throughout this paper because while it is immensely powerful, it is very technical and unintuitive for the wide audience that this paper is intended for. Moreover, it lacks the clear connections to gravitational and general chaotic systems that the replica trick has. Alas, for exact calculations of the trace distance and Uhlmann fidelity, we were unable to push our diagrammatic techniques far enough so we address distinguishability measures using free probability in this appendix.

We begin a minimal introduction of the tools we will use from applications of free probability theory as applied to random matrices. The interested reader is encouraged to consult Refs.~\cite{nica_speicher_2006,MingoSpeicher_2017}. First, we define a unital linear form, $\tau$, acting on a vector space. For $N\times N$ matrices, $X$, this can be taken to be the normalized trace $\tau(X) := \frac{1}{N}\Tr\left[ X \right]$. Random variables $X$ and $Y$ are \textit{free} if they satisfy
\begin{align}
    \tau\left(P_1(X)Q_1(Y)\dots P_k(X)Q_k(Y)\right) = 0,
\end{align}
for any set of polynomials $\{P,Q\}$ when $\tau\left(P_i(X) \right)=\tau\left(Q_i(Y) \right) =0$. This, in turn, implies a factorization property
\begin{align}
    \tau\left( P(X)Q(Y)\right) = \tau\left( P(X)\right)\tau\left(Q(Y)\right),
    \label{free_factor}
\end{align}
which can be seen by taking $P_1(X) = P(X) -\tau(P(X))$ and $Q_1(X) = Q(X) -\tau(Q(X))$.
There are several useful ways to package the moments of a free random variable, $m_n := \tau(X^n)$. These enable the evaluation of empirical spectral measures for random matrices. For a free random variable $X$, we define the moment function as a formal power series
\begin{align}
    M_X(z) := \sum_{m = 0}^{\infty}\tau(X^n)z^m,\quad z \in \mathbb{C}.
\end{align}
This is related to the Green's function or \textit{Cauchy transform} as
\begin{align}
    G_X(z) := \frac{1+M_X(z)}{z}.
\end{align}
From the Cauchy transform, one may extract the spectral measure using a Stieltjes transformation
\begin{align}
    \mu_X(\lambda) = -\frac{1}{\pi}\lim_{\epsilon\rightarrow 0}\Im\left[G_X(\lambda + i\epsilon) \right].
\end{align}
Sometimes, we know that random variables $X$ and $Y$ have spectral measures $\rho_X$ and $\rho_Y$ but want to know the spectral measure of their sum, $X+Y$, or product, $XY$. The spectral measure of their sum is defined as the free convolution $\mu_{X+Y} := \mu_X \boxplus \mu_Y$ while the spectral measure of their product is defined as the free multiplicative product $\mu_{XY} := \mu_X \boxtimes \mu_Y$. In order to obtain the free convolution, it is convenient to introduce the \textit{R-transform}
\begin{align}
    R_X(G_X(z)) + \frac{1}{G_X(z)} :=  z.
\end{align}
The R-transform of a sum of free random variables is given by the sum of their individual R-transforms
\begin{align}
    R_{X+Y}(z) =  R_{X}(z) + R_{Y}(z) .
\end{align}
For the free multiplicative product, it is convenient to introduce the \textit{S-tranform}
\begin{align}
    z M_X(z) S_X(M_X(z)) := 1 + M_X(z) .
    \label{stransform_def}
\end{align}
The S-tranform of a product of free random variables is given by the product of their individual S-transforms
\begin{align}
    S_{XY}(z) =  S_{X}(z) + S_{Y}(z) .
\end{align}
The key to the usefulness of free probability for us is that Wishart random matrices are free random variables asymptotically, as the dimensions become large. Their empirical spectral measure is given by the Marchenko–Pastur distribution
\begin{align}
    \mu^{(c)}_{MP}(x) = \max\left[1- \frac{1}{c},0 \right]\delta(x)+ \frac{\sqrt{(x-c(1-{c}^{-1/2})^2)(c(1+{c}^{-1/2})^2-x)} }{2\pi c x},
\end{align}
where $c :=d_A/d_B$ is the rectangular parameter and $x := d_A \lambda$.

\paragraph{Petz R\'enyi relative entropy}
For the PRRE, we need to evaluate $\overline{\Tr\left[\rho_A^{\alpha}\sigma^{1-\alpha}_A\right]}$. Because $\rho_A$ and $\sigma_A$ are asymptotically free random variables with Marchenko–Pastur distributions, $\mu^{(c)}_{MP}$, this factorizes according to \eqref{free_factor} 
\begin{align}
    \overline{\Tr\left[\rho_A^{\alpha}\sigma^{1-\alpha}_A\right]} = \frac{\overline{\Tr\left[\rho_A^{\alpha}\right]\Tr\left[\sigma^{1-\alpha}_A\right]}}{d_A} = \int \mu^{(c)}_{MP}(x) x^{\alpha}\int \mu^{(c)}_{MP}(x) x^{1-\alpha},
\end{align}
These integrals may be evaluated to reproduce \eqref{PRRE_pre_eq_replica}, the result from the replica trick.

\begin{figure}
    \centering
    \includegraphics[width = .48\textwidth]{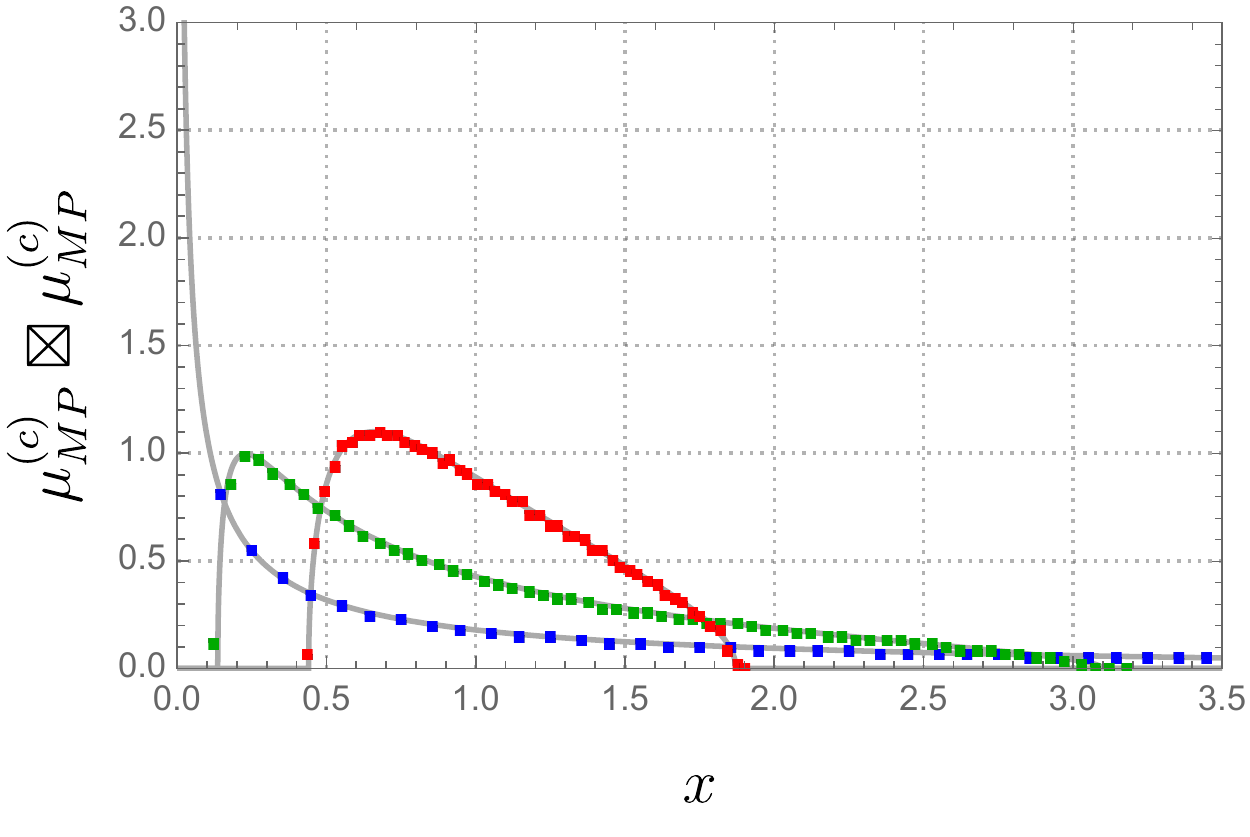}
    \includegraphics[width = .48\textwidth]{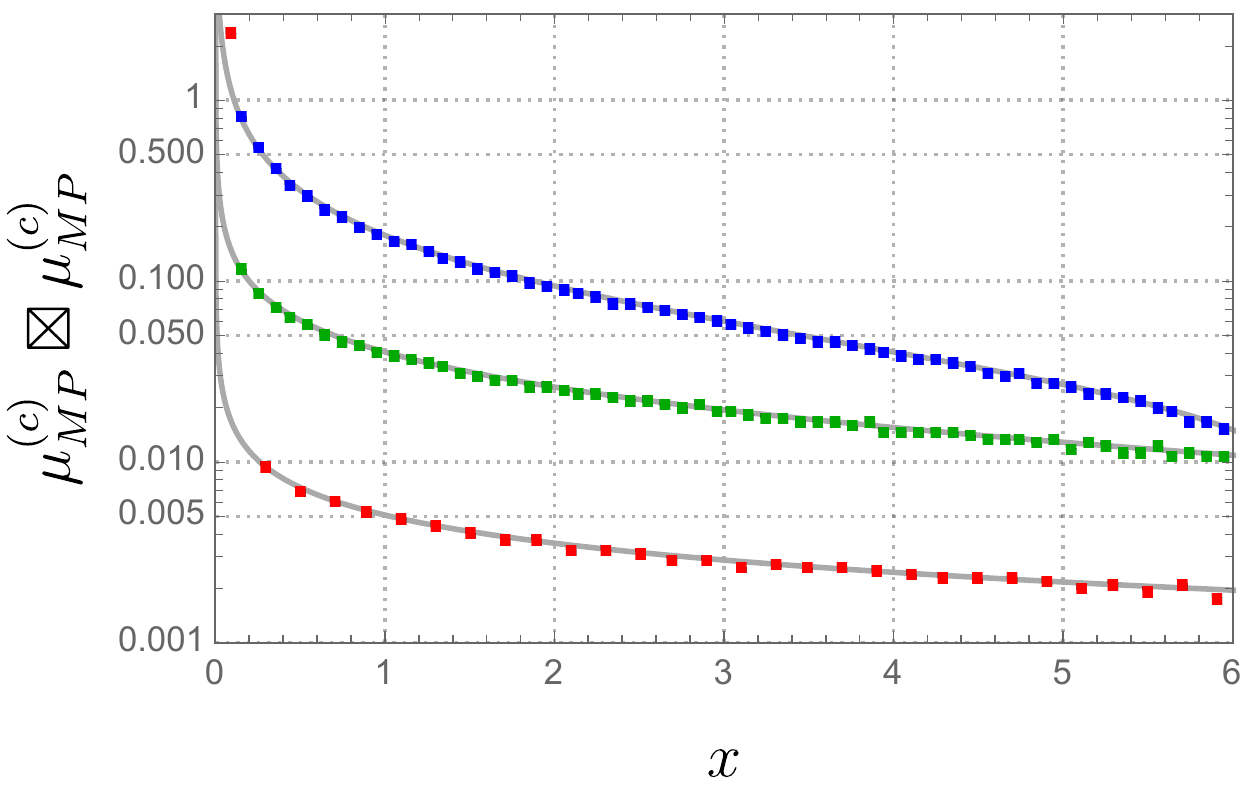}
    \caption{The probability density function for the generalized Fuss-Catalan distribution is shown (grey lines) with comparison to numerics. On the left, the blue, green and red dots correspond to $c = {2^0,2^{-2}, 2^{-4}}$ respectively. On the left, the blue, green and red dots correspond to $c = {2^0,2^{2}, 2^{4}}$ respectively. The total system size is $2^{16}$ and we disorder average over $10^3$ realizations.}
    \label{mu_square_distrib_numerics}
\end{figure}

\paragraph{Sandwiched R\'enyi relative entropy}
For the SRRE, we need to evaluate the averaged moments of $\sigma_A^{\frac{1-\alpha}{2\alpha}}\rho_A\sigma_A^{\frac{1-\alpha}{2\alpha}}$. Using the replica trick, we succeeded for integer $\alpha$, but were unable to analytically continue to real-valued $\alpha$ in order to evaluate the Uhlmann fidelity at $\alpha = \frac{1}{2}$. Here, we evaluate the spectrum of $\sigma_A^{\frac{1}{2}}\rho_A \sigma_A^{\frac{1}{2}}$ (equivalently $\rho_A \sigma_A$) to accomplish this goal. This is the free multiplicative convolution of Marchenko–Pastur laws $\mu^{(c)}_{MP} \boxtimes \mu^{(c)}_{MP}$ and has been called a generalized Fuss-Catalan distribution \cite{2015PhRvE..92a2121M}. The Fuss-Catalan distributions themselves were first derived in Ref.~\cite{2011PhRvE..83f1118P}. The S-tranform for the Marchenko–Pastur distribution is given by
\begin{align}
    S_{\rho_A}(z) = \frac{1}{1+cM_{\rho_A}(z)}. 
\end{align}
Therefore, the S-transform for $\rho_A \sigma_A$ is
\begin{align}
    S_{\rho_A \sigma_A}(z) = \frac{1}{\left(1+cM_{\rho_A}(z)\right)^2}. 
\end{align}
Plugging this into \eqref{stransform_def}, we find
\begin{align}
    z M_{\rho_A \sigma_A}(z) = (1+M_{\rho_A \sigma_A}(z))(1+cM_{\rho_A \sigma_A}(z))^2.
\end{align}
Taking the correct root of this cubic equation and taking the Stieltjes transformation, we find the spectral measure
\begin{align}
    \mu^{(c)}_{MP} \boxtimes \mu^{(c)}_{MP}(x) &= \max\left[0,\frac{c-1}{c}\right]\delta (x)\nonumber\\&- \frac{1}{\pi } \begin{cases}
\frac{1}{\sqrt{3} \cdot 2^{2/3}x } \frac{3x+(c-1)^2}{\left( \sqrt{P}+A\right) ^{1/3} }-\frac{\sqrt{3} \cdot 2^{2/3} }{12c^2x} \left( \sqrt{P}+A\right) ^{1/3} ,&(x_1,x_2) \\
0, & \text{otherwise}
\end{cases},
\label{MP_squared_eq}
\end{align}
where we have defined
\begin{equation}
\begin{split}
& A = c^3 \left[ 9(c+2)x+2(c-1)^3\right]  \\
& P = 27c^6 x\left[ \left( 8-(c-20)c\right) x+4(c-1)^3-4x^2\right] 
\end{split}
\end{equation}
and end points
\begin{equation}
\begin{split}
x_1 &= \max\left[0,\frac{8+20c-c^2-\sqrt{c}(8+c)^{3/2}}{8} \right]\\
x_2 &= \frac{8+20c-c^2+\sqrt{c}(8+c)^{3/2}}{8}
\end{split} .
\end{equation}
The Uhlmann fidelity is then given by the following integral
\begin{align}
    F(\rho_A ||\sigma_A) = \left(\int_{x_1}^{x_2} \mu^{(c)}_{MP} \boxtimes \mu^{(c)}_{MP}(x) \sqrt{x} \right)^2.
\end{align}
We compare the derived distribution with numerics in Fig.~\ref{mu_square_distrib_numerics}.

\paragraph{Trace distance}

The trace distance is defined using the trace norm of the difference of $\rho_A$ and $\sigma_A$. This is tailor-made for a computation in free probability because of free convolution. The R-transform for the Marchenko–Pastur distribution is 
\begin{align}
    \mathcal{R}_{\rho_A}(z) =  \frac{1}{1-c z}.
\end{align}
Because we are taking the difference, we must rescale the second R-transform
\begin{align}
    \mathcal{R}_{-\sigma_A}(z) = -\mathcal{R}_{\sigma_A}(-z)= -\frac{1}{1+c z}.
\end{align}
Therefore, the R-transform of the difference is
\begin{align}
    \mathcal{R}_{\rho_A-\sigma_A}(z) = \frac{2cz}{1-c^2z^2}.
\end{align}
Inverting the R-transform, one finds the Cauchy transform
\begin{align}
    \frac{2cG(z)}{1-c^2G(z)^2}+\frac{1}{G(z)}=z.
\end{align}
Once again, one can take the correct root of the cubic equation and take the Stieltjes transformation to find the spectral measure \cite{2016PhRvA..93f2112P, 2015arXiv151107278M}
\begin{align}
    \mu_{MP}^{(c)}\boxplus\mathcal{D}_{-1}\left[\mu_{MP}^{(c)}\right](x)= \begin{cases}
    \frac{\sqrt{(2-c)^2+3x^2}}{\sqrt{3}\pi c |x|} \sinh\left[\frac{\log\left[ \eta(x)+\sqrt{\eta^2(x)-1}\right]}{3} \right],& |x| \in  (x_-,x_+)
    \\
    \max\left[0,1-\frac{2}{c} \right]\delta(x),& \text{otherwise}
    \end{cases},
\end{align}
where $\mathcal{D}_{-1}$ represents the rescaling, the function $\eta(x)$ is given by
\begin{align}
    \eta(x):= \frac{9(c+1)^2x^2+(2-c)^2}{((2-c)^2+3x^2)^{3/2}},
\end{align}
and the endpoints of the spectrum are
\begin{align}
    x_\pm = \max\left[0, \frac{1}{4}\left( \sqrt{4c+1}\pm3\right)^{3/2}\left( \sqrt{4c+1}\mp 1\right)^{1/2} \right].
\end{align}
One can then evaluate the trace distance from the integral
\begin{align}
    T(\rho_A ||\sigma_A) = \frac{1}{2}\int \mu_{MP}^{(c)}\boxplus\mathcal{D}_{-1}\left[\mu_{MP}^{(c)}\right](x) |x|,
\end{align}
leading to \eqref{tracedistance_exact}.

\section{Commutation of ensemble average and logarithm}
\label{commute_app}

We have been using the replica trick throughout the paper to compute relative entropies. This has involved evaluating ensemble averages of traces of powers of density matrices and then taking a logarithm. In general, the ensemble average and logarithm do not commute. In this appendix, we show that the two operations approximately commute in the large-$N$ limit. To properly take the average of a logarithm, we need an additional replica trick
\begin{align}
    \overline{\log \left[\Tr\left[ \rho_A^{\alpha} \sigma_A^{m}\right]\right]} = \lim_{q\rightarrow 0}\frac{ \overline{\left(\Tr\left[ \rho_A^{\alpha} \sigma_A^{m}\right]\right)^q}-1}{q}.
    \label{replica_log}
\end{align}
For illustration, we work with the PRRE though the argument is the same for all other quantities. In diagrams, the necessary moments are 
\begin{align}
    \overline{\left(\Tr\left[ \rho_A^{\alpha} \sigma_A^{m}\right]\right)^q} =     \tikz[scale=0.35,baseline=-0.5ex]{
    \draw[dashed] (0,0) -- (1,0);
    \draw (0,0)-- (0,.15);
    \draw (1,0)-- (1,.15);
    \draw (-0.2,0.15)-- (-0.2,-0.35);
    \draw (1.2,-0.15)-- (1.2,0.15);
    \draw[dashed,black] (2,0) -- (3,0);
    \draw[black] (2,0)-- (2,0.15);
    \draw[black] (3,0)-- (3,0.15);
    \draw[black] (1.8,-.15)-- (1.8,0.15);
    \draw[black] (3.2,-0.15)-- (3.2,0.15);
    \draw[dashed,black] (5,0) -- (6,0);
    \draw[black] (5,0)-- (5,0.15);
    \draw[black] (6,0)-- (6,0.15);
    \draw[black] (4.8,0.15)-- (4.8,-0.15);
    \draw[black] (6.2,-0.15)-- (6.2,0.15);
    \draw[red,dashed] (7+0,0) -- (7+1,0);
    \draw[red] (7+0,0)-- (7+0,.15);
    \draw[red] (7+1,0)-- (7+1,.15);
    \draw[red] (7-0.2,0.15)-- (7-0.2,-0.15);
    \draw[red] (7+1.2,-0.15)-- (7+1.2,0.15);
    \draw[dashed,red] (7+2,0) -- (7+3,0);
    \draw[red] (7+2,0)-- (7+2,0.15);
    \draw[red] (7+3,0)-- (7+3,0.15);
    \draw[red] (7+1.8,-.15)-- (7+1.8,0.15);
    \draw[red] (7+3.2,-0.15)-- (7+3.2,0.15);
    \draw[dashed,red] (7+5,0) -- (7+6,0);
    \draw[red] (7+5,0)-- (7+5,0.15);
    \draw[red] (7+6,0)-- (7+6,0.15);
    \draw[red] (7+4.8,0.15)-- (7+4.8,-0.15);
    \draw[red] (7+6.2,-0.35)-- (7+6.2,0.15);
    \draw (1.2,-0.15)--(1.8,-0.15);
    \draw[black] (3.2,-0.15)--(3.45,-0.15);
    \draw (-0.2,-0.35)-- (6.2,-0.35);
    \draw[black] (4.5,-0.15)--(4.8,-0.15);
    \node[] at (4.,0.2) {$\cdots$};
    \draw[black] (7-2+1.2,-0.15)--(7-2+1.8,-0.15);
    \draw[black] (7-2+1.2,-0.35)--(7-2+1.8,-0.35);
    \draw[red] (7+1.2,-0.15)--(7+1.8,-0.15);
    \draw[red] (7+3.2,-0.15)--(7+3.45,-0.15);
    \draw[red] (7-0.2,-0.35)-- (7+6.2,-0.35);
    \draw[red] (7+4.5,-0.15)--(7+4.8,-0.15);
    \node[] at (7+4.,0.2) {$\cdots$};
    }
    \hspace{.25cm}
    \dots
    \hspace{.25cm}
    \tikz[scale=0.35,baseline=-0.5ex]{
    \draw[dashed] (0,0) -- (1,0);
    \draw (0,0)-- (0,.15);
    \draw (1,0)-- (1,.15);
    \draw (-0.2,0.15)-- (-0.2,-0.35);
    \draw (1.2,-0.15)-- (1.2,0.15);
    \draw[dashed,black] (2,0) -- (3,0);
    \draw[black] (2,0)-- (2,0.15);
    \draw[black] (3,0)-- (3,0.15);
    \draw[black] (1.8,-.15)-- (1.8,0.15);
    \draw[black] (3.2,-0.15)-- (3.2,0.15);
    \draw[dashed,black] (5,0) -- (6,0);
    \draw[black] (5,0)-- (5,0.15);
    \draw[black] (6,0)-- (6,0.15);
    \draw[black] (4.8,0.15)-- (4.8,-0.15);
    \draw[black] (6.2,-0.15)-- (6.2,0.15);
    \draw[red,dashed] (7+0,0) -- (7+1,0);
    \draw[red] (7+0,0)-- (7+0,.15);
    \draw[red] (7+1,0)-- (7+1,.15);
    \draw[red] (7-0.2,0.15)-- (7-0.2,-0.15);
    \draw[red] (7+1.2,-0.15)-- (7+1.2,0.15);
    \draw[dashed,red] (7+2,0) -- (7+3,0);
    \draw[red] (7+2,0)-- (7+2,0.15);
    \draw[red] (7+3,0)-- (7+3,0.15);
    \draw[red] (7+1.8,-.15)-- (7+1.8,0.15);
    \draw[red] (7+3.2,-0.15)-- (7+3.2,0.15);
    \draw[dashed,red] (7+5,0) -- (7+6,0);
    \draw[red] (7+5,0)-- (7+5,0.15);
    \draw[red] (7+6,0)-- (7+6,0.15);
    \draw[red] (7+4.8,0.15)-- (7+4.8,-0.15);
    \draw[red] (7+6.2,-0.35)-- (7+6.2,0.15);
    \draw (1.2,-0.15)--(1.8,-0.15);
    \draw[black] (3.2,-0.15)--(3.45,-0.15);
    \draw (-0.2,-0.35)-- (6.2,-0.35);
    \draw[black] (4.5,-0.15)--(4.8,-0.15);
    \node[] at (4.,0.2) {$\cdots$};
    \draw[black] (7-2+1.2,-0.15)--(7-2+1.8,-0.15);
    \draw[black] (7-2+1.2,-0.35)--(7-2+1.8,-0.35);
    \draw[red] (7+1.2,-0.15)--(7+1.8,-0.15);
    \draw[red] (7+3.2,-0.15)--(7+3.45,-0.15);
    \draw[red] (7-0.2,-0.35)-- (7+6.2,-0.35);
    \draw[red] (7+4.5,-0.15)--(7+4.8,-0.15);
    \node[] at (7+4.,0.2) {$\cdots$};
    },
\end{align}
where there are $q$ total blocks. As a sum over permutations, this is
\begin{align}
    \overline{\left(\Tr\left[ \rho_A^{\alpha} \sigma_A^{m}\right]\right)^q} = \frac{1}{(d_A d_B)^{q(\alpha+m)}}\sum_{\tau \in (S_{\alpha} \times S_{m})^{\times q}}d_A^{C((\eta^{-1})^{\times q}\circ \tau)}d_B^{C(\tau)},
\end{align}
where in cycle notation
\begin{align}
    (\eta^{-1})^{\times q} = \prod_{i = 0}^{q-1}(1+q,2+q,\dots, \alpha +m+q).
\end{align}
The leading terms come from noncrossing contractions within each block independently, leading to $C((\eta^{-1})^{\times q}\circ \tau)+ C(\tau) = q(\alpha + m + 1)$. Contractions that connect the blocks will be subleading with $C((\eta^{-1})^{\times q}\circ \tau)+ C(\tau) \leq q(\alpha + m + 1)-2$. Therefore, the ensemble average factorizes at leading order
\begin{align}
    \overline{\left(\Tr\left[ \rho_A^{\alpha} \sigma_A^{m}\right]\right)^q}  \simeq \overline{\left(\Tr\left[ \rho_A^{\alpha} \sigma_A^{m}\right]\right)}^q .
\end{align}
Using \eqref{replica_log}, this implies that the ensemble average and logarithm approximately commute at large-$N$.

\section{Equivalence with Haar unitary tensor networks}
\label{equiv_haar_wick_app}

Frequently, random tensor networks are constructed by projected Haar unitary states. This is in fact equivalent to the Gaussian random networks we use.
The reason is the following. In the Haar random construction, every vertex of degree $k$ is a state of $k$ qudits projected to a random state $U|0\rangle $ where $U$ is a Haar random unitary and $|0\rangle $ is any state. This gives the state $\langle 0|^{\otimes k} U^{\dagger}  |i_1\rangle \cdots |i_k\rangle $. Denoting the set of $k$ indices by one index $i$, this exactly corresponds to the Gaussian tensor network with the identification $U_{i,0} \leftrightarrow X^*_i$.
Every edge corresponds to a maximally entangled pair in the projected Haar random network, which is just the index contraction in the Gaussian network. The projected unitaries indeed have a Gaussian distribution, since (see e.g.~Ref.~\cite{2002math.ph...5010C})
\begin{equation}
\begin{split}
    & \langle X^*_{i_1} \cdots X^*_{i_n} X_{j_1} \cdots X_{j_n}\rangle =\langle U_{i_1,0} \cdots U_{i_n,0} U^{\dagger} _{0,j_1} \cdots U ^{\dagger} _{0,j_n}\rangle =\\
    &= \sum _{\sigma ,\tau \in S_n} \delta _{i_1,j_{\sigma(1)}} \cdots \delta _{i_n,j_{\sigma (n)}} \text{Wg}(n,\tau \circ \sigma ^{-1}) \propto \sum _{\sigma } \delta _{i_1,j_{\sigma(1)}} \cdots \delta _{i_n,j_{\sigma (n)}} 
\end{split}
\end{equation}
and zero for a different number of $X$ and $X^*$'s, just as for Gaussian variables.

\section{Interpolating between QSD and QHT}
\label{interp_app}

\begin{figure}
    \centering
    \includegraphics[width = .6\textwidth]{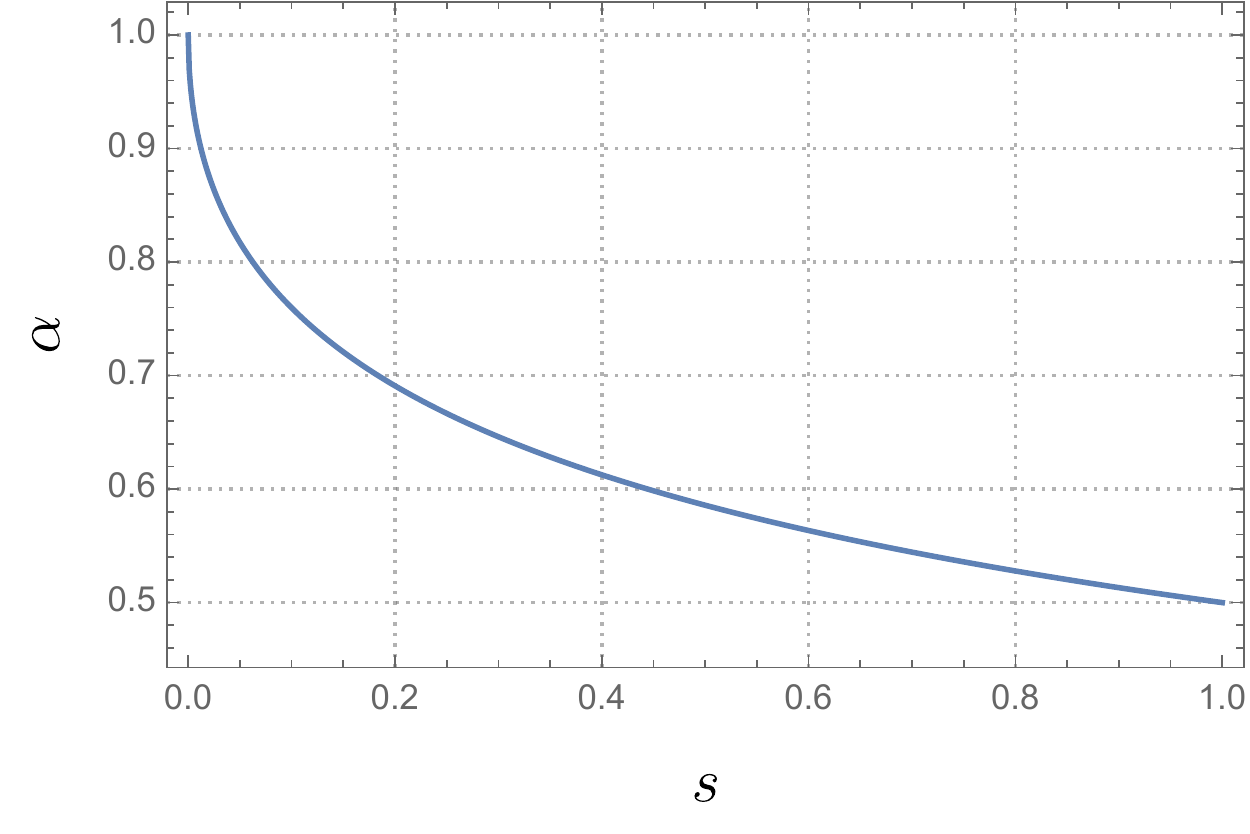}
    \caption{The interpolation between QHT ($s=0$) and QSD ($s=1$) is shown for $d_A < d_B$.}
    \label{interp_fig}
\end{figure}

In the main text, we characterized the asymptotic error rates in distinguishing states in the totally symmetric (QSD) and totally asymmetric (QHT) cases with the quantum Chernoff distance and relative entropy respectively. It is natural to ask if the there is a way to interpolate between the two. This was addressed in Ref.~\cite{2021arXiv210409553S} where the type II error $\beta(A)$ was optimized given the constraint that $\alpha(A) \leq \varepsilon^{1-s} \beta(A)^s$ for $s \geq 0$ and $\varepsilon > 0$. Note that this coincides with QHT for $s = 0$ and QSD for $s=1$. This is referred to as $s$-hypothesis testing and the error rate was proven to be given by the $s$-quantum divergence defined as
\begin{align}
    \xi_s(\rho||\sigma) := \max_{0\leq \alpha \leq 1} \frac{\log \left[ \Tr \left[ \rho^{\alpha}\sigma^{1-\alpha}\right]\right]}{\alpha(1-s)-1}.
\end{align}
It is instructive to examine the two familiar limits. When $s = 0$, the quantity being maximized is the PRRE. We know that the PRRE is monotonically increasing with $\alpha$ so $\xi_0(\rho||\sigma)$ is given by the relative entropy in accordance with our expectation. When $s = 1$, the RHS becomes the definition of the quantum Chernoff distance. Using \eqref{PRRE_eq}, we can evaluate the $s$-quantum divergence for random states. We plot the value of $\alpha$ that maximizes the RHS as a function of $s$ in Fig.~\ref{interp_fig}. This function monotonically decreases from one at $s = 0$ to zero at $s = \infty$, passing through $\alpha = 1/2$ at $s = 1$.

\section{Entanglement plateau}

Ref.~\cite{2021PhRvL.126q1603K} was not the first attempt in the literature to use related information theoretic quantities to try to distinguish black hole microstates. A selection of previous works include Refs.~\cite{2017PhRvD..96f6017B,2016arXiv161000302L,2017JHEP...02..060S,2018JHEP...07..179M,2018PhRvL.121y1603G,2020arXiv201012565A}. In particular, Ref.~\cite{2017PhRvD..96f6017B} used the Holevo information to distinguish black hole microstates. The Holevo information of $A$ for an ensemble of density matrices $\{ \rho_i \}$ is given by the average relative entropy from each microstate to the ensemble average
\begin{align}
    \chi(A) := \sum_i p_i D\left(\rho_{A,i} ||\sum_j p_j \rho_{A,j}\right) = S_{vN}\left(\sum_i p_i\rho_{A,i}\right) - \sum_i p_i S_{vN}(\rho_{A,i}).
\end{align}
This is bounded above by the Shannon entropy, $-\sum_i p_i \log p_i$, which happens to be the black hole entropy, $S_{BH}$, in this case.

\begin{figure}
    \centering
    \includegraphics[width = .95\textwidth]{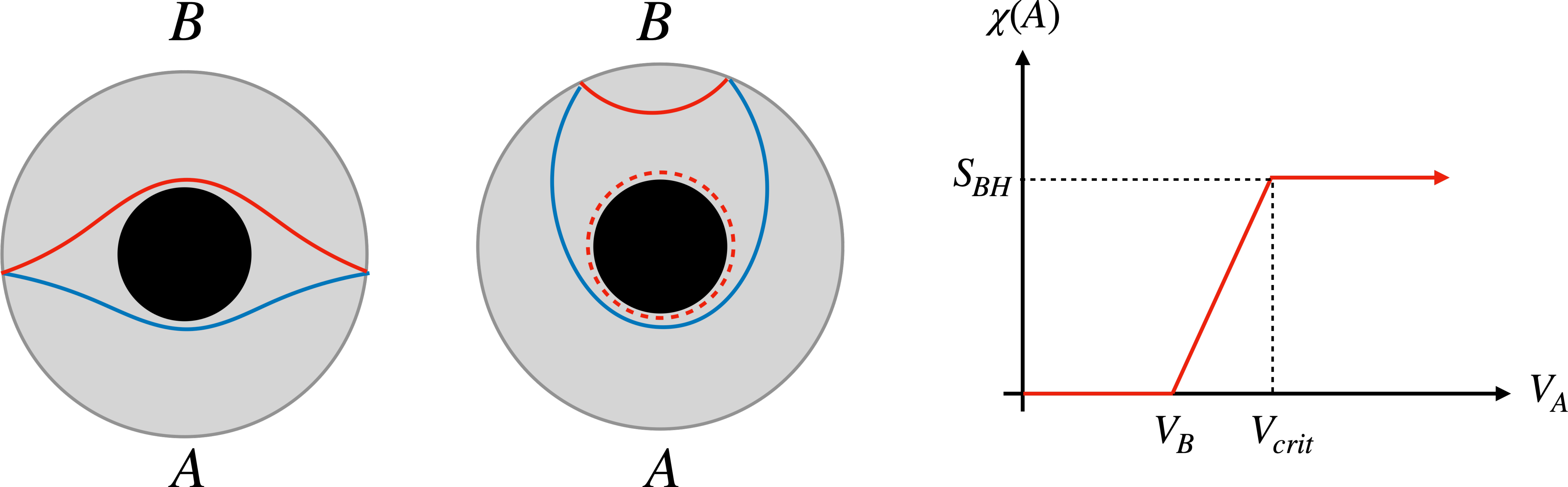}
    \caption{Left: Two competing extremal surfaces are shown for region $A$. In this case, there is no bulk entropy term because the black hole is in a pure state. Center: When the volume of $A$, $V_A$, becomes sufficiently large ($>V_{crit}$), the area of the red curve plus the black hole entropy is smaller than the area of the blue curve. The black hole entropy can either be viewed as a bulk entropy term or an additional area term. Right: The Holevo information up to nonperturbative corrections as a function of the volume of $A$.}
    \label{holevo_fig}
\end{figure}

The authors computed each piece holographically to leading order in Newton's constant using the Ryu-Takayanagi formula \cite{2006PhRvL..96r1602R,2006JHEP...08..045R} and determined that there are three phases of $\chi(A)$. The microstate entropies were computed from the two extremal surfaces of the previous section, so the transition occurred when $A$ was at half the total system size. In contrast, the ensemble averaged state (Gibbs state) entropy does not transition to the second extremal surface at the same time. While this was not the perspective taken in the original paper, we attribute this to the black hole contributing to the bulk entropy in the FLM formula \cite{2013JHEP...11..074F}. This transition occurs when $A$ is much larger than its complement and discussed thoroughly in Ref.~\cite{2013JHEP...08..092H} where it is called the ``entanglement plateaux.'' In summary, they found the Holevo information to be identically zero until the halfway point, then linearly increase until a critical size when it saturates to its maximal value of the black hole entropy. The relevant extremal surfaces and behavior of the Holevo information are shown in Fig.~\ref{holevo_fig}.

Using fixed-area states, we can improve upon this analysis by computing nonperturbative corrections. As already reviewed in the previous section, all pure black hole microstates will have entropy
\begin{align}
    S_{vN}(\rho_{A,i}) = \frac{A_{1,2}}{4G_N}  -\frac{e^{(A_{1,2} - A_{2,1})/4G_N}}{2},
\end{align}
where the first (second) subscript occurs when $A_1$ ($A_2$) is minimal. 

For the entropy in the Gibbs state, we must account for the bulk entropy
\begin{align}
    \Tr\left[ \left(\sum_i p_i\rho_{A,i} \right)^{n}\right]= \sum_{\tau \in  S_{n}}\frac{e^{\left(C(\eta^{-1} \circ \tau) A_1 + C(\tau) A_2\right)/4G_N}}{e^{n( A_1+A_2)/4G_N }}\Tr\left[\rho_b^{n_1}\right]\dots\Tr\left[\rho_b^{n_{C(\tau)}}\right],
\end{align}
where $n_i$ are the lengths of the cycles of $\tau$. The bulk entropy terms manifestly go away when the black hole is in a pure state. For simplicity, let us take $\rho_b$ to be the maximally mixed state with dimension $e^{S_{BH}}$. Finite temperature corrections are not immediately important to demonstrate our main conclusions but may be interesting. In this case, the sum simplifies to 
\begin{align}
    \Tr\left[ \left(\sum_i p_i\rho_{A,i} \right)^{n}\right]= \sum_{\tau \in  S_{n}}\frac{e^{\left(C(\eta^{-1} \circ \tau) A_1/4G_N + C(\tau) (A_2/4G_N+S_{BH})\right)}}{e^{n(( A_1+A_2)/4G_N + S_{BH})}}.
\end{align}
This is just a renormalization of $A_2/4G_N$ to  $A_2/4G_N + S_{BH}$. This makes sense because the alternative perspective would be to fix the area of the new extremal surface that includes the black hole horizon. Computing the sum as before, we find
\begin{align}
    S_{vN} \left(\sum_i p_i\rho_{A,i} \right) = \begin{cases}
    \frac{A_{1}}{4G_N}  -\frac{e^{(A_{1} - A_{2})/4G_N-S_{BH}}}{2}, & \frac{A_{1}}{4G_N} < \frac{A_{2}}{4G_N} + S_{BH}
    \\
    \frac{A_{2}}{4G_N} + S_{BH} -\frac{e^{(A_{2} - A_{1})/4G_N+S_{BH}}}{2}, & \frac{A_{1}}{4G_N} > \frac{A_{2}}{4G_N} + S_{BH}
    \end{cases}
\end{align}
In total, the Holevo information is
\begin{align}
    \chi(A) = \begin{cases}
    \frac{e^{(A_{1} - A_{2})/4G_N}(1-e^{-S_{BH}})}{2} , & A_1 < A_2
    \\
    \frac{A_1 - A_2}{4G_N} + \frac{e^{(A_{1} - A_{2})/4G_N-S_{BH}}-e^{(A_{2} - A_{1})/4G_N}}{2}, & A_2 < A_1 < A_2 + 4G_NS_{BH}
    \\
    S_{BH}+\frac{e^{(A_{2} - A_{1})/4G_N}(1+e^{S_{BH}})}{2} ,& A_1 > A_2 + 4G_NS_{BH}
    \end{cases}.
\end{align}
We see that this is the same answer as in Ref.~\cite{2017PhRvD..96f6017B} up to nonperturbative corrections\footnote{This result seems to be in tension with Ref.~\cite{2018PhRvL.121y1603G} where it was argued that there are $O(G_N^0)$ corrections to the Holevo information in AdS$_3$/CFT$_2$. It is not clear to us how to resolve this tension due to the very different flavors of assumptions that went into both calculations. This may be worth examining further.}. The most important improvements are the first regime where we show the Holevo information to be nonzero $\Rightarrow$ microstates are distinguishable and the last regime where we show the Holevo information to be submaximal $\Rightarrow$ microstates are not perfectly distinguishable.


\begin{thebibliography}{100}

\bibitem{Bekenstein:1973ur}
J.~D. Bekenstein, {\it {Black holes and entropy}},  {\em Phys. Rev. D} {\bf 7}
  (1973) 2333--2346.

\bibitem{cmp/1103899181}
S.~W. Hawking, {\it {Particle creation by black holes}},  {\em Communications
  in Mathematical Physics} {\bf 43} (1975), no.~3 199 -- 220.

\bibitem{2021PhRvL.126q1603K}
J.~{Kudler-Flam}, {\it {Relative Entropy of Random States and Black Holes}},
  {\em \prl} {\bf 126} (Apr., 2021) 171603,
  [\href{http://arxiv.org/abs/2102.05053}{{\tt arXiv:2102.05053}}].

\bibitem{10.2307/1970079}
E.~P. Wigner, {\it Characteristic vectors of bordered matrices with infinite
  dimensions},  {\em Annals of Mathematics} {\bf 62} (1955), no.~3 548--564.

\bibitem{2016JMP....57a5215C}
B.~{Collins} and I.~{Nechita}, {\it {Random matrix techniques in quantum
  information theory}},  {\em Journal of Mathematical Physics} {\bf 57} (Jan.,
  2016) 015215, [\href{http://arxiv.org/abs/1509.04689}{{\tt
  arXiv:1509.04689}}].

\bibitem{2016AdPhy..65..239D}
L.~{D'Alessio}, Y.~{Kafri}, A.~{Polkovnikov}, and M.~{Rigol}, {\it {From
  quantum chaos and eigenstate thermalization to statistical mechanics and
  thermodynamics}},  {\em Advances in Physics} {\bf 65} (May, 2016) 239--362,
  [\href{http://arxiv.org/abs/1509.06411}{{\tt arXiv:1509.06411}}].

\bibitem{1993PhRvL..71.3743P}
D.~N. {Page}, {\it {Information in black hole radiation}},  {\em \prl} {\bf 71}
  (Dec., 1993) 3743--3746, [\href{http://arxiv.org/abs/hep-th/9306083}{{\tt
  hep-th/9306083}}].

\bibitem{2007JHEP...09..120H}
P.~{Hayden} and J.~{Preskill}, {\it {Black holes as mirrors: quantum
  information in random subsystems}},  {\em Journal of High Energy Physics}
  {\bf 2007} (Sept., 2007) 120, [\href{http://arxiv.org/abs/0708.4025}{{\tt
  arXiv:0708.4025}}].

\bibitem{2019JHEP...10..240D}
X.~{Dong}, D.~{Harlow}, and D.~{Marolf}, {\it {Flat entanglement spectra in
  fixed-area states of quantum gravity}},  {\em Journal of High Energy Physics}
  {\bf 2019} (Oct., 2019) 240, [\href{http://arxiv.org/abs/1811.05382}{{\tt
  arXiv:1811.05382}}].

\bibitem{2019JHEP...05..052A}
C.~{Akers} and P.~{Rath}, {\it {Holographic Renyi entropy from quantum error
  correction}},  {\em Journal of High Energy Physics} {\bf 2019} (May, 2019)
  52, [\href{http://arxiv.org/abs/1811.05171}{{\tt arXiv:1811.05171}}].

\bibitem{2019arXiv191111977P}
G.~{Penington}, S.~H. {Shenker}, D.~{Stanford}, and Z.~{Yang}, {\it {Replica
  wormholes and the black hole interior}},  {\em arXiv e-prints} (Nov., 2019)
  arXiv:1911.11977, [\href{http://arxiv.org/abs/1911.11977}{{\tt
  arXiv:1911.11977}}].

\bibitem{2010NJPh...12g5021D}
J.~M. {Deutsch}, {\it {Thermodynamic entropy of a many-body energy
  eigenstate}},  {\em New Journal of Physics} {\bf 12} (July, 2010) 075021,
  [\href{http://arxiv.org/abs/0911.0056}{{\tt arXiv:0911.0056}}].

\bibitem{2019PhRvE.100b2131M}
C.~{Murthy} and M.~{Srednicki}, {\it {Structure of chaotic eigenstates and
  their entanglement entropy}},  {\em \pre} {\bf 100} (Aug., 2019) 022131,
  [\href{http://arxiv.org/abs/1906.04295}{{\tt arXiv:1906.04295}}].

\bibitem{2017arXiv170908784L}
T.-C. {Lu} and T.~{Grover}, {\it {Renyi Entropy of Chaotic Eigenstates}},  {\em
  arXiv e-prints} (Sept., 2017) arXiv:1709.08784,
  [\href{http://arxiv.org/abs/1709.08784}{{\tt arXiv:1709.08784}}].

\bibitem{2018PhRvE..97a2140D}
A.~{Dymarsky}, N.~{Lashkari}, and H.~{Liu}, {\it {Subsystem eigenstate
  thermalization hypothesis}},  {\em \pre} {\bf 97} (Jan., 2018) 012140,
  [\href{http://arxiv.org/abs/1611.08764}{{\tt arXiv:1611.08764}}].

\bibitem{lindblad1975}
G.~Lindblad, {\it Completely positive maps and entropy inequalities},  {\em
  Comm. Math. Phys.} {\bf 40} (1975), no.~2 147--151.

\bibitem{10020820209}
A.~RENYI, {\it On measures of entropy and information},  {\em Proc. of the
  Fourth Berkeley Symp. on Math. Statist. and Prob.} {\bf 1} (1961) 547--561.

\bibitem{1986RpMP...23...57P}
D.~{Petz}, {\it {Quasi-entropies for finite quantum systems}},  {\em Reports on
  Mathematical Physics} {\bf 23} (Feb., 1986) 57--65.

\bibitem{LIEB1973267}
E.~H. Lieb, {\it Convex trace functions and the wigner-yanase-dyson
  conjecture},  {\em Advances in Mathematics} {\bf 11} (1973), no.~3 267--288.

\bibitem{cmp/1103900757}
A.~Uhlmann, {\it {Relative entropy and the Wigner-Yanase-Dyson-Lieb concavity
  in an interpolation theory}},  {\em Communications in Mathematical Physics}
  {\bf 54} (1977), no.~1 21 -- 32.

\bibitem{2018arXiv180102800W}
M.~M. {Wilde}, {\it {Recoverability for Holevo's just-as-good fidelity}},  {\em
  arXiv e-prints} (Jan., 2018) arXiv:1801.02800,
  [\href{http://arxiv.org/abs/1801.02800}{{\tt arXiv:1801.02800}}].

\bibitem{1972TMP....13.1071K}
A.~S. {Kholevo}, {\it {On quasiequivalence of locally normal states}},  {\em
  Theoretical and Mathematical Physics} {\bf 13} (Nov., 1972) 1071--1082.

\bibitem{doi:10.1080/09500349414552171}
R.~Jozsa, {\it Fidelity for mixed quantum states},  {\em Journal of Modern
  Optics} {\bf 41} (1994), no.~12 2315--2323,
  [\href{http://arxiv.org/abs/https://doi.org/10.1080/09500349414552171}{{\tt
  https://doi.org/10.1080/09500349414552171}}].

\bibitem{2015JMP....56b2202A}
K.~M.~R. {Audenaert} and N.~{Datta}, {\it {{\ensuremath{\alpha}}-z-R{\'e}nyi
  relative entropies}},  {\em Journal of Mathematical Physics} {\bf 56} (Feb.,
  2015) 022202, [\href{http://arxiv.org/abs/1310.7178}{{\tt arXiv:1310.7178}}].

\bibitem{2013JMP....54l2203M}
M.~{M{\"u}ller-Lennert}, F.~{Dupuis}, O.~{Szehr}, S.~{Fehr}, and
  M.~{Tomamichel}, {\it {On quantum R{\'e}nyi entropies: A new generalization
  and some properties}},  {\em Journal of Mathematical Physics} {\bf 54} (Dec.,
  2013) 122203--122203, [\href{http://arxiv.org/abs/1306.3142}{{\tt
  arXiv:1306.3142}}].

\bibitem{2014CMaPh.331..593W}
M.~M. {Wilde}, A.~{Winter}, and D.~{Yang}, {\it {Strong Converse for the
  Classical Capacity of Entanglement-Breaking and Hadamard Channels via a
  Sandwiched R{\'e}nyi Relative Entropy}},  {\em Communications in Mathematical
  Physics} {\bf 331} (Oct., 2014) 593--622,
  [\href{http://arxiv.org/abs/1306.1586}{{\tt arXiv:1306.1586}}].

\bibitem{ohya2004quantum}
M.~Ohya and D.~Petz, {\em Quantum Entropy and Its Use}.
\newblock Theoretical and Mathematical Physics. Springer Berlin Heidelberg,
  2004.

\bibitem{1997quant.ph.12042F}
C.~A. {Fuchs} and J.~{van de Graaf}, {\it {Cryptographic Distinguishability
  Measures for Quantum Mechanical States}},  {\em arXiv e-prints} (Dec., 1997)
  quant--ph/9712042, [\href{http://arxiv.org/abs/quant-ph/9712042}{{\tt
  quant-ph/9712042}}].

\bibitem{Hayashi:1338967}
M.~Hayashi, {\em {Quantum Information: An Introduction}}.
\newblock Springer, Berlin, Heidelberg, 2006.

\bibitem{2020arXiv201104672K}
S.~{Khatri} and M.~M. {Wilde}, {\it {Principles of Quantum Communication
  Theory: A Modern Approach}},  {\em arXiv e-prints} (Nov., 2020)
  arXiv:2011.04672, [\href{http://arxiv.org/abs/2011.04672}{{\tt
  arXiv:2011.04672}}].

\bibitem{2021arXiv210409553S}
R.~{Salzmann} and N.~{Datta}, {\it {Interpolating between symmetric and
  asymmetric hypothesis testing}},  {\em arXiv e-prints} (Apr., 2021)
  arXiv:2104.09553, [\href{http://arxiv.org/abs/2104.09553}{{\tt
  arXiv:2104.09553}}].

\bibitem{1969JSP.....1..231H}
C.~W. {Helstrom}, {\it {Quantum detection and estimation theory}},  {\em
  Journal of Statistical Physics} {\bf 1} (June, 1969) 231--252.

\bibitem{Helstrom:110988}
C.~W. Helstrom, {\em {Quantum detection and estimation theory}}.
\newblock Mathematics in science and engineering. Academic Press, New York, NY,
  1976.

\bibitem{2007PhRvL..98p0501A}
K.~M.~R. {Audenaert}, J.~{Calsamiglia}, R.~{Mu{\~n}oz-Tapia}, E.~{Bagan},
  L.~{Masanes}, A.~{Acin}, and F.~{Verstraete}, {\it {Discriminating States:
  The Quantum Chernoff Bound}},  {\em \prl} {\bf 98} (Apr., 2007) 160501,
  [\href{http://arxiv.org/abs/quant-ph/0610027}{{\tt quant-ph/0610027}}].

\bibitem{2006quant.ph..7216N}
M.~{Nussbaum} and A.~{Szko{\l}a}, {\it {The Chernoff lower bound for symmetric
  quantum hypothesis testing}},  {\em arXiv e-prints} (July, 2006)
  quant--ph/0607216, [\href{http://arxiv.org/abs/quant-ph/0607216}{{\tt
  quant-ph/0607216}}].

\bibitem{cmp/1104248844}
F.~Hiai and D.~Petz, {\it {The proper formula for relative entropy and its
  asymptotics in quantum probability}},  {\em Communications in Mathematical
  Physics} {\bf 143} (1991), no.~1 99 -- 114.

\bibitem{2005atqs.book...28O}
T.~{Ogawa} and H.~{Nagaoka}, {\em {Strong Converse and Stein's Lemma in Quantum
  Hypothesis Testing}}, pp.~28--42.
\newblock 2005.

\bibitem{2007PhRvA..76f2301H}
M.~{Hayashi}, {\it {Error exponent in asymmetric quantum hypothesis testing and
  its application to classical-quantum channel coding}},  {\em \pra} {\bf 76}
  (Dec., 2007) 062301, [\href{http://arxiv.org/abs/quant-ph/0611013}{{\tt
  quant-ph/0611013}}].

\bibitem{2006quant.ph.11289N}
H.~{Nagaoka}, {\it {The Converse Part of The Theorem for Quantum Hoeffding
  Bound}},  {\em arXiv e-prints} (Nov., 2006) quant--ph/0611289,
  [\href{http://arxiv.org/abs/quant-ph/0611289}{{\tt quant-ph/0611289}}].

\bibitem{2009arXiv0912.1286M}
M.~{Mosonyi} and F.~{Hiai}, {\it {On the quantum Renyi relative entropies and
  related capacity formulas}},  {\em arXiv e-prints} (Dec., 2009)
  arXiv:0912.1286, [\href{http://arxiv.org/abs/0912.1286}{{\tt
  arXiv:0912.1286}}].

\bibitem{2015CMaPh.334.1617M}
M.~{Mosonyi} and T.~{Ogawa}, {\it {Quantum Hypothesis Testing and the
  Operational Interpretation of the Quantum R{\'e}nyi Relative Entropies}},
  {\em Communications in Mathematical Physics} {\bf 334} (Mar., 2015)
  1617--1648, [\href{http://arxiv.org/abs/1309.3228}{{\tt arXiv:1309.3228}}].

\bibitem{2010JMP....51g2203N}
M.~{Nussbaum} and A.~{Szko{\l}a}, {\it {Exponential error rates in multiple
  state discrimination on a quantum spin chain}},  {\em Journal of Mathematical
  Physics} {\bf 51} (July, 2010) 072203--072203,
  [\href{http://arxiv.org/abs/1001.2651}{{\tt arXiv:1001.2651}}].

\bibitem{10.1007/978-3-642-18073-6_1}
M.~Nussbaum and A.~Szko{\l}a, {\it Asymptotically optimal discrimination
  between pure quantum states},  in {\em Theory of Quantum Computation,
  Communication, and Cryptography} (W.~van Dam, V.~M. Kendon, and S.~Severini,
  eds.), (Berlin, Heidelberg), pp.~1--8, Springer Berlin Heidelberg, 2011.

\bibitem{2011arXiv1112.1529N}
M.~{Nussbaum} and A.~{Szko{\l}a}, {\it {An asymptotic error bound for testing
  multiple quantum hypotheses}},  {\em arXiv e-prints} (Dec., 2011)
  arXiv:1112.1529, [\href{http://arxiv.org/abs/1112.1529}{{\tt
  arXiv:1112.1529}}].

\bibitem{2014JMP....55j2201A}
K.~M.~R. {Audenaert} and M.~{Mosonyi}, {\it {Upper bounds on the error
  probabilities and asymptotic error exponents in quantum multiple state
  discrimination}},  {\em Journal of Mathematical Physics} {\bf 55} (Oct.,
  2014) 102201, [\href{http://arxiv.org/abs/1401.7658}{{\tt arXiv:1401.7658}}].

\bibitem{2002JPhA...3510759H}
M.~{Hayashi}, {\it {Optimal sequence of quantum measurements in the sense of
  Stein's lemma in quantum hypothesis testing}},  {\em Journal of Physics A
  Mathematical General} {\bf 35} (Dec., 2002) 10759--10773,
  [\href{http://arxiv.org/abs/quant-ph/0208020}{{\tt quant-ph/0208020}}].

\bibitem{2005CMaPh.260..659B}
I.~{Bjelakovi{\'c}}, J.-D. {Deuschel}, T.~{Kr{\"u}ger}, R.~{Seiler},
  R.~{Siegmund-Schultze}, and A.~{Szko{\l}a}, {\it {A Quantum Version of
  Sanov's Theorem}},  {\em Communications in Mathematical Physics} {\bf 260}
  (Dec., 2005) 659--671, [\href{http://arxiv.org/abs/quant-ph/0412157}{{\tt
  quant-ph/0412157}}].

\bibitem{2021arXiv210309893F}
K.~{Furuya}, N.~{Lashkari}, and S.~{Ouseph}, {\it {Monotonic multi-state
  quantum f-divergences}},  {\em arXiv e-prints} (Mar., 2021) arXiv:2103.09893,
  [\href{http://arxiv.org/abs/2103.09893}{{\tt arXiv:2103.09893}}].

\bibitem{2015arXiv150806624L}
K.~{Li}, {\it {Discriminating quantum states: the multiple Chernoff distance}},
   {\em arXiv e-prints} (Aug., 2015) arXiv:1508.06624,
  [\href{http://arxiv.org/abs/1508.06624}{{\tt arXiv:1508.06624}}].

\bibitem{1055351}
H.~{Yuen}, R.~{Kennedy}, and M.~{Lax}, {\it Optimum testing of multiple
  hypotheses in quantum detection theory},  {\em IEEE Transactions on
  Information Theory} {\bf 21} (1975), no.~2 125--134.

\bibitem{2001JPhA...34.7111Z}
K.~{Zyczkowski} and H.-J. {Sommers}, {\it {Induced measures in the space of
  mixed quantum states}},  {\em Journal of Physics A Mathematical General} {\bf
  34} (Sept., 2001) 7111--7125,
  [\href{http://arxiv.org/abs/quant-ph/0012101}{{\tt quant-ph/0012101}}].

\bibitem{2004JPhA...37.8457S}
H.-J. {Sommers} and K.~{Zyczkowski}, {\it {Statistical properties of random
  density matrices}},  {\em Journal of Physics A Mathematical General} {\bf 37}
  (Sept., 2004) 8457--8466, [\href{http://arxiv.org/abs/quant-ph/0405031}{{\tt
  quant-ph/0405031}}].

\bibitem{2011JMP....52f2201Z}
K.~{{\.Z}yczkowski}, K.~A. {Penson}, I.~{Nechita}, and B.~{Collins}, {\it
  {Generating random density matrices}},  {\em Journal of Mathematical Physics}
  {\bf 52} (June, 2011) 062201--062201,
  [\href{http://arxiv.org/abs/1010.3570}{{\tt arXiv:1010.3570}}].

\bibitem{2007AnHP....8.1521N}
I.~{Nechita}, {\it {Asymptotics of Random Density Matrices}},  {\em Annales
  Henri Poincar\&eacute;} {\bf 8} (Nov., 2007) 1521--1538,
  [\href{http://arxiv.org/abs/quant-ph/0702154}{{\tt quant-ph/0702154}}].

\bibitem{2009arXiv0910.1768C}
B.~{Collins} and I.~{Nechita}, {\it {Gaussianization and eigenvalue statistics
  for random quantum channels (III)}},  {\em arXiv e-prints} (Oct., 2009)
  arXiv:0910.1768, [\href{http://arxiv.org/abs/0910.1768}{{\tt
  arXiv:0910.1768}}].

\bibitem{2010JPhA...43A5303C}
B.~{Collins}, I.~{Nechita}, and K.~{{\.Z}yczkowski}, {\it {Random graph states,
  maximal flow and Fuss-Catalan distributions}},  {\em Journal of Physics A
  Mathematical General} {\bf 43} (July, 2010) 275303,
  [\href{http://arxiv.org/abs/1003.3075}{{\tt arXiv:1003.3075}}].

\bibitem{2016JHEP...11..009H}
P.~{Hayden}, S.~{Nezami}, X.-L. {Qi}, N.~{Thomas}, M.~{Walter}, and Z.~{Yang},
  {\it {Holographic duality from random tensor networks}},  {\em Journal of
  High Energy Physics} {\bf 2016} (Nov., 2016) 9,
  [\href{http://arxiv.org/abs/1601.01694}{{\tt arXiv:1601.01694}}].

\bibitem{1995NuPhB.453..531B}
E.~{Br{\'e}zin} and A.~{Zee}, {\it {Universal relation between Green functions
  in random matrix theory}},  {\em Nuclear Physics B} {\bf 453} (Feb., 1995)
  531--551, [\href{http://arxiv.org/abs/cond-mat/9507032}{{\tt
  cond-mat/9507032}}].

\bibitem{2008AcPPB..39..799J}
J.~{Jurkiewicz}, G.~{{\L}ukaszewski}, and M.~A. {Nowak}, {\it {Diagrammatic
  Approach to Fluctuations in the Wishart Ensemble}},  {\em Acta Physica
  Polonica B} {\bf 39} (Apr., 2008) 799.

\bibitem{2020arXiv201101277S}
H.~{Shapourian}, S.~{Liu}, J.~{Kudler-Flam}, and A.~{Vishwanath}, {\it
  {Entanglement negativity spectrum of random mixed states: A diagrammatic
  approach}},  {\em arXiv e-prints} (Nov., 2020) arXiv:2011.01277,
  [\href{http://arxiv.org/abs/2011.01277}{{\tt arXiv:2011.01277}}].

\bibitem{2016PhRvL.117d1601L}
N.~{Lashkari}, {\it {Modular Hamiltonian for Excited States in Conformal Field
  Theory}},  {\em \prl} {\bf 117} (July, 2016) 041601,
  [\href{http://arxiv.org/abs/1508.03506}{{\tt arXiv:1508.03506}}].

\bibitem{1993PhRvL..71.1291P}
D.~N. {Page}, {\it {Average entropy of a subsystem}},  {\em \prl} {\bf 71}
  (Aug., 1993) 1291--1294, [\href{http://arxiv.org/abs/gr-qc/9305007}{{\tt
  gr-qc/9305007}}].

\bibitem{2020PhRvA.102a2405K}
S.~{Kumar}, {\it {Wishart and random density matrices: Analytical results for
  the mean-square Hilbert-Schmidt distance}},  {\em \pra} {\bf 102} (July,
  2020) 012405, [\href{http://arxiv.org/abs/2008.05153}{{\tt
  arXiv:2008.05153}}].

\bibitem{2020JPhA...53X5202K}
S.~{Kumar} and S.~{Sai Charan}, {\it {Spectral statistics for the difference of
  two Wishart matrices}},  {\em Journal of Physics A Mathematical General} {\bf
  53} (Nov., 2020) 505202, [\href{http://arxiv.org/abs/2011.07362}{{\tt
  arXiv:2011.07362}}].

\bibitem{2021arXiv210502743L}
A.~{Laha}, A.~{Aggarwal}, and S.~{Kumar}, {\it {Random density matrices:
  analytical results for mean root fidelity and mean square Bures distance}},
  {\em arXiv e-prints} (May, 2021) arXiv:2105.02743,
  [\href{http://arxiv.org/abs/2105.02743}{{\tt arXiv:2105.02743}}].

\bibitem{2016PhRvA..93f2112P}
Z.~{Pucha{\l}a}, {\L}.~{Pawela}, and K.~{{\.Z}yczkowski}, {\it
  {Distinguishability of generic quantum states}},  {\em \pra} {\bf 93} (June,
  2016) 062112, [\href{http://arxiv.org/abs/1507.05123}{{\tt
  arXiv:1507.05123}}].

\bibitem{KREWERAS1972333}
G.~Kreweras, {\it Sur les partitions non croisees d'un cycle},  {\em Discrete
  Mathematics} {\bf 1} (1972), no.~4 333 -- 350.

\bibitem{SIMION2000367}
R.~Simion, {\it Noncrossing partitions},  {\em Discrete Mathematics} {\bf 217}
  (2000), no.~1 367 -- 409.

\bibitem{2005PhRvA..71c2313Z}
K.~{{\.Z}yczkowski} and H.-J. {Sommers}, {\it {Average fidelity between random
  quantum states}},  {\em \pra} {\bf 71} (Mar., 2005) 032313,
  [\href{http://arxiv.org/abs/quant-ph/0311117}{{\tt quant-ph/0311117}}].

\bibitem{2005PhDT.......176R}
R.~{Renner}, {\em {Security of Quantum Key Distribution}}.
\newblock PhD thesis, -, Dec., 2005.

\bibitem{2019PhRvL.122n1602Z}
J.~{Zhang}, P.~{Ruggiero}, and P.~{Calabrese}, {\it {Subsystem Trace Distance
  in Quantum Field Theory}},  {\em \prl} {\bf 122} (Apr., 2019) 141602,
  [\href{http://arxiv.org/abs/1901.10993}{{\tt arXiv:1901.10993}}].

\bibitem{2015arXiv151107278M}
J.~{Mej{\'\i}a}, C.~{Zapata}, and A.~{Botero}, {\it {The difference between two
  random mixed quantum states: exact and asymptotic spectral analysis}},  {\em
  arXiv e-prints} (Nov., 2015) arXiv:1511.07278,
  [\href{http://arxiv.org/abs/1511.07278}{{\tt arXiv:1511.07278}}].

\bibitem{2018arXiv180100862P}
J.~{Preskill}, {\it {Quantum Computing in the NISQ era and beyond}},  {\em
  arXiv e-prints} (Jan., 2018) arXiv:1801.00862,
  [\href{http://arxiv.org/abs/1801.00862}{{\tt arXiv:1801.00862}}].

\bibitem{1994NuPhB.424..443H}
C.~{Holzhey}, F.~{Larsen}, and F.~{Wilczek}, {\it {Geometric and renormalized
  entropy in conformal field theory}},  {\em Nuclear Physics B} {\bf 424}
  (Aug., 1994) 443--467, [\href{http://arxiv.org/abs/hep-th/9403108}{{\tt
  hep-th/9403108}}].

\bibitem{2004JSMTE..06..002C}
P.~{Calabrese} and J.~{Cardy}, {\it {Entanglement entropy and quantum field
  theory}},  {\em Journal of Statistical Mechanics: Theory and Experiment} {\bf
  2004} (June, 2004) 06002, [\href{http://arxiv.org/abs/hep-th/0405152}{{\tt
  hep-th/0405152}}].

\bibitem{2013JHEP...08..090L}
A.~{Lewkowycz} and J.~{Maldacena}, {\it {Generalized gravitational entropy}},
  {\em Journal of High Energy Physics} {\bf 2013} (Aug., 2013) 90,
  [\href{http://arxiv.org/abs/1304.4926}{{\tt arXiv:1304.4926}}].

\bibitem{2016NatCo...712472D}
X.~{Dong}, {\it {The gravity dual of R{\'e}nyi entropy}},  {\em Nature
  Communications} {\bf 7} (Aug., 2016) 12472,
  [\href{http://arxiv.org/abs/1601.06788}{{\tt arXiv:1601.06788}}].

\bibitem{2013JHEP...11..074F}
T.~{Faulkner}, A.~{Lewkowycz}, and J.~{Maldacena}, {\it {Quantum corrections to
  holographic entanglement entropy}},  {\em Journal of High Energy Physics}
  {\bf 2013} (Nov., 2013) 74, [\href{http://arxiv.org/abs/1307.2892}{{\tt
  arXiv:1307.2892}}].

\bibitem{2015JHEP...04..163A}
A.~{Almheiri}, X.~{Dong}, and D.~{Harlow}, {\it {Bulk locality and quantum
  error correction in AdS/CFT}},  {\em Journal of High Energy Physics} {\bf
  2015} (Apr., 2015) 163, [\href{http://arxiv.org/abs/1411.7041}{{\tt
  arXiv:1411.7041}}].

\bibitem{2017CMaPh.354..865H}
D.~{Harlow}, {\it {The Ryu-Takayanagi Formula from Quantum Error Correction}},
  {\em Communications in Mathematical Physics} {\bf 354} (Sept., 2017)
  865--912, [\href{http://arxiv.org/abs/1607.03901}{{\tt arXiv:1607.03901}}].

\bibitem{2016JHEP...06..004J}
D.~L. {Jafferis}, A.~{Lewkowycz}, J.~{Maldacena}, and S.~J. {Suh}, {\it
  {Relative entropy equals bulk relative entropy}},  {\em Journal of High
  Energy Physics} {\bf 2016} (June, 2016) 4,
  [\href{http://arxiv.org/abs/1512.06431}{{\tt arXiv:1512.06431}}].

\bibitem{2021arXiv210700009H}
P.~{Hayden}, O.~{Parrikar}, and J.~{Sorce}, {\it {The Markov gap for geometric
  reflected entropy}},  {\em arXiv e-prints} (June, 2021) arXiv:2107.00009,
  [\href{http://arxiv.org/abs/2107.00009}{{\tt arXiv:2107.00009}}].

\bibitem{2020JHEP...09..002P}
G.~{Penington}, {\it {Entanglement wedge reconstruction and the information
  paradox}},  {\em Journal of High Energy Physics} {\bf 2020} (Sept., 2020) 2,
  [\href{http://arxiv.org/abs/1905.08255}{{\tt arXiv:1905.08255}}].

\bibitem{2019JHEP...12..063A}
A.~{Almheiri}, N.~{Engelhardt}, D.~{Marolf}, and H.~{Maxfield}, {\it {The
  entropy of bulk quantum fields and the entanglement wedge of an evaporating
  black hole}},  {\em Journal of High Energy Physics} {\bf 2019} (Dec., 2019)
  63, [\href{http://arxiv.org/abs/1905.08762}{{\tt arXiv:1905.08762}}].

\bibitem{2020JHEP...05..013A}
A.~{Almheiri}, T.~{Hartman}, J.~{Maldacena}, E.~{Shaghoulian}, and
  A.~{Tajdini}, {\it {Replica wormholes and the entropy of Hawking radiation}},
   {\em Journal of High Energy Physics} {\bf 2020} (May, 2020) 13,
  [\href{http://arxiv.org/abs/1911.12333}{{\tt arXiv:1911.12333}}].

\bibitem{2015JHEP...01..073E}
N.~{Engelhardt} and A.~C. {Wall}, {\it {Quantum extremal surfaces: holographic
  entanglement entropy beyond the classical regime}},  {\em Journal of High
  Energy Physics} {\bf 2015} (Jan., 2015) 73,
  [\href{http://arxiv.org/abs/1408.3203}{{\tt arXiv:1408.3203}}].

\bibitem{2020arXiv200606872A}
A.~{Almheiri}, T.~{Hartman}, J.~{Maldacena}, E.~{Shaghoulian}, and
  A.~{Tajdini}, {\it {The entropy of Hawking radiation}},  {\em arXiv e-prints}
  (June, 2020) arXiv:2006.06872, [\href{http://arxiv.org/abs/2006.06872}{{\tt
  arXiv:2006.06872}}].

\bibitem{2020arXiv201106005C}
Y.~{Chen} and H.~W. {Lin}, {\it {Signatures of global symmetry violation in
  relative entropies and replica wormholes}},  {\em arXiv e-prints} (Nov.,
  2020) arXiv:2011.06005, [\href{http://arxiv.org/abs/2011.06005}{{\tt
  arXiv:2011.06005}}].

\bibitem{2015JHEP...06..149P}
F.~{Pastawski}, B.~{Yoshida}, D.~{Harlow}, and J.~{Preskill}, {\it {Holographic
  quantum error-correcting codes: toy models for the bulk/boundary
  correspondence}},  {\em Journal of High Energy Physics} {\bf 2015} (June,
  2015) 149, [\href{http://arxiv.org/abs/1503.06237}{{\tt arXiv:1503.06237}}].

\bibitem{ford_fulkerson_1956}
L.~R. Ford and D.~R. Fulkerson, {\it Maximal flow through a network},  {\em
  Canadian Journal of Mathematics} {\bf 8} (1956) 399–404.

\bibitem{1056816}
P.~Elias, A.~Feinstein, and C.~Shannon, {\it A note on the maximum flow through
  a network},  {\em IRE Transactions on Information Theory} {\bf 2} (1956),
  no.~4 117--119.

\bibitem{2017PhRvX...7c1016N}
A.~{Nahum}, J.~{Ruhman}, S.~{Vijay}, and J.~{Haah}, {\it {Quantum Entanglement
  Growth under Random Unitary Dynamics}},  {\em Physical Review X} {\bf 7}
  (July, 2017) 031016, [\href{http://arxiv.org/abs/1608.06950}{{\tt
  arXiv:1608.06950}}].

\bibitem{2018PhRvX...8b1013V}
C.~W. {von Keyserlingk}, T.~{Rakovszky}, F.~{Pollmann}, and S.~L. {Sondhi},
  {\it {Operator Hydrodynamics, OTOCs, and Entanglement Growth in Systems
  without Conservation Laws}},  {\em Physical Review X} {\bf 8} (Apr., 2018)
  021013, [\href{http://arxiv.org/abs/1705.08910}{{\tt arXiv:1705.08910}}].

\bibitem{2018PhRvX...8b1014N}
A.~{Nahum}, S.~{Vijay}, and J.~{Haah}, {\it {Operator Spreading in Random
  Unitary Circuits}},  {\em Physical Review X} {\bf 8} (Apr., 2018) 021014,
  [\href{http://arxiv.org/abs/1705.08975}{{\tt arXiv:1705.08975}}].

\bibitem{2018arXiv180300089J}
C.~{Jonay}, D.~A. {Huse}, and A.~{Nahum}, {\it {Coarse-grained dynamics of
  operator and state entanglement}},  {\em arXiv e-prints} (Feb., 2018)
  arXiv:1803.00089, [\href{http://arxiv.org/abs/1803.00089}{{\tt
  arXiv:1803.00089}}].

\bibitem{2019arXiv190512053H}
N.~{Hunter-Jones}, {\it {Unitary designs from statistical mechanics in random
  quantum circuits}},  {\em arXiv e-prints} (May, 2019) arXiv:1905.12053,
  [\href{http://arxiv.org/abs/1905.12053}{{\tt arXiv:1905.12053}}].

\bibitem{2020JHEP...01..031K}
J.~{Kudler-Flam}, M.~{Nozaki}, S.~{Ryu}, and M.~T. {Tan}, {\it {Quantum vs.
  classical information: operator negativity as a probe of scrambling}},  {\em
  Journal of High Energy Physics} {\bf 2020} (Jan., 2020) 31,
  [\href{http://arxiv.org/abs/1906.07639}{{\tt arXiv:1906.07639}}].

\bibitem{2019JHEP...12..020W}
H.~{Wang} and T.~{Zhou}, {\it {Barrier from chaos: operator entanglement
  dynamics of the reduced density matrix}},  {\em Journal of High Energy
  Physics} {\bf 2019} (Dec., 2019) 20,
  [\href{http://arxiv.org/abs/1907.09581}{{\tt arXiv:1907.09581}}].

\bibitem{2019arXiv191208918L}
H.~{Liu} and S.~{Vardhan}, {\it {Void formation in operator growth,
  entanglement, and unitarity}},  {\em arXiv e-prints} (Dec., 2019)
  arXiv:1912.08918, [\href{http://arxiv.org/abs/1912.08918}{{\tt
  arXiv:1912.08918}}].

\bibitem{2020JHEP...04..074K}
J.~{Kudler-Flam}, Y.~{Kusuki}, and S.~{Ryu}, {\it {Correlation measures and the
  entanglement wedge cross-section after quantum quenches in two-dimensional
  conformal field theories}},  {\em Journal of High Energy Physics} {\bf 2020}
  (Apr., 2020) 74, [\href{http://arxiv.org/abs/2001.05501}{{\tt
  arXiv:2001.05501}}].

\bibitem{2020arXiv200514243K}
J.~{Kudler-Flam}, M.~{Nozaki}, S.~{Ryu}, and M.~{Tian Tan}, {\it {Entanglement
  of Local Operators and the Butterfly Effect}},  {\em arXiv e-prints} (May,
  2020) arXiv:2005.14243, [\href{http://arxiv.org/abs/2005.14243}{{\tt
  arXiv:2005.14243}}].

\bibitem{1991PhRvA..43.2046D}
J.~M. {Deutsch}, {\it {Quantum statistical mechanics in a closed system}},
  {\em \pra} {\bf 43} (Feb., 1991) 2046--2049.

\bibitem{1994PhRvE..50..888S}
M.~{Srednicki}, {\it {Chaos and quantum thermalization}},  {\em \pre} {\bf 50}
  (Aug., 1994) 888--901, [\href{http://arxiv.org/abs/cond-mat/9403051}{{\tt
  cond-mat/9403051}}].

\bibitem{2018RPPh...81h2001D}
J.~M. {Deutsch}, {\it {Eigenstate thermalization hypothesis}},  {\em Reports on
  Progress in Physics} {\bf 81} (Aug., 2018) 082001,
  [\href{http://arxiv.org/abs/1805.01616}{{\tt arXiv:1805.01616}}].

\bibitem{2018PhRvX...8b1026G}
J.~R. {Garrison} and T.~{Grover}, {\it {Does a Single Eigenstate Encode the
  Full Hamiltonian?}},  {\em Physical Review X} {\bf 8} (Apr., 2018) 021026,
  [\href{http://arxiv.org/abs/1503.00729}{{\tt arXiv:1503.00729}}].

\bibitem{2016arXiv161000302L}
N.~{Lashkari}, A.~{Dymarsky}, and H.~{Liu}, {\it {Eigenstate Thermalization
  Hypothesis in Conformal Field Theory}},  {\em arXiv e-prints} (Oct., 2016)
  arXiv:1610.00302, [\href{http://arxiv.org/abs/1610.00302}{{\tt
  arXiv:1610.00302}}].

\bibitem{2020JHEP...11..007D}
X.~{Dong} and H.~{Wang}, {\it {Enhanced corrections near holographic
  entanglement transitions: a chaotic case study}},  {\em Journal of High
  Energy Physics} {\bf 2020} (Nov., 2020) 7,
  [\href{http://arxiv.org/abs/2006.10051}{{\tt arXiv:2006.10051}}].

\bibitem{2013JHEP...05..014H}
T.~{Hartman} and J.~{Maldacena}, {\it {Time evolution of entanglement entropy
  from black hole interiors}},  {\em Journal of High Energy Physics} {\bf 2013}
  (May, 2013) 14, [\href{http://arxiv.org/abs/1303.1080}{{\tt
  arXiv:1303.1080}}].

\bibitem{2017arXiv170702325K}
I.~{Kourkoulou} and J.~{Maldacena}, {\it {Pure states in the SYK model and
  nearly-$AdS_2$ gravity}},  {\em arXiv e-prints} (July, 2017)
  arXiv:1707.02325, [\href{http://arxiv.org/abs/1707.02325}{{\tt
  arXiv:1707.02325}}].

\bibitem{2018arXiv180304434A}
A.~{Almheiri}, A.~{Mousatov}, and M.~{Shyani}, {\it {Escaping the Interiors of
  Pure Boundary-State Black Holes}},  {\em arXiv e-prints} (Mar., 2018)
  arXiv:1803.04434, [\href{http://arxiv.org/abs/1803.04434}{{\tt
  arXiv:1803.04434}}].

\bibitem{2019JHEP...07..065C}
S.~{Cooper}, M.~{Rozali}, B.~{Swingle}, M.~{Van Raamsdonk}, C.~{Waddell}, and
  D.~{Wakeham}, {\it {Black hole microstate cosmology}},  {\em Journal of High
  Energy Physics} {\bf 2019} (July, 2019) 65,
  [\href{http://arxiv.org/abs/1810.10601}{{\tt arXiv:1810.10601}}].

\bibitem{2021arXiv210306893M}
M.~{Miyaji}, T.~{Takayanagi}, and T.~{Ugajin}, {\it {Spectrum of End of the
  World Branes in Holographic BCFTs}},  {\em arXiv e-prints} (Mar., 2021)
  arXiv:2103.06893, [\href{http://arxiv.org/abs/2103.06893}{{\tt
  arXiv:2103.06893}}].

\bibitem{2017arXiv170204924H}
S.~{Hollands} and K.~{Sanders}, {\it {Entanglement measures and their
  properties in quantum field theory}},  {\em arXiv e-prints} (Feb., 2017)
  arXiv:1702.04924, [\href{http://arxiv.org/abs/1702.04924}{{\tt
  arXiv:1702.04924}}].

\bibitem{2018arXiv180304993W}
E.~{Witten}, {\it {Notes on Some Entanglement Properties of Quantum Field
  Theory}},  {\em arXiv e-prints} (Mar., 2018) arXiv:1803.04993,
  [\href{http://arxiv.org/abs/1803.04993}{{\tt arXiv:1803.04993}}].

\bibitem{2017PhRvL.119v0603V}
L.~{Vidmar} and M.~{Rigol}, {\it {Entanglement Entropy of Eigenstates of
  Quantum Chaotic Hamiltonians}},  {\em \prl} {\bf 119} (Dec., 2017) 220603,
  [\href{http://arxiv.org/abs/1708.08453}{{\tt arXiv:1708.08453}}].

\bibitem{2020JHEP...12..084M}
D.~{Marolf}, S.~{Wang}, and Z.~{Wang}, {\it {Probing phase transitions of
  holographic entanglement entropy with fixed area states}},  {\em Journal of
  High Energy Physics} {\bf 2020} (Dec., 2020) 84,
  [\href{http://arxiv.org/abs/2006.10089}{{\tt arXiv:2006.10089}}].

\bibitem{nica_speicher_2006}
A.~Nica and R.~Speicher, {\em Lectures on the Combinatorics of Free
  Probability}.
\newblock London Mathematical Society Lecture Note Series. Cambridge University
  Press, 2006.

\bibitem{MingoSpeicher_2017}
J.~A. Mingo and R.~Speicher, {\em Free Probability and Random Matrices}.
\newblock Springer, 2017.

\bibitem{2015PhRvE..92a2121M}
W.~{M{\l}otkowski}, M.~A. {Nowak}, K.~A. {Penson}, and K.~{{\.Z}yczkowski},
  {\it {Spectral density of generalized Wishart matrices and free
  multiplicative convolution}},  {\em \pre} {\bf 92} (July, 2015) 012121,
  [\href{http://arxiv.org/abs/1407.1282}{{\tt arXiv:1407.1282}}].

\bibitem{2011PhRvE..83f1118P}
K.~A. {Penson} and K.~{{\.Z}yczkowski}, {\it {Product of Ginibre matrices:
  Fuss-Catalan and Raney distributions}},  {\em \pre} {\bf 83} (June, 2011)
  061118, [\href{http://arxiv.org/abs/1103.3453}{{\tt arXiv:1103.3453}}].

\bibitem{2002math.ph...5010C}
B.~{Collins}, {\it {Moments and Cumulants of Polynomial random variables on
  unitary groups, the Itzykson-Zuber integral and free probability}},  {\em
  arXiv e-prints} (May, 2002) math--ph/0205010,
  [\href{http://arxiv.org/abs/math-ph/0205010}{{\tt math-ph/0205010}}].

\bibitem{2017PhRvD..96f6017B}
N.~{Bao} and H.~{Ooguri}, {\it {Distinguishability of black hole microstates}},
   {\em \prd} {\bf 96} (Sept., 2017) 066017,
  [\href{http://arxiv.org/abs/1705.07943}{{\tt arXiv:1705.07943}}].

\bibitem{2017JHEP...02..060S}
G.~{S{\'a}rosi} and T.~{Ugajin}, {\it {Relative entropy of excited states in
  conformal field theories of arbitrary dimensions}},  {\em Journal of High
  Energy Physics} {\bf 2017} (Feb., 2017) 60,
  [\href{http://arxiv.org/abs/1611.02959}{{\tt arXiv:1611.02959}}].

\bibitem{2018JHEP...07..179M}
B.~{Michel} and A.~{Puhm}, {\it {Corrections in the relative entropy of black
  hole microstates}},  {\em Journal of High Energy Physics} {\bf 2018} (July,
  2018) 179, [\href{http://arxiv.org/abs/1801.02615}{{\tt arXiv:1801.02615}}].

\bibitem{2018PhRvL.121y1603G}
W.-z. {Guo}, F.-L. {Lin}, and J.~{Zhang}, {\it {Distinguishing Black Hole
  Microstates using Holevo Information}},  {\em \prl} {\bf 121} (Dec., 2018)
  251603, [\href{http://arxiv.org/abs/1808.02873}{{\tt arXiv:1808.02873}}].

\bibitem{2020arXiv201012565A}
I.~{Akal}, {\it {Universality, intertwiners and black hole information}},  {\em
  arXiv e-prints} (Oct., 2020) arXiv:2010.12565,
  [\href{http://arxiv.org/abs/2010.12565}{{\tt arXiv:2010.12565}}].

\bibitem{2006PhRvL..96r1602R}
S.~{Ryu} and T.~{Takayanagi}, {\it {Holographic Derivation of Entanglement
  Entropy from the anti de Sitter Space/Conformal Field Theory
  Correspondence}},  {\em \prl} {\bf 96} (May, 2006) 181602,
  [\href{http://arxiv.org/abs/hep-th/0603001}{{\tt hep-th/0603001}}].

\bibitem{2006JHEP...08..045R}
S.~{Ryu} and T.~{Takayanagi}, {\it {Aspects of holographic entanglement
  entropy}},  {\em Journal of High Energy Physics} {\bf 2006} (Aug., 2006) 045,
  [\href{http://arxiv.org/abs/hep-th/0605073}{{\tt hep-th/0605073}}].

\bibitem{2013JHEP...08..092H}
V.~E. {Hubeny}, H.~{Maxfield}, M.~{Rangamani}, and E.~{Tonni}, {\it
  {Holographic entanglement plateaux}},  {\em Journal of High Energy Physics}
  {\bf 2013} (Aug., 2013) 92, [\href{http://arxiv.org/abs/1306.4004}{{\tt
  arXiv:1306.4004}}].

\end{thebibliography}
\providecommand{\href}[2]{#2}\begingroup\raggedright\endgroup

\end{document}